\documentclass[reprint,amsmath,amssymb,aps,superscriptaddress,nofootinbib,twocolumn,10pt]{revtex4-1}

\usepackage[colorlinks,linkcolor=blue,citecolor=blue,urlcolor=blue]{hyperref}
\usepackage{graphicx}
\usepackage{dcolumn}
\usepackage{subfigure}
\usepackage{bm}
\usepackage{amsmath,amsfonts,amssymb,graphicx,multirow,dcolumn,bm,latexsym,soul,nicefrac}
\usepackage{acronym}
\usepackage{enumitem}
\usepackage{array} 
\usepackage{multirow}
\usepackage{appendix}
\usepackage{longtable}
\usepackage{footnote}

\newcommand{\TRC}{MOE Key Laboratory of TianQin Mission, TianQin Research Center for
Gravitational Physics \& School of Physics and Astronomy, Frontiers
Science Center for TianQin, Gravitational Wave Research Center of
CNSA, Sun Yat-sen University (Zhuhai Campus), Zhuhai 519082, China.}

\newacro{DWD}{double white dwarf}
\newacro{GR}{general relativity}
\newacro{GW}{gravitational wave}
\newacro{GWs}{gravitational waves}
\newacro{CO}{compact object}
\newacro{PE}{parameter estimation}
\newacro{SNR}{signal-to-noise ratio}
\newacro{PN}{post Newtonian}
\newacro{FIM}{Fisher information matrix}
\newacro{BBH}{binary black hole}
\newacro{BH}{black hole}
\newacro{BNS}{binary neutron star}
\newacro{NS}{neutron star}
\newacro{VB}{verification binary}
\newacro{SGWB}{stochastic gravitational wave background}
\newacro{TDI}{time delay interferometry}
\newacro{PSD}{power spectral density}
\newacro{CDF}{cumulative distribution function}
\newacro{EM}{electromagnetic}
\newacro{CW}{continuous wave}
\newacro{CB}{compact binary}
\newacro{EMRI}{extreme mass ratio inspiral}
\newacro{CBC}{compact binary coalescence}

\begin{document}

\title{Constraining the Extra Polarization Modes of Gravitational Waves with Double White Dwarfs}

\author{Ning Xie}
\affiliation{\TRC}

\author{Jian-dong Zhang}
\email{zhangjd9@mail.sysu.edu.cn}
\affiliation{\TRC}

\author{Shun-Jia Huang}
\affiliation{\TRC}

\author{Yi-Ming Hu}
\affiliation{\TRC}

\author{Jianwei Mei}
\affiliation{\TRC}

\date{\today}
   
\begin{abstract}
The detection of \acp{GW} has opened a new window to test the theory of gravity in the strong field regime.
In \ac{GR}, \ac{GW} can only possess two tensor polarization modes,
which are known as the $+$ and $\times$ modes.
However, vector and scalar modes can exist in some modified theories of gravity,
and we can test the gravitational theories by probing these extra polarization modes.
As a space-borne \ac{GW} detector, TianQin will be launched in the 2030s,
and it is expected to observe plenty of GW signals,
including those from nearly ten thousand \acp{DWD} in our Galaxy.
This offers an excellent chance for testing the existence of extra polarization modes. 
In this article, we analyze the capability of detecting the extra polarization modes with \acp{DWD}.
For the extra modes, we consider both the leading order dipole radiation and the sub-leading quadrupole radiation.
We find that the capability of TianQin has very strong dependency on the source orientation.
We also analyze all the verification binaries with determined position and frequency,
and find that ZTF J1539 can give the best constraint on the extra polarization modes.
We have also considered the case of TianQin twin constellation, and the joint observation with LISA.
\end{abstract}

\maketitle

\section{\label{sec:1}Introduction\protect} 

As a prediction of general relativity (GR), gravitational waves (GWs) were first observed by aLIGO on September 14, 2015 \cite{2016LV}, and 90 \ac{GW} events have been reported by the LVK Collaboration up until the present day~\cite{2016LIGO,2019LIGO,2020LIGO,2021LIGOE,2021LIGO}. 
The analysis of GW150914 and all the GWTC events \cite{GW150914,GW170817,GWTC,2020GWTC2,GWTC3} shows that no deviation of \ac{GR} has been detected with current observations.
However, more sensitive observatories such as ET \cite{2010ET} and CE \cite{2016CE} on the Earth, or TianQin \cite{2016TQ} and LISA \cite{2017LISA} in space, will be built in the future.
So the deviation of \ac{GR} may be observed, or be constrained to a higher level with the data which is more accurate.
Most of the constraint is obtained by model dependent method, which means that we need to calculate the waveform of the sources in \ac{GR} or in the alternative theory, thus the inaccuracy of the waveform model may cause misinterpretation of the observed data.
However, the polarization mode, an intrinsic feature of \ac{GW}, can be used to test \ac{GR} in a model independent way.

In \ac{GR}, \acp{GW} have only two tensor polarization modes.
But in some modified theories of gravity, vector and scalar polarization modes can also exist, and there could be at most six polarization modes in the most general cases~\cite{1973LightmanLee,1973L2,Gong}.
For example, in the massless Scalar-Tensor theory, \acp{GW} have an additional scalar breathing mode.
In a general Scalar-Tensor theory \cite{WillLiv,2000Magg,2006Cap} and the $f(R)$ theory \cite{2016Kausar,2018Gong,2019Katsuragawa,2019Moretti}, \acp{GW} both have two scalar mode: the breathing mode and the longitude mode.
Moreover, in the Einstein-Æther theory \cite{2018Gong2}, TeVeS theory \cite{2010Sagi}, and bimetric theory \cite{2004Paula}, all the six polarization modes could exist.
Thus by testing whether these extra modes exist or not in the detected \ac{GW} signal, we can test \ac{GR}.

For the \ac{CBC} signals detected by ground-based detectors like LIGO and Virgo,
a Bayesian model selection method has been used by comparing the assumptions of
the \ac{GW} is constituted of pure vector or scalar mode against pure tensor mode,
and the obtained results still prefer the pure tensor mode assumption \cite{2017BHGW,GWTC}.
{
Among all the events, GW170817 gives the best constraint, with the Bayes factor larger than $10^{20}$~\cite{GW170817}.}
For the sources detected by at least three detectors,
the null-stream method \cite{1989Guersel,2005Wen,2007Wen,2006Chatterji,2018Hagihara,2019Hagihara,2019HagiharaY} can be used to test if there exist non-tensor modes.
This method has been used for the O2, O3a and O3b GW events \cite{2020ns,2021ns}, 
but the data is still consistent with the pure tensor mode hypothesis and other polarization hypotheses beyond \ac{GR} are relatively disfavoured. 
{Moreover, using the position information of GW170817 from \ac{EM} observation, Ref.~\cite{2019Hagihara} puts the upper limit on the amplitude of vector modes, and Ref.~\cite{2021Takeda} constrains the relative amplitude of scalar modes in a specific scalar-tensor mixed polarization model.}
However, in order to distinguish all the extra modes, we need a network with more detectors to get enough degree of freedom~\cite{2012Chatziioannou,2018Takeda}. 

For the \ac{CBC} signals shown in ground-based detectors  the duration will be very short, and the motion of the Earth can be neglected. 
Nevertheless, for \ac{CW} generated by fast spinning \ac{NS}, the monochromatic signal can last all over the observation period, which means the rotation of the Earth and the movement of the Earth around the Sun need to be considered in the signal modelling.
Then different polarization modes can be distinguished by the observational data of single detector \cite{2015Isi,2017Isi,2017Abbott,2020puls}.
However, \ac{CW} has not been observed  up to date, but this method is also applicable for space-borne detectors, since the signals in mHz band also last for a long time.

On the other hand, the extra polarization modes of \acp{GW} can also be scouted using pulsar timing arrays (PTAs) \cite{2011daSilvaAlves,2008Lee,2018Zhao,2019OBeirne,2020Boitier}.
North American Nanohertz Observatory for Gravitational Waves (NANOGrav) has searched for evidence of \ac{SGWB} in the 12.5 yr pulsar-timing data in \cite{NANOG}, 
while it is found that the result shows very strong prefer of a pure breathing mode instead of tensor mode \cite{2021chen}. 
The result from \cite{2021chen} is proven by the NANOGrav group, but the conclusion seems to be strongly related with single pulsar in catalogue \cite{2021NANOGrav}. 
They figure out that if one of the pulsar is removed from the data set, the tensor mode is preferred instead of the transverse scalar mode, and also the subject is still under debate. 

For a space-borne \ac{GW} observatory, the extra mode can be detected by considering the motion of detector, with similar method used for \ac{CW} by ground-based detector. 
The space-borne detector, such as TianQin \cite{2016TQ}, is expected to be sufficient to detect \ac{GW} from binary of supermassive black hole \cite{2016enrico,2019Feng,2019Wang}, \ac{EMRI} \cite{2017Babak,2020Fan}, galactic \ac{CB} \cite{2018Cornish2,2019Sesana,2020Liu,2020Huang} and cosmic background \cite{2016Cornish,2021Liang}. 
During the operation period, the detector can discover many non-transient signals, which can be used to distinguish different polarizations.
It has been studied that there exist $10^8$ \acp{CB} in our Galaxy, and most of them are double white dwarfs (DWDs). 
According to previous calculation, TianQin and LISA could detect about $10^4$ pairs of \acp{DWD}~\cite{2019Sesana,2020Huang}. 
Moreover, we have already observed some \acp{VB} which could be definitely detected by TianQin and LISA.
On the other hand, the \ac{GW} generated by these \acp{DWD} can be regarded as quasi-monochromatic signals~\cite{2019Na}, since the evolution of their frequencies is quite small.
So the waveform model for these \acp{DWD} will be very simple, and the consideration of the extra modes for them is also very easy.
So \acp{DWD} will be very ideal source to test \ac{GR} by testing the existence of the extra polarization modes. 

In this work, we focus on the capability of constraining the extra polarization modes of \ac{GW} with \acp{DWD}. 
The signals are assumed to be monochromatic, and the Doppler effect caused by the motion of the detector is considered in the response of the detectors. 
For the extra modes, we consider both the leading order dipole radiation, and the sub-leading order quadrupole radiation, separately.
For the dipole radiation, its amplitude is stronger, but its frequency is half of the tensor mode, so it causes some difficulty to identify the common origin with the tensor mode. 
However, for the quadrupole radiation, although its amplitude is weaker, its frequency is the same as the tensor mode.
We calculate the capability of constraining the extra modes for \acp{DWD} at different location, since the location governs the result.
We consider both TianQin and its twin constellation, we also consider the joint detection with LISA.
Furthermore, we also perform the analysis for \acp{VB} with certain positions \cite{2020Huang}. 

This paper is organized as following. In Sec.~\ref{sec:2}, we introduce some basic methods, including the waveform of extra polarization mode, the response of the GW detector, and the parameter estimation method.
In Sec.~\ref{sec:3}, we focus on the source location dependency of the \ac{PE} accuracy for different detector configurations{, and we also discuss the effect of the source inclination angle on the \ac{PE} accuracy of extra polarization amplitude. }
In Sec.~\ref{sec:4}, we analyze the capability of extra mode with the verification binaries.
Finally, we give a brief summary and discussion in Sec.~\ref{sec:5}.
In this paper, we adopt $G=c=1$.

\section{Model and Method}\label{sec:2}

For a general metric theory of gravity, the metric tensor has 6 propagation d.o.f.,
and so there exist 6 polarization modes.
More explicitly, there are two tensor modes, plus ($+$) and cross ($\times$),
 two vector modes, x (\textit{x}) and y ({\it{y}}),
and two scalar modes, breathing (\textit{b}) and longitudinal (\textit{l}).

In general relativity, only $+$ and $\times$ modes survive, and all the other modes vanish.
For this reason, if the existence of the extra polarization modes are confirmed in the detected \ac{GW} signal,
it means that \ac{GR} definitely needs to be modified. 

Although there already exist several waveforms including the extra polarizations for some specific theories \cite{1994Will,2012Chatziioannou,2016Sennett},
we keep our attention to the parameterized model in this work.
Due to the radiation of the extra modes, more energy is carried out compared to \ac{GR},   
so the evolution of the binary system is faster.
However, for most of the \acp{DWD} in our Galaxy, the evolution of their frequencies is very slow.
Therefore they can be regarded as monochromatic source, and the energy loss caused by extra modes can also be neglected.
Then the correction due to the existence of the extra modes can be characterised by their amplitude.

In general, the detected signal can be written as a linear combination of the responses of all polarizations: 
\begin{equation}\label{eq:signal}
h(t)=\sum_P F_P h_P(t),
\end{equation}
where $P$ stands for different polarizations.
$h_P(t)$ are the waveform for each polarization, and $F_P$ are the antenna pattern function \cite{2014gravi}, which described the response of the detector.
In the following part of this section, a more detailed introduction of the waveform and the response will be presented.

\subsection{Waveform}\label{sec:2A}

For a \ac{DWD} system, its \ac{GW} can be regarded as a nearly monochromatic wave,
since the evolution of its orbit is extremely small \cite{2019Na}.
For example, the \ac{GW} frequency $f$ of J0806 is about $6$mHz,
while the changing rate of the frequency $\dot{f}$ is about $10^{-16}$Hz/s \cite{2010Roelofs}.
Thus for a 5 year observation, the variation of $f$ is about $10^{-8}$Hz,
about the same order as the \ac{PE} accuracy of $f$.      
{Note that for \ac{DWD} with eccentricity, the \ac{GW} frequency can change with a faster rate due to the fact that the orbital energy of \ac{DWD} brought by the \ac{GW} radiation increases~\cite{1963PM}. However, most of the \acp{DWD} circularize during the common envelope phase, and we expect the binary system would have a low eccentricity~\cite{2001Nelemans}. In this situation, the eccentricity effect is insignificant and the monochromaticity approximation is still applicable. 
Therefore we will not consider the $\dot{f}$ term in the waveform in this work.
The eccentricity will also introduce higher order harmonics in the \ac{GW} waveform, but these will not change our method. 
}

In \ac{GR}, the leading order of \ac{GW} comes from the quadrupole radiation,
and the corresponding waveform for the $+$ and $\times$ modes in the source frame is:
\begin{align}
h_{+}(t)&=\mathcal{A} [(1+\cos^2{\iota})/2] \cos(\Phi),\\
h_{\times}(t)&=\mathcal{A} \cos\iota \sin(\Phi).
\end{align}
The amplitude $\mathcal{A}$ is given by
\begin{align}
\mathcal{A}=\frac{4\mathcal{M}}{D_L}(\pi\mathcal{M}f)^{2/3},
\end{align}
where $\mathcal{M}=(m_1m_2)^{3/5}/(m_2+m_2)^{1/5}$ is the chirp mass of the \ac{DWD} system
constitute by two white dwarfs with mass $m_1$ and $m_2$,
and $D_L$ is the luminosity distance of the system.
The inclination angle $\iota$ is the angle between the direction of angular momentum of the source $\hat{L}$ and the direction of GW propagation $\hat{k} $, 
i.e., $\cos\iota=\hat{k}\cdot\hat{L}$, 
where $\hat{k}$ is the unit vector along the direction of GW propagation. 
Since we focus on the detectability of space-borne gravitational wave observatory, it is convenient to discuss in the ecliptic coordinate system. 
For a \ac{GW} located at $(\theta_s, \phi_s)$ in the heliocentric ecliptic coordinate, the GW phase $\Phi$ is given by
\begin{equation}
\Phi=2\pi f t+2\pi f R \sin\theta_s \cos(2\pi f_m t-\phi_s+\phi_m)+\phi_0,
\end{equation}
where $f$ is the frequency of {\it tensor modes}, which is twice of the \ac{DWD}'s orbital frequency.
The initial GW phase $\phi_0$ is chosen to be 0 in this work, and we will not consider it in the Fisher information matrix analysis (see Sec.~\ref{sec:2C}).
The second term arises from the Doppler effect since the space-borne detector is moving around the Sun following the Earth.
Thus $\phi_m$ represents the initial position of detector,
$R=1~\text{AU}$ is the radius of the Earth orbit, and $f_m=1~\text{yr}^{-1}$ is the frequency of the Earth motion around the Sun.

To the leading order, the extra polarization modes are dipole radiation,
and the waveform can be written as~\cite{WillLiv}
\begin{align}
h_{x}(t)&=\mathcal{A}_v \sin(\Phi/2),\\
h_{y}(t)&=\mathcal{A}_v \cos\iota \cos(\Phi/2),\\
\label{hbd}h_{b}(t)&=\mathcal{A}_b \sin\iota \cos(\Phi/2),\\
\label{hld}h_{l}(t)&=\mathcal{A}_l \sin\iota \cos(\Phi/2).
\end{align}
Note that the frequency is half that of the tensor modes. 
The amplitudes of the two vector modes $x$ and $y$ have the same value, since they can be transformed between each other by a rotation.

However, since we can detect about ten-thousands of \acp{DWD},
it is difficult to distinguish whether it is the dipole radiation of an extra polarization mode,
or it is the tensor mode of another \ac{DWD} which coincidentally has the half frequency of the previous one. 
Also, the misinterpretation of polarization mode of a GW signal can lead to a incorrect source location, owing to the fact that each polarization has an unique response function. 
The wrong sky location information brings more difficulty to pinpoint the common origin of \acp{DWD}. 

For the sub-leading quadrupole radiation, the waveform can be written as:
\begin{align}
\label{hxq}	h_{x}(t)&=\mathcal{A}'_{v} \sin\iota\sin(\Phi),\\
	h_{y}(t)&=\mathcal{A}'_{v} \sin\iota\cos\iota \cos(\Phi),\\
	h_{b}(t)&=\mathcal{A}'_{b} \sin^2\iota \cos(\Phi),\\
\label{hlq}	h_{l}(t)&=\mathcal{A}'_{l} \sin^2\iota \cos(\Phi).
\end{align}
Here, the amplitudes are smaller than that for the leading dipole modes.

We have assumed that all modes are massless and propagate at the speed of light.
However, the non-transverse mode are in fact massive.
But due to the monochromaticity, the dispersion only causes a constant phase shift.
Thus the approximation will not influence the final result.
We will take the dispersion effect into consideration in our future work.

\subsection{Detector Response}\label{sec:2B}

Space-borne \ac{GW} detectors such as TianQin and LISA have equilateral triangle configuration, from which we can construct two independent orthogonal effective Michelson interferometers.
We will not consider the \ac{TDI} in this work, since it will not change the result significantly~\cite{2019Zhang}.

The antenna pattern function $F_P$ in Eq.~\eqref{eq:signal} for the two  Michelson interferometers $I$ and $II$ are
\begin{equation}\label{eq:Fp}
F_{{P}}^{I,II}(f)=D_{ij}^{I,II}e_{P,ij}.
\end{equation}
In the low frequency limit, the detector tensor $D_{ij}$ for an interferometer with equal arms is independent of frequency, and only depends on the position of three spacecrafts, thus it can be expressed as
\begin{align}
D^I&=\dfrac{1}{2}(\hat{r}_{12}\otimes\hat{r}_{12}-\hat{r}_{13}\otimes\hat{r}_{13}),\\
D^{II}&=\frac{1}{2\sqrt{3}}\left(\hat{r}_{12}\otimes\hat{r}_{12}+\hat{r}_{13}\otimes\hat{r}_{13}-2\hat{r}_{23}\otimes\hat{r}_{23}\right)
\end{align}
where $\hat r_{ij}$ is the unit vector from the $i$-th spacecraft to the $j$-th spacecraft.
Once the design scheme of space-based observatory is given, the ecliptic coordinates of each spacecraft can be derived from the motion of constellation. 
The analytical formulae of spacecraft coordinates can be found in \cite{2018Hu,2020Zhang} for TianQin and \cite{2004Rubbo} for LISA.

In order to express the six polarizations mathematically, we define a source coordinate $(\hat{p},\,\hat{q},\,\hat{k})$. 
The basis vectors $\hat{p}$ and $\hat{q}$ are defined as
\begin{align}
\hat{p}&=\dfrac{\hat{L}\times\hat{k}}{|{\hat{L}}\times{\hat{k}}|},\\
\hat{q}&={\hat{k}}\times{\hat{p}},
\end{align}
Therefore, the basis tensors can be written as 
\begin{equation}
\begin{aligned}
\textbf{e}_{+}&=\hat{p}\otimes\hat{p}-\hat{q}\otimes\hat{q},\\
\textbf{e}_{\times}&=\hat{p}\otimes\hat{q}+\hat{q}\otimes\hat{p},\\
\textbf{e}_{\textit{x}\,} &=\hat{p}\otimes\hat{k}+\hat{k}\otimes\hat{p},\\
\textbf{e}_{\textit{y}\,} &=\hat{q}\otimes\hat{k}+\hat{k}\otimes\hat{q},\\
\textbf{e}_{\textit{b}\,} &=\hat{p}\otimes\hat{p}+\hat{q}\otimes\hat{q},\\
\textbf{e}_{\textit{l}\ } &=\hat{k}\otimes\hat{k}.
\end{aligned}
\end{equation}
Here $\otimes$ denotes tensor product of two vectors.

Moreover, in another frame $(\hat{x},\hat{y},\hat{z})$, the polarization angle $\psi$ is determined as
\begin{equation}\label{eq:psi}
\psi=-\arctan\Big[\frac{\hat{L}\cdot\hat{z}-(\hat{k}\cdot\hat{L})(\hat{k}\cdot\hat{z})}
{\hat{k}\cdot(\hat{L}\times\hat{z})}\Big],
\end{equation}
Different from the inclination angle, the value of the polarization angle dependents on the choice of coordinate.
In the detector frame, $\hat{z}$ is along the direction of the detector' plane, and the polarization angle is noted as $\bar{\psi}$. 
In the ecliptic frame, $\hat{z}$ is along the direction the ecliptic pole, and the polarization angle is noted as $\psi_s$.
It's convenient to use the detector frame to describe the response, but the polarization angle will change due to the motion of the detector.
However, in the ecliptic frame, the polarization angle can be treated as a constant, since the precession of the \acp{DWD} is negligible.
So we will define the source parameter in the ecliptic frame, and transform to the detector frame to calculate the response.

In the detector frame, the antenna pattern function for each polarization is:
\begin{equation}
\begin{aligned}
F_+&= \frac{\sqrt{3}}{2} \left(\frac{1 + \cos^2\bar{\theta}}{2} \cos2\bar{\phi} \cos2\bar{\psi} - \cos\bar{\theta}\sin2\bar{\phi}  \sin2\bar{\psi} \right),\\
F_\times&=\frac{\sqrt{3}}{2} \left(\frac{1 + \cos^2\bar{\theta}}{2} \cos2\bar{\phi} \sin2\bar{\psi} + \cos\bar{\theta}\sin2\bar{\phi}  \cos2\bar{\psi} \right),\\
F_x&= -\frac{\sqrt{3}}{2} \sin\bar{\theta} \left(\cos\bar{\theta}\cos2\bar{\phi} \cos\bar{\psi} - \sin2\bar{\phi} \sin\bar{\psi} \right),\\
F_y&= -\frac{\sqrt{3}}{2} \sin\bar{\theta} \left(\cos\bar{\theta}\cos2\bar{\phi} \sin\bar{\psi} + \sin2\bar{\phi} \cos\bar{\psi} \right),\\
F_b&=-F_l= - \frac{\sqrt{3}}{4} \sin^2\bar{\theta}\cos2\bar{\phi}.
\end{aligned}
\end{equation}
The last equality means that the response of the breathing mode and the longitude mode are degenerate.
The degeneracy can be broken if we consider the differences in their waveform, such as the propagation speed due to their mass.
But we will not consider this feature in this work, and thus we can't distinguish the two scalar modes in this waveform model.
So we have
\begin{equation}
F_bh_b(t)+F_lh_l(t)=F_b(h_b(t)-h_l(t))=F_bh_s(t),
\end{equation}
the we can define the scalar mode amplitude as 
\begin{equation}
\mathcal{A}_s=\mathcal{A}_b-\mathcal{A}_l,~~~
\mathcal{A}'_{s}=\mathcal{A}'_{b}-\mathcal{A}'_{l}.
\end{equation}
If $\mathcal{A}_s,\mathcal{A}'_{s}\neq0$, it can be inferred that there must exist at least one kind of scalar mode. 

\subsection{Parameterization and Fisher Information Matrix}\label{sec:2C}
 
To evaluate the \ac{PE} accuracy, we adopt \acf{FIM} \cite{2007FIM} to deduce the uncertainty of the involved parameters. 
The definition of \ac{FIM} is 
\begin{equation}
\Gamma_{ij}=\left(\frac{\partial h}{\partial\lambda_i}\bigg|\frac{\partial h}{\partial\lambda_j}\right),
\end{equation}
where $\lambda _i$ is the $i$-th parameter and $(~|~)$ denotes the noise-weighted inner product \cite{1992Finn,1994Cutler},
which is defined as 
\begin{equation}
\left(h_1|h_2\right) =2\int_{0}^{\infty}\mathrm{d}f\dfrac{\tilde{h}_1(f)\tilde{h}_2^*(f)+\tilde{h}_1^*(f)\tilde{h}_2(f)}{S_n(f)},
\end{equation}
Here $\tilde{h}_i$ is the Fourier transformations of \(h_i\),
and \(S_N(f)\) is the one-sided \ac{PSD} for the instrument noise of the detector, 
whose explicit formula for TianQin and LISA can be found in \cite{2018Hu} and \cite{2018Cornish}, respectively. 
For the joint detection with multiple detectors,  the total FIM is the sum of each one. 
The \ac{PE} accuracy for each parameter is then given by
\begin{equation}
\Delta \lambda_i=\sqrt{\Gamma^{-1}_{ii}}.
\end{equation}
 
In this work, in order to characterise the magnitude of the extra polarization modes, 
we define
\begin{equation}
\alpha_{v,s}=\mathcal{A}_{v,s}/\mathcal{A},~~~
\alpha'_{v,s}=\mathcal{A}'_{v,s}/\mathcal{A}
\end{equation}
as the amplitudes of the extra polarization modes relative to the tensor mode. 
The parameters we choose are 
\begin{equation}
\bm\lambda=(\theta_s,\phi_s,\psi_s,\iota,f,\mathcal{A},\alpha_v,\alpha_s),
\end{equation}
or
\begin{equation}
\bm\lambda'=(\theta_s,\phi_s,\psi_s,\iota,f,\mathcal{A},\alpha'_{v},\alpha'_{s}).
\end{equation}
For verification binaries, the position of the source $(\theta_s,\phi_s)$ is fixed.
The frequency can also be determined by electromagnetic observation, but since its accuracy is worse than \ac{GW} observation, we will not fix the frequencies.

In this work, the capabilities of detectors to probe extra polarizations are described by \(\Delta\alpha_v\) and 
\(\Delta\alpha_s\) (or \(\Delta\alpha'_{v}\) and \(\Delta\alpha'_{s}\)).
The value for the amplitude of the extra modes is chosen as $0$ in the following \ac{FIM} analysis, since we assume \ac{GR} as the standard theory.
Thus the value of the \ac{PE} accuracy means that we can only probe the extra modes beyond these result, otherwise it will be confused with the statistic error caused by the noise.

\section{\label{sec:3}Constraint with general DWDs}

For the capability towards extra polarizations, the influence of the amplitude and frequency is almost trivial.
On the other hand, the response of different polarization modes and the correction due to the Doppler effect is significantly different for the sources at different positions.
So the most non-trivial parameters are the position of source, and thus the result will be influenced by the configuration of the detector.
TianQin is a geocentric orbit detector with fixed orientation, and LISA is a heliocentric orbit detector with its orientation moving around the ecliptic pole.
So their capability for the sources at different positions could be totally different.

\begin{figure}[h]
	\includegraphics[width=1\linewidth]{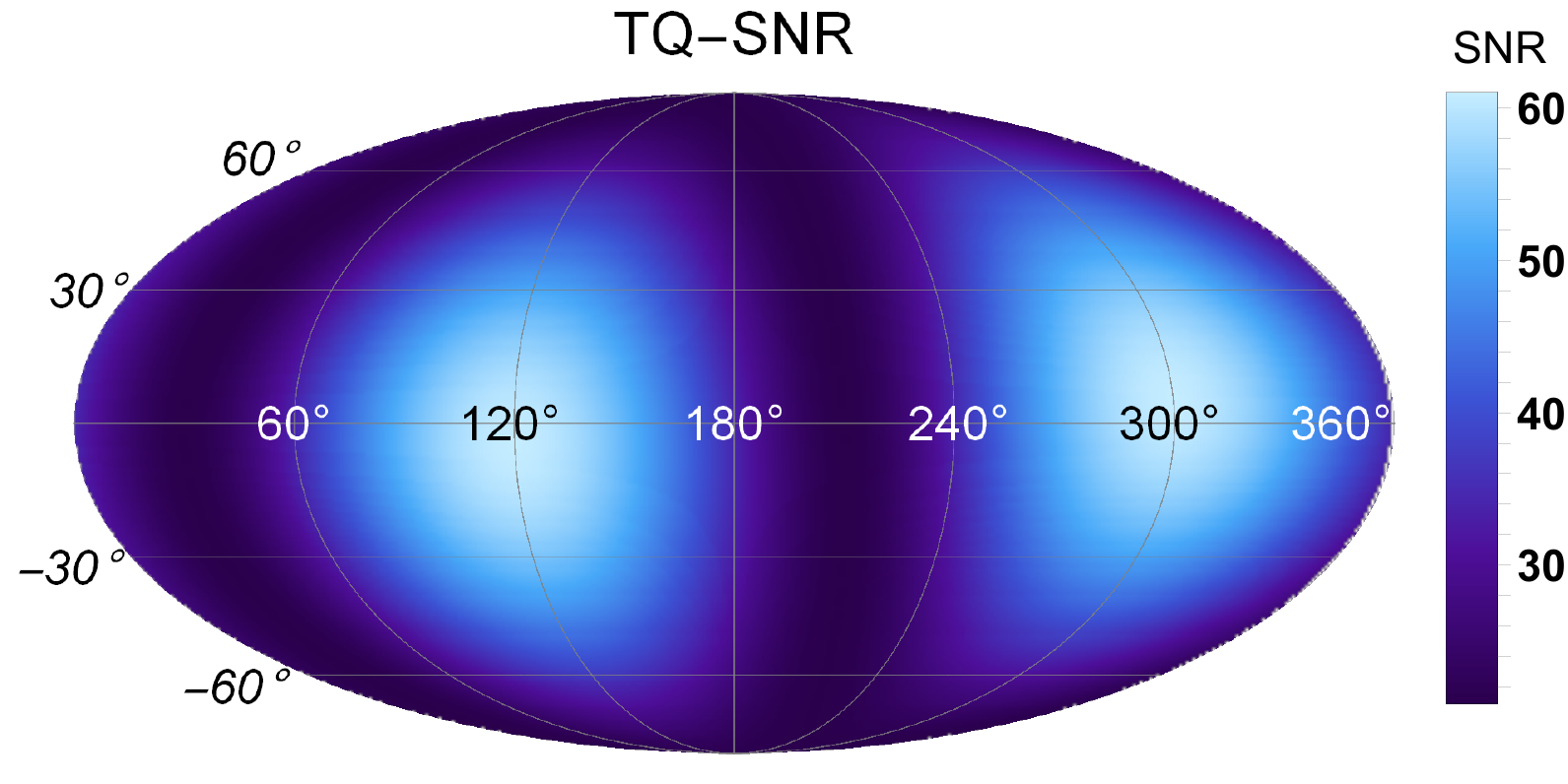}
	\includegraphics[width=1\linewidth]{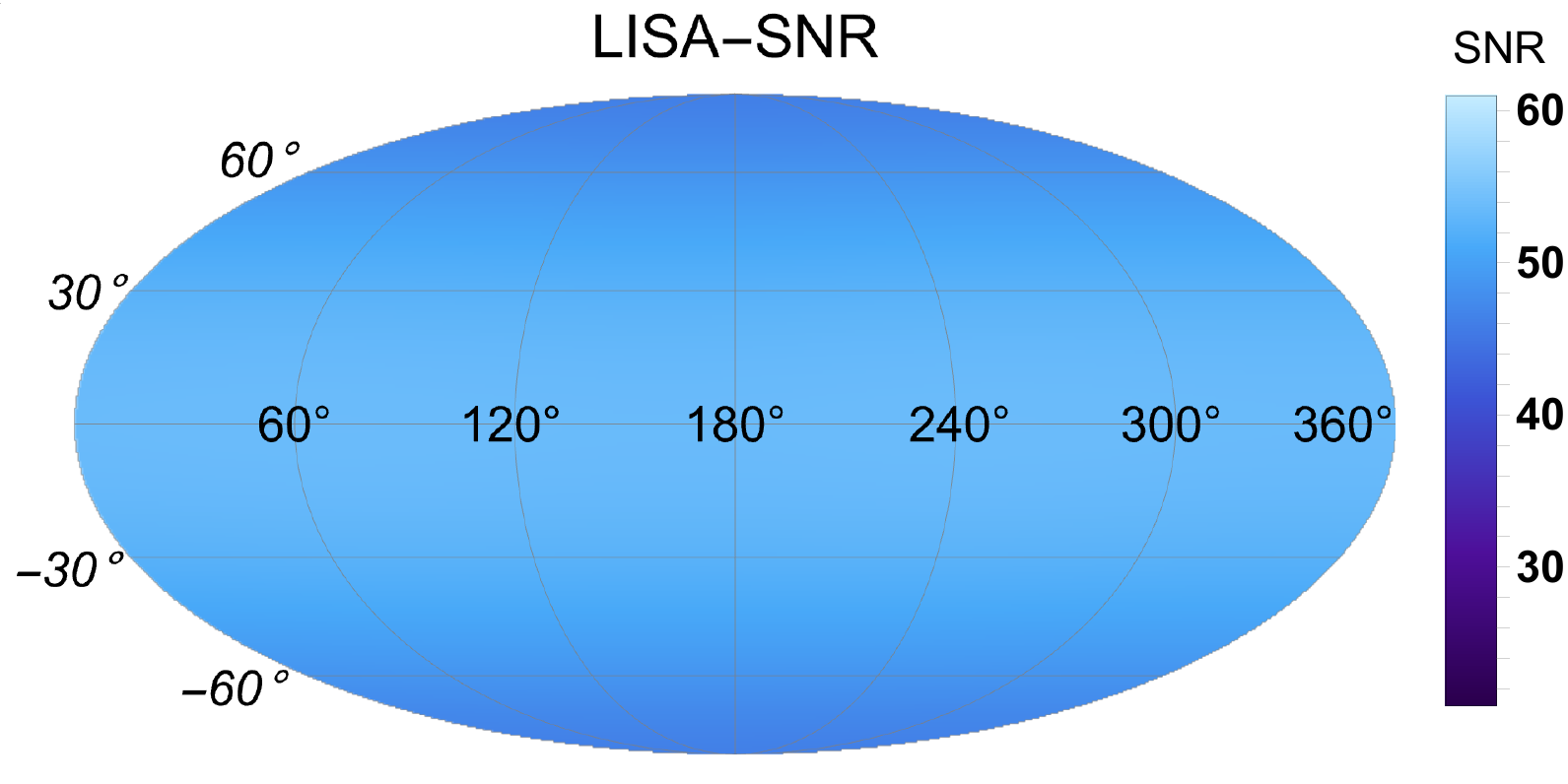}
	\caption{{ The distribution of \ac{SNR} on the celestial sphere in ecliptic coordinate for TianQin and LISA. }
			}
	\label{snr}
\end{figure} 

In the following calculation, the position of the source will vary all over the celestial sphere,
and the other parameters of the source are chosen as:
$f=0.02\text{Hz},~\mathcal{A}=10^{-22},~\iota=\pi/4,~\psi_s=\pi/4$.
The observation time is 1 year, and the results for $n$ years are just divided by $\sqrt{n}$ due to the monochromaticity of the source and the periodicity of the detector's orbit.
{
The \ac{SNR} for the source at different position is given in Fig.~\ref{snr}, with upper panel for TianQin and bottom panel for LISA. 
We can see that for the chosen parameters, the \ac{SNR} for the sources all over the sky is far beyond the threshold for detection, and thus satisfy the high \ac{SNR} condition in \ac{FIM} analysis. 
}

TianQin will point to the \ac{DWD} J0806 (the coordinate in ecliptic frame is $\theta_s=-4.7^\circ$ and $\phi_s=120.4^\circ$) and operate in the so-called ``3+3'' mode \cite{2016TQ}, which means that the detector will follow the “three months on and three months off” working scheme, and thus the effective observation time ratio is $50\%$. 
So, we will also consider the twin constellation, which has another three crafts move along the orbit orthogonal to the primary TianQin constellation, so it will work during the primary constellation is off.
We denote the two constellations as ``TQ'' and ``TQ II'', respectively. The TianQin twin constellation will be denoted as ``TQ I+II''. 
We will also calculate the capability of LISA and its joint detection with TianQin and the twin constellation.

\subsection{Constraint on the Dipole Radiation}\label{sec:3A}

In this subsection, we will only consider the leading order dipole radiation of the extra modes.

\begin{figure}[h]
	\includegraphics[width=1\linewidth]{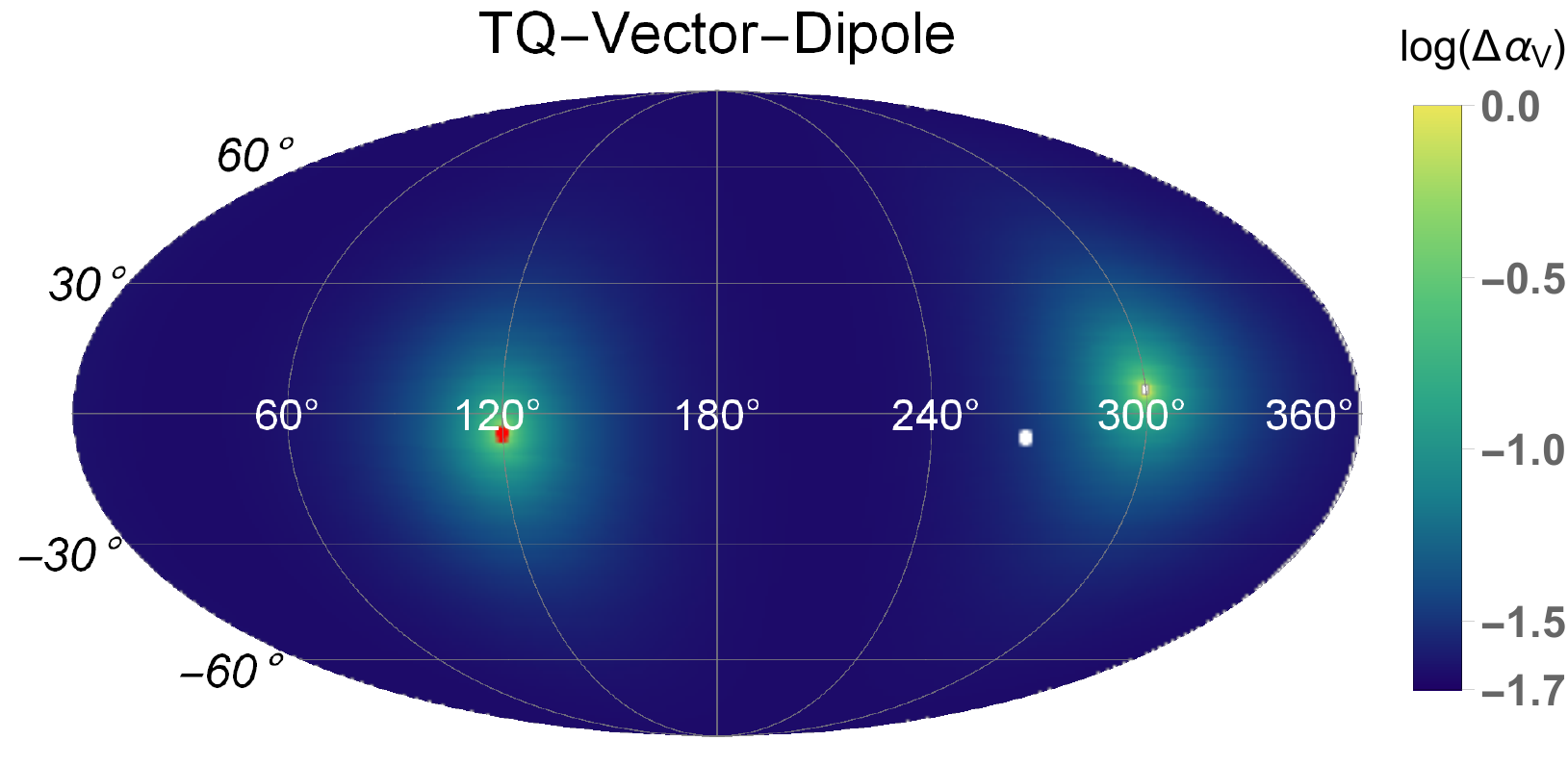}
	\includegraphics[width=1\linewidth]{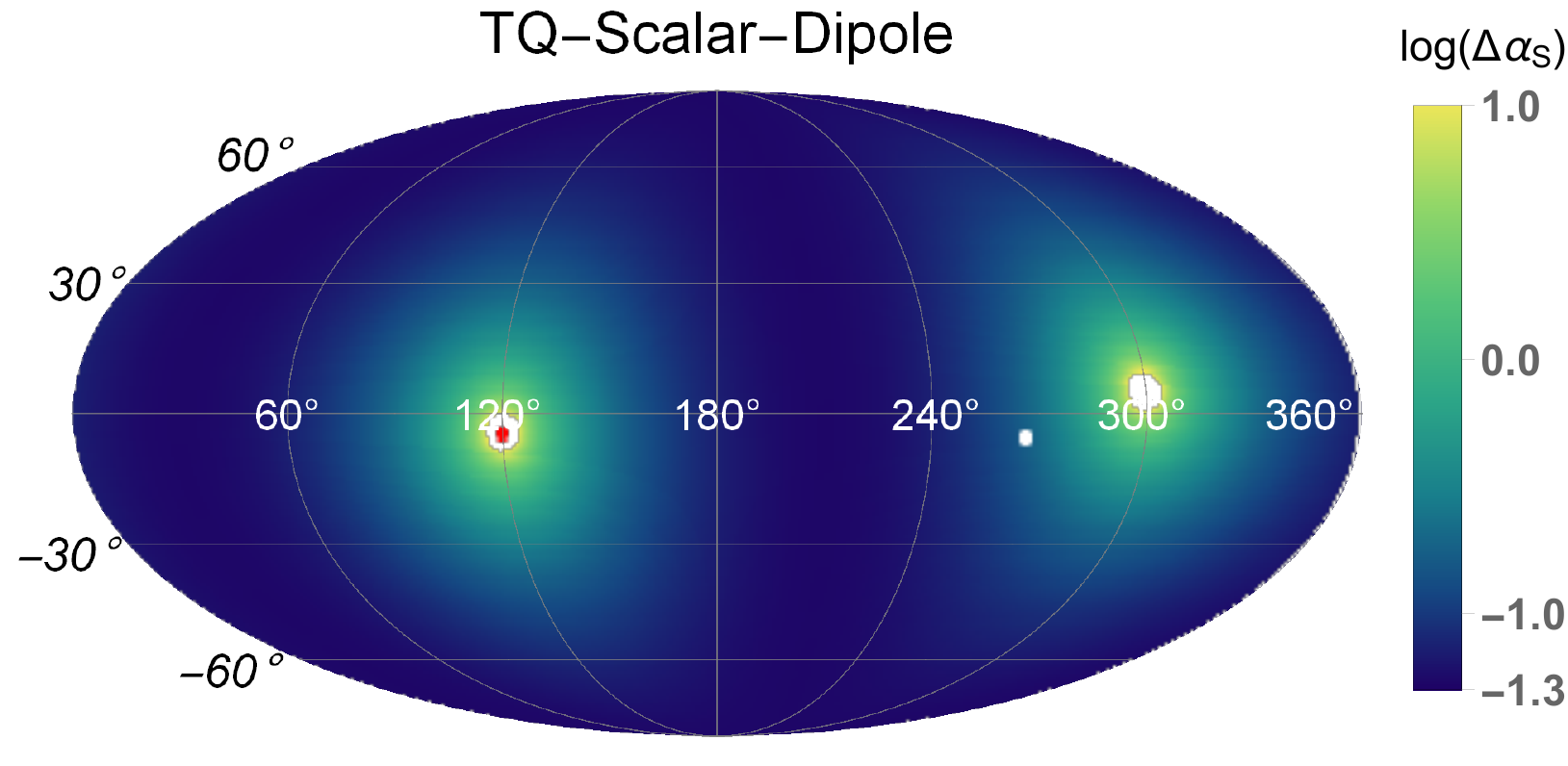}
	\caption{ The distribution of \(\Delta\alpha_v\) and \(\Delta\alpha_s\) on the celestial sphere in ecliptic coordinate for TianQin.
	The red dot is J0806, and the white dot is the galactic center.		}
	\label{dT}
\end{figure}

For TianQin, the result is shown in Fig.~\ref{dT}.
The result shows that the \ac{PE} accuracy of $\alpha_v$ and $\alpha_s$ are strongly dependent on the location of the source.
Especially, the result diverges for the direction of J0806 and its antipodal point.
This feature is caused by the fact that the antenna pattern $F_P$ will be 0 for the sources at $\bar{\theta}=0,\pi$, and thus in the response signal, the components for the extra modes will be 0.
However, for the other regions, the accuracy can be a few percent for vector and scalar mode,
and the capability for vector mode is better than scalar mode.

\begin{figure}[h]
	\includegraphics[width=1\linewidth]{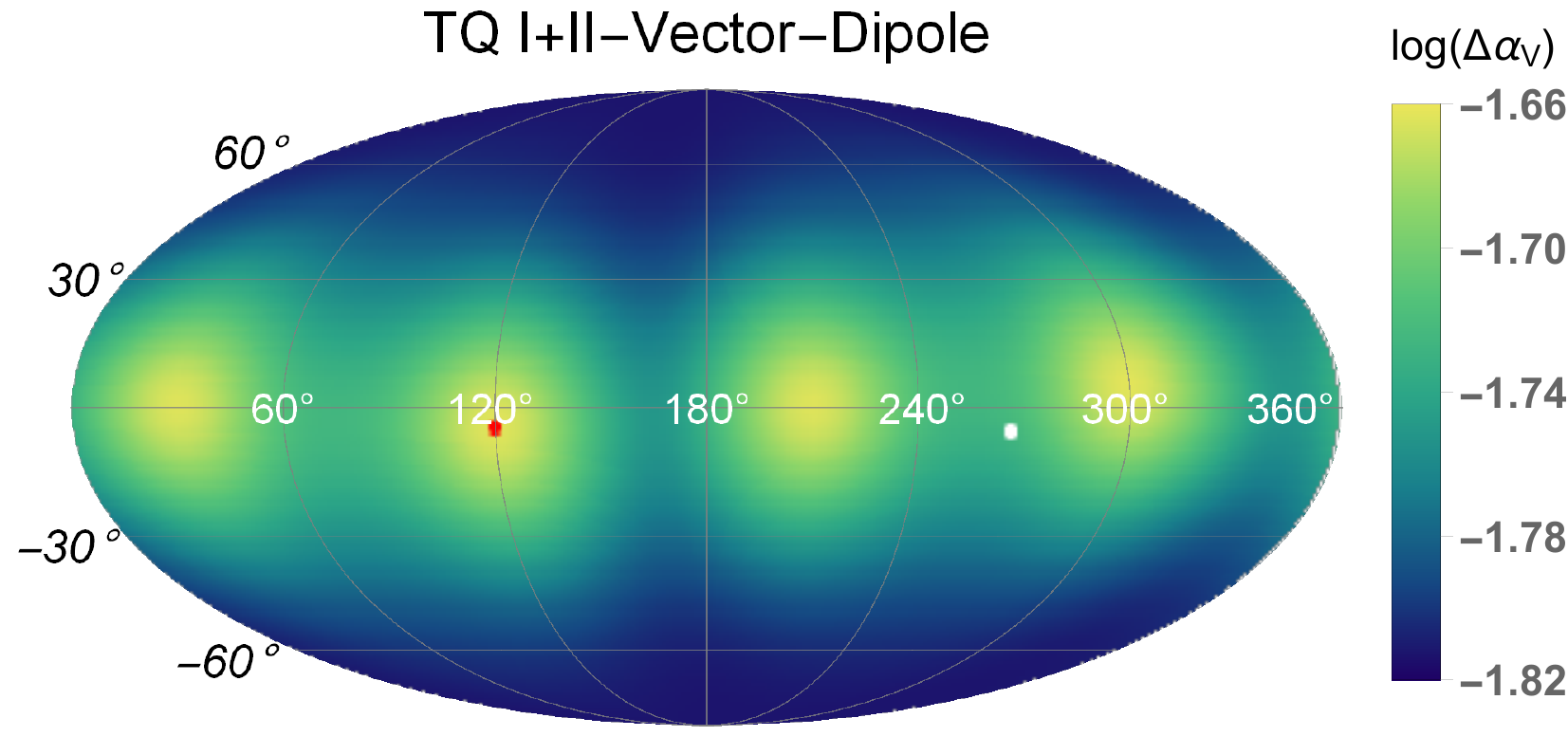}
	\includegraphics[width=1\linewidth]{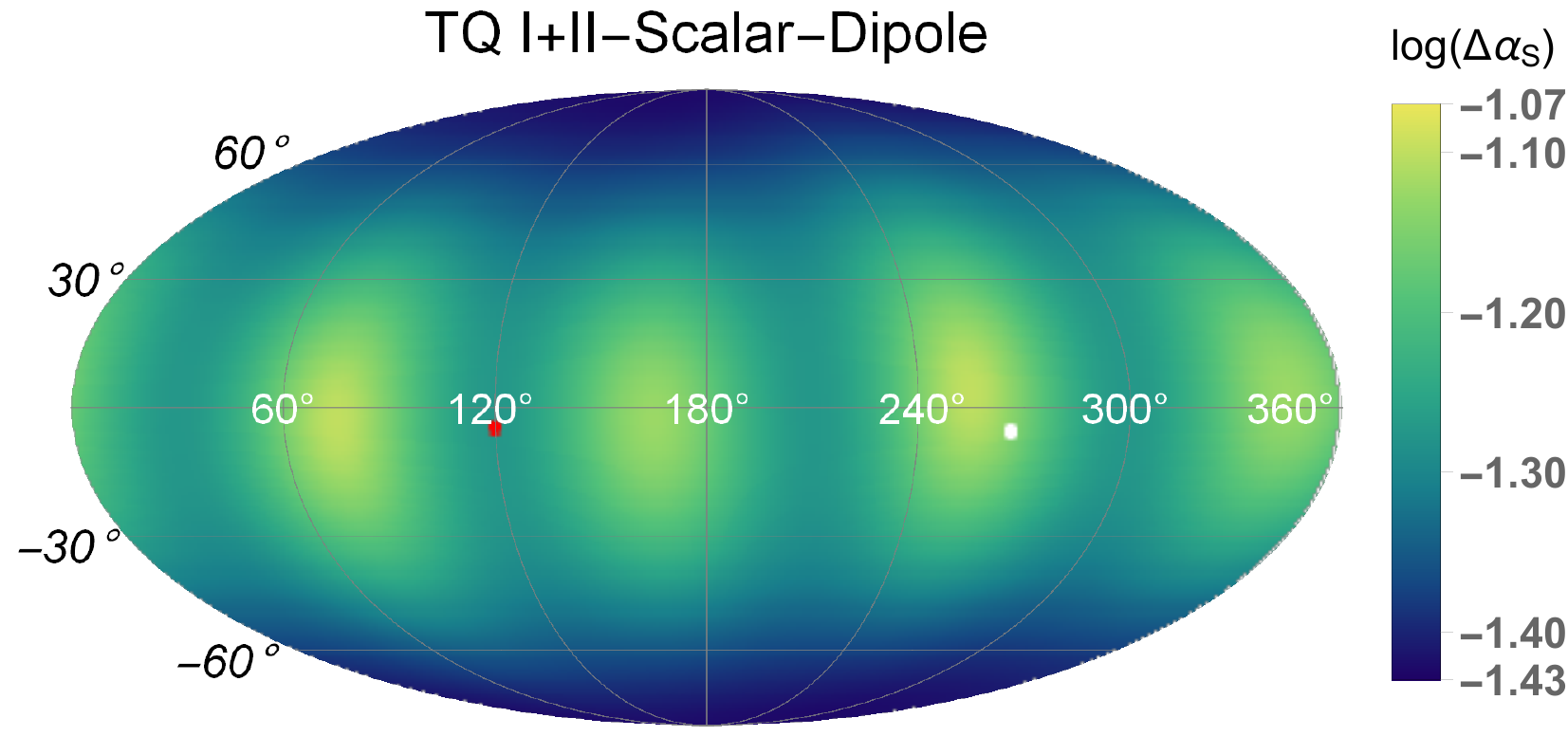}
	\caption{The distribution of \(\Delta\alpha_v\) and \(\Delta\alpha_s\) on the celestial sphere in ecliptic coordinate for the TianQin  twin constellation.
	}\label{dtT} 
\end{figure}

The TianQin twin constellation is useful to avoid the divergent. 
Thus the results are a few percent for both extra modes all over the sky, as plotted in Fig.~\ref{dtT}.
However, we can see that the contour plot will be different for vector and scalar modes.
For vector modes, the region near the orientations of the two constellations will be worse.
But for scalar modes, the region between the orientations will be worse.
This caused by the difference on the dependence of $\bar{\theta}$ for the response of the extra modes.
For vector modes, as $\bar{\theta}$ approaches to 0, the response approaches to 0 as $\sin\bar{\theta}$.
But the response for scalar modes will approaches to 0 as $\sin^2\bar{\theta}$.
So although the result for both modes will divergent at $\bar{\theta}=0,\pi$, but the region near these points will be larger for scalar modes, since the response will grow slower as the sources move away from the divergent points.

\begin{figure}[h]
	\includegraphics[width=1\linewidth]{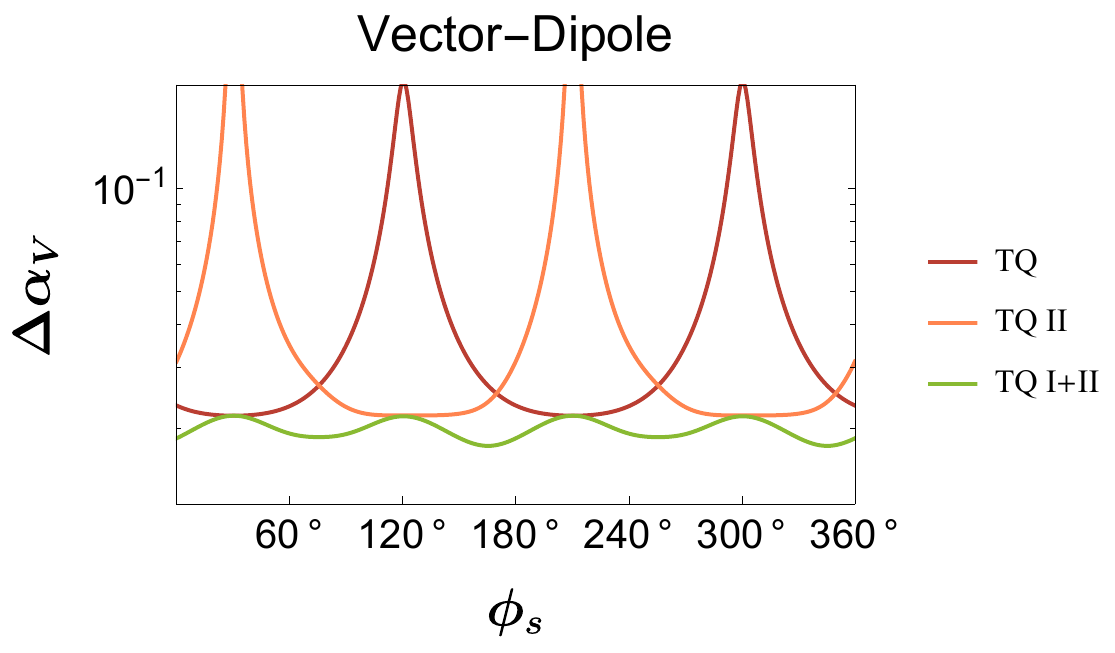}
	\includegraphics[width=1\linewidth]{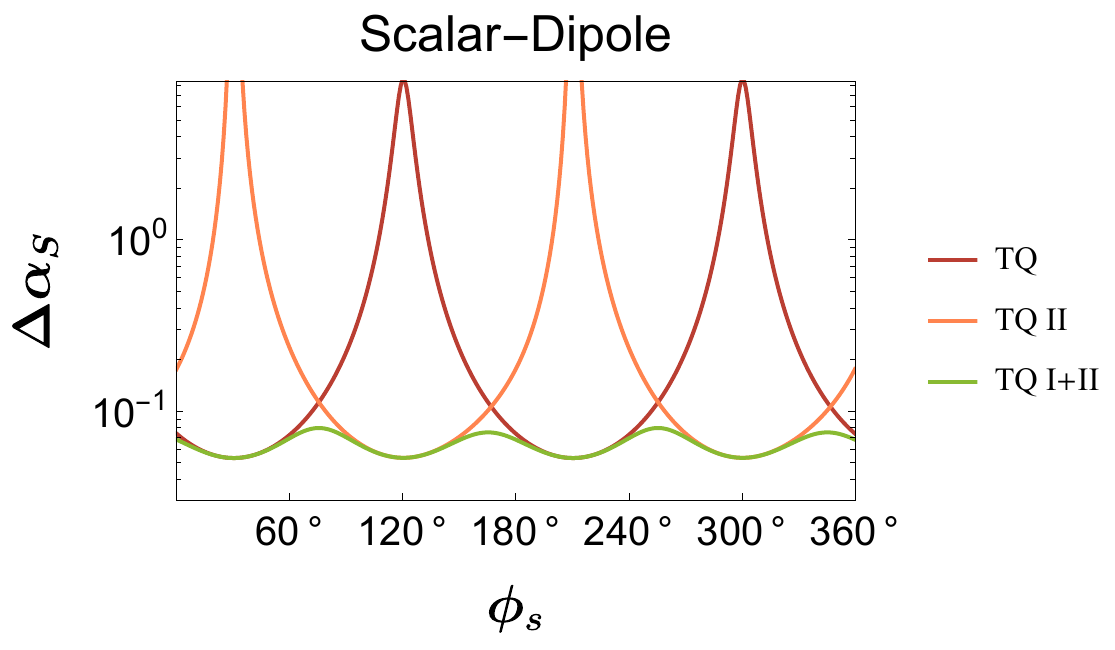}
	\caption{The distribution of \(\Delta\alpha_v\) and \(\Delta\alpha_s\) along the ecliptic plane for TQ (red line), TQ II (orange line) and TQ I+II (green line). Note that for TQ I+II, the better regions for the vector modes and the scalar modes located at different longitude. 
	}\label{dtTecliptic} 
\end{figure}

In Fig.~\ref{dtTecliptic}, we plot the accuracy of each mode for different detectors for the sources at the ecliptic plane.
We should notice that the divergent point for TQ II located just on the ecliptic plane, and for TQ the result will be worst for the points has the same longitude as the divergent point.
At the worst point for one detector, the accuracy will be the best for the other detector, and the result will be almost exact the value for the best one.
But for the intermediate region between the divergent points, the accuracy will be almost the same for both detectors, and the joint detection will have an improvement for about $\sqrt{2}$ times.
So for the vector modes, the \ac{PE} result will decrease faster, and although the accuracy for the intermediate points will be worse than the best point for each detector, an improvement of $\sqrt{2}$ times will make it better than the best value.
But for the scalar modes, the \ac{PE} result will decrease slower, and an improvement of $\sqrt{2}$ times is still worse than the best value.
Although the shape of the contours will be different, the results are still at the same order for the twin constellation all over the sky. 

\begin{figure}[h]
	\includegraphics[width=\linewidth]{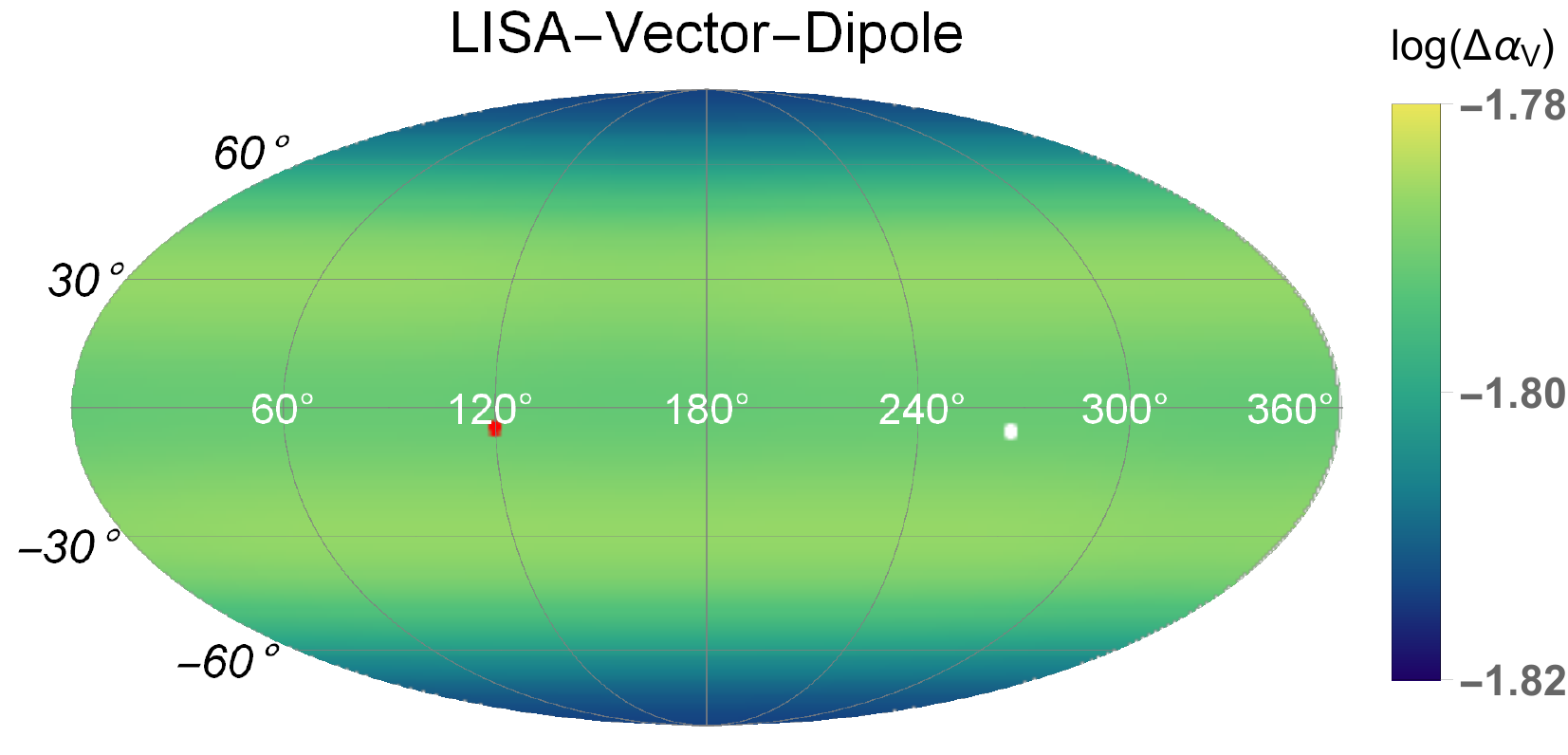}
	\quad
	\includegraphics[width=\linewidth]{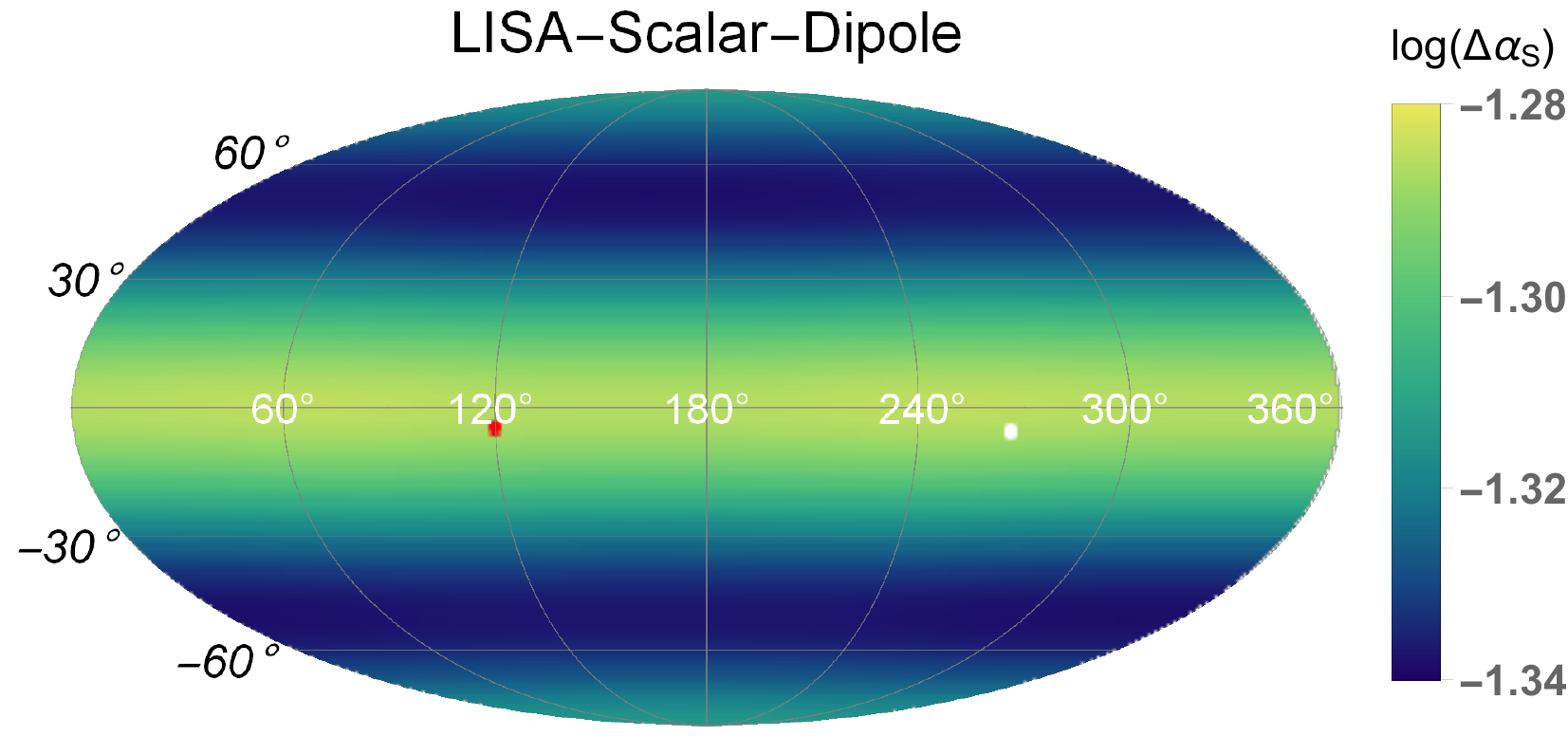}
	\caption{The distribution of \(\Delta\alpha_v\) and \(\Delta\alpha_s\) on the celestial sphere in ecliptic coordinate for LISA.
	}\label{dL}
\end{figure}

{For LISA, as showed in Fig.~\ref{dL}},
we can find that the accuracy is mainly determined by the latitude due to the periodic motion of LISA, and the result is about a few percent all over the sky.
Additionally, we find that the distribution is different for vector and scalar modes.
For the vector modes, the regions with better accuracy are located near the north and south pole.         
For the scalar modes, the accuracy will be worse for the region at the ecliptic plane and the poles. 
For the joint detection for LISA with TianQin or TianQin twin constellation, the accuracy will be improved, and the results are shown in Fig.~\ref{dTL} and Fig.~\ref{dtTL} at Appendix~\ref{figure}.

\begin{figure}[h]
	\includegraphics[width=1\linewidth]{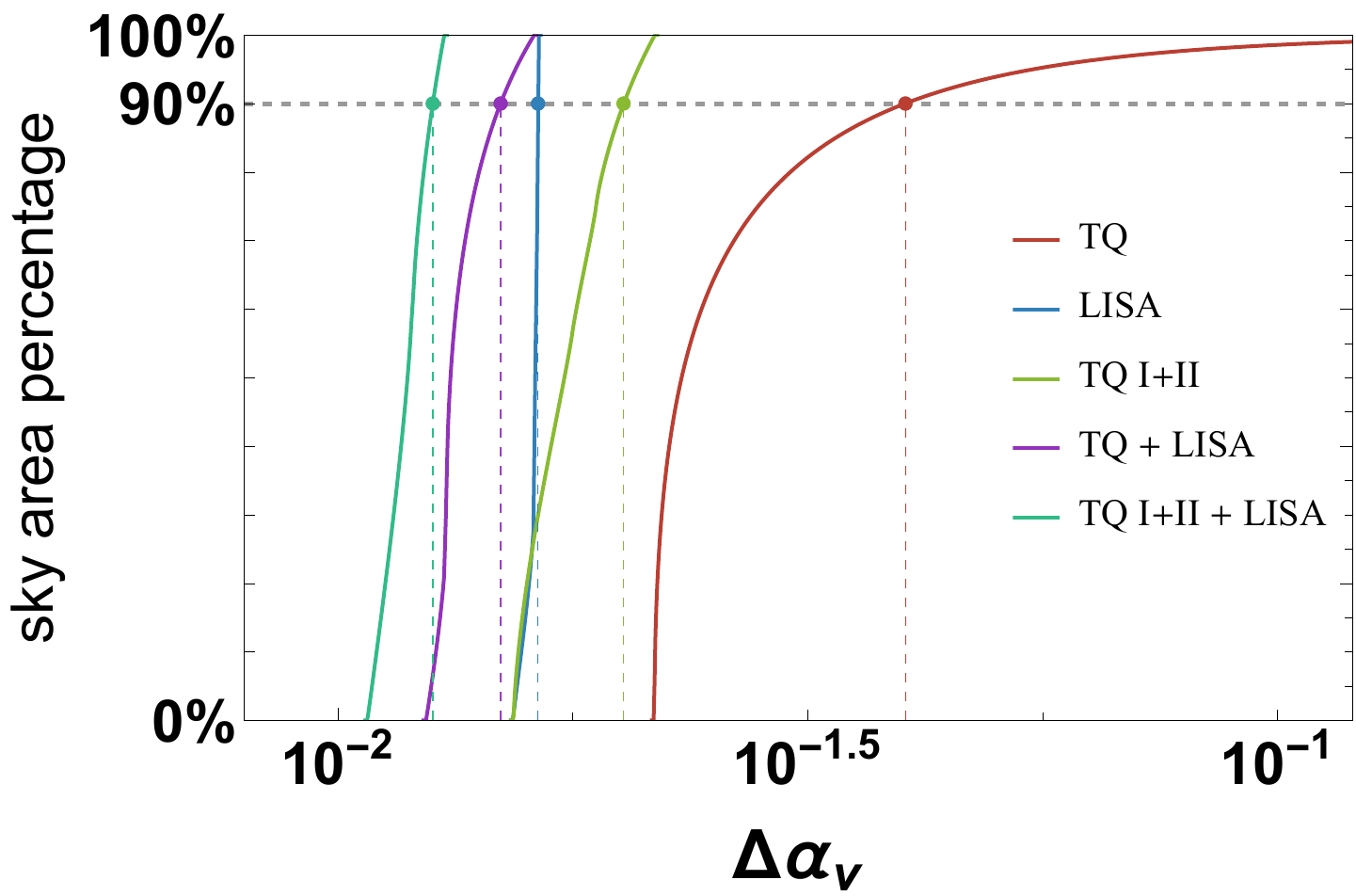}
	\includegraphics[width=1\linewidth]{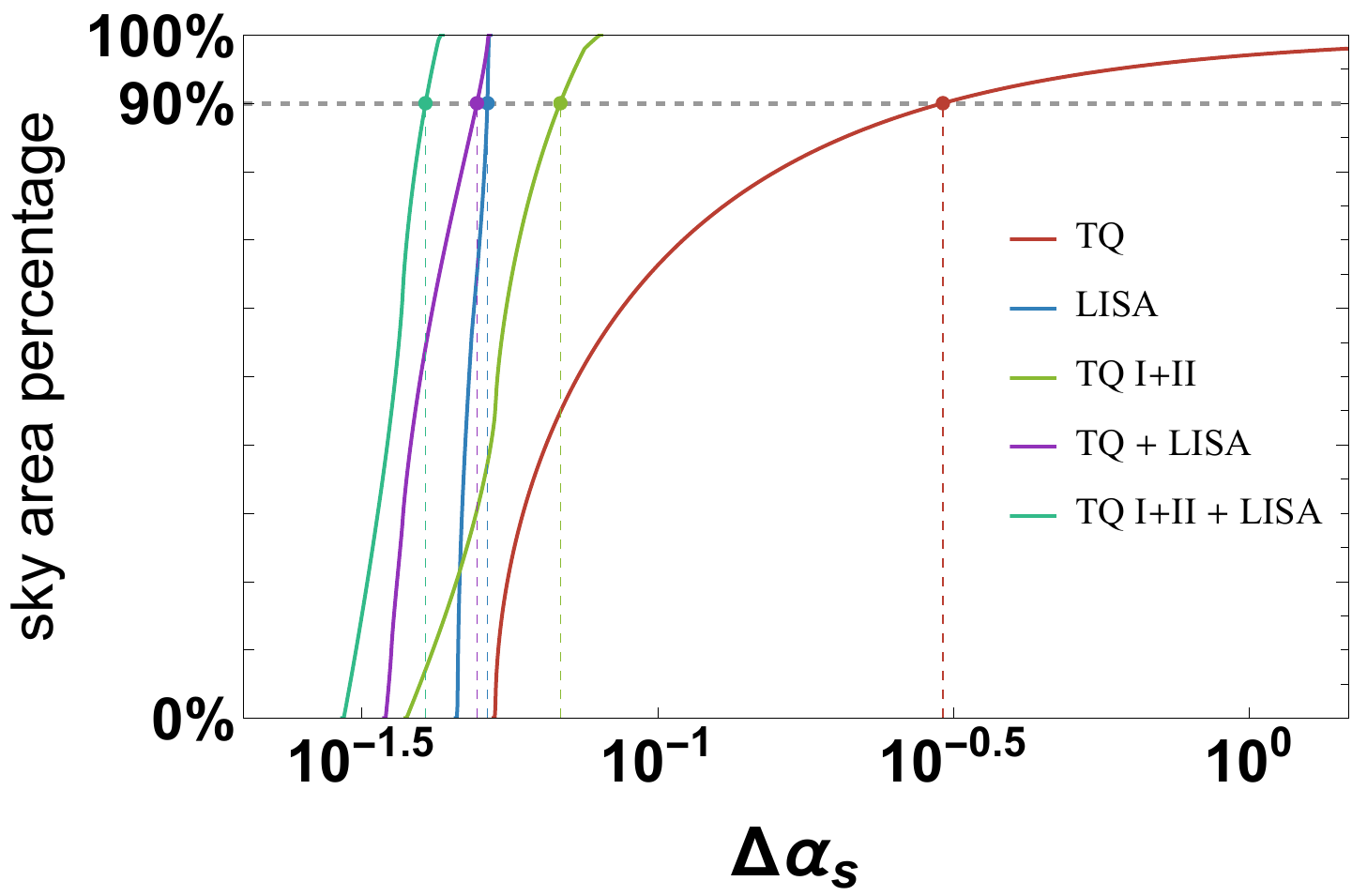}
	\caption{Top panel: the area ratio of sky region where the $\Delta\alpha_v$ always takes on a value less than 
	a certain quantity. Bottom panel: same as the former one but for $\Delta\alpha_s$. 
	The configurations considered include TQ, LISA, TQ I+II, TQ + LISA and TQ I+II + LISA.
	The dashed lines corresponding to the 90\% threshold.
	}\label{dc} 
\end{figure}

\begin{table}[h]
	\caption{The representative detecting accuracy for the 90\% level and best of $\alpha_v$ and $\alpha_s$ for each configuration. 
	}\label{tdc}
	\begin{ruledtabular}
		\begin{tabular}{lccccc}
			&
			\textrm{\footnotesize TQ}&
			\textrm{\footnotesize TQ I+II}&
			\textrm{\footnotesize LISA}&
			\textrm{\footnotesize TQ + LISA}&
			\textrm{\footnotesize TQ I+II + LISA}\\
			\colrule
			$\Delta\alpha_v(90\%)$ & 0.040 & 0.020 & 0.016 &0.015& 0.013\\
			$\Delta\alpha_v(\text{best})$ & 0.022 & 0.015 & 0.015 & 0.013 & 0.010\\
			
			$\Delta\alpha_s(90\%)$ & 0.304 & 0.069 & 0.052 &0.050& 0.041\\
			$\Delta\alpha_s(\text{best})$ & 0.054 & 0.040 & 0.048 & 0.036 & 0.030\\
		\end{tabular}
	\end{ruledtabular}
\end{table}

For a more intuitive exhibition, we also plot the \ac{CDF} of the accuracy distribution on the sky for each detector configuration in Fig. \ref{dc}.
The dashed line corresponds to the accuracy which can be achieved by 90\% area in celestial sphere.
The detailed results corresponding to this 90\% threshold and the best accuracy for each configuration are shown in Table~\ref{tdc}.
From these figures, we can find that for TianQin, the distribution of the accuracy will have a very wide range due to the existence of the divergent point, although the best value can be comparable with other configurations.
For all the other configurations, the distribution of the accuracy will be narrow, and 
Moreover, the accuracy for vector modes will be a few times better than the scalar modes in general.

\begin{figure}[!h]
	\includegraphics[width=1\linewidth]{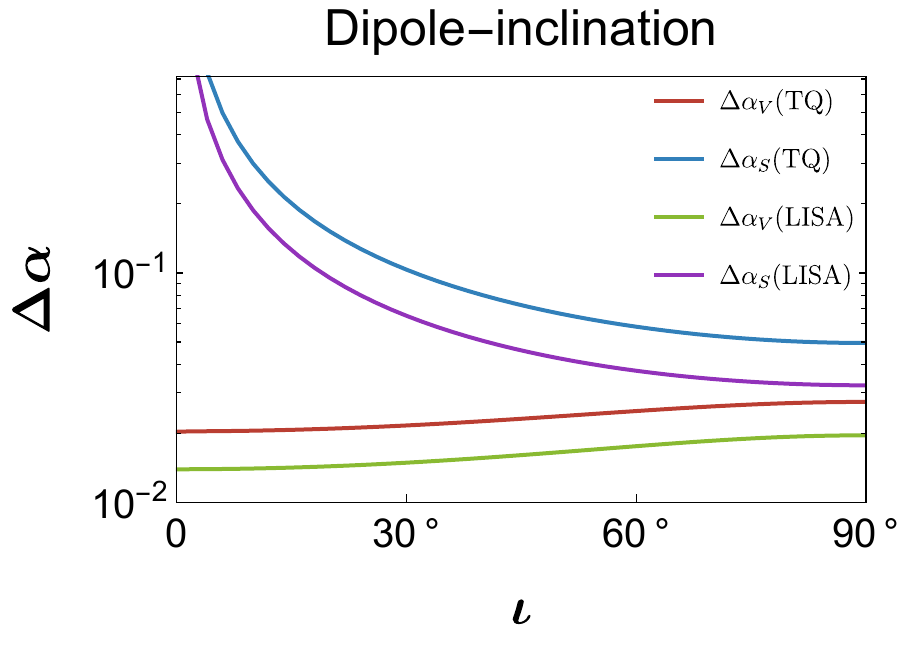}
	\caption{{ The distribution of \(\Delta\alpha_v\) and \(\Delta\alpha_s\) with different inclination angle for TianQin and LISA.}}
	\label{di} 
\end{figure}

{Despite the position of the source, the inclination angle will also have significant influence on the \ac{PE} accuracy for the amplitude of extra polarizations.
So here we consider the \ac{DWD} source with different inclination angle, while the source position is chosen as $\theta_s=30^\circ$ and $\phi_s=60^\circ$. 
The dependence on inclination angle $\iota$ for $\Delta\alpha_v$ and $\Delta\alpha_s$ are shown in Fig.~\ref{di}. 
If we choose another pair of position parameter, the result of $\Delta\alpha_v$ and $\Delta\alpha_s$ might have a moderate difference quantitatively, but the tendency of their dependence on $\iota$ will be similar. 
We find that $\Delta\alpha_s$ diverges when $\iota$ approaches to zero (face-on case) since the amplitude of scalar modes vanish for this case (cf. Eqs.~\eqref{hbd} and \eqref{hld}). 
Moreover, due to the different on the factor of $\iota$, the \ac{PE} of the amplitude for extra modes will reach its best value at different $\iota$, which is $\pi/2$ for scalar modes, and $0$ for vector modes. 
}

\subsection{\label{sec:3B}Constraint on the Quadrupole Radiation}

For the sub-leading quadrupole radiation, the calculations are almost the same,
except that we only consider these sub-leading term, and ignore the leading order dipole radiation.

\begin{figure}[h]
	\includegraphics[width=1\linewidth]{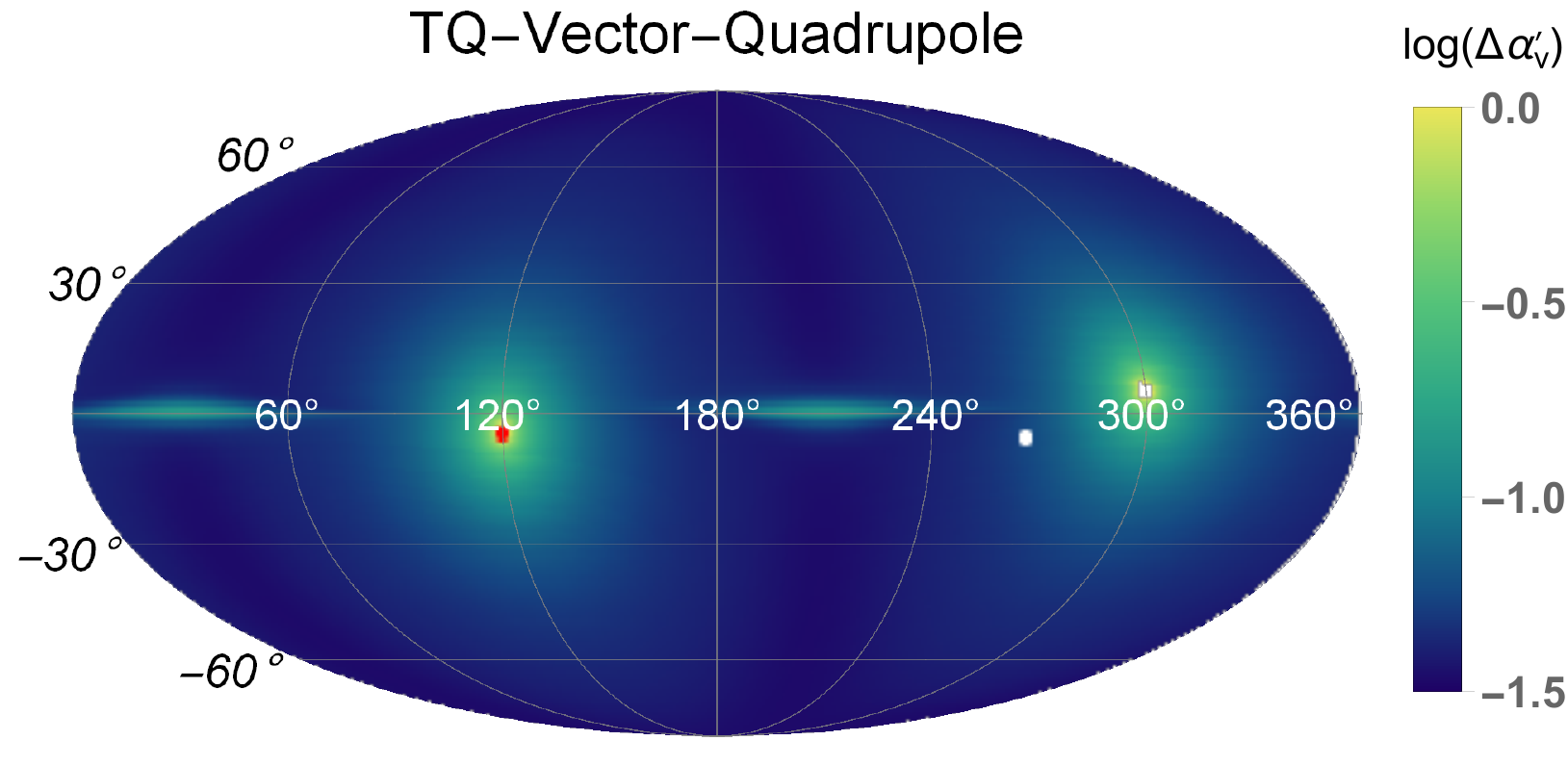}
	\includegraphics[width=1\linewidth]{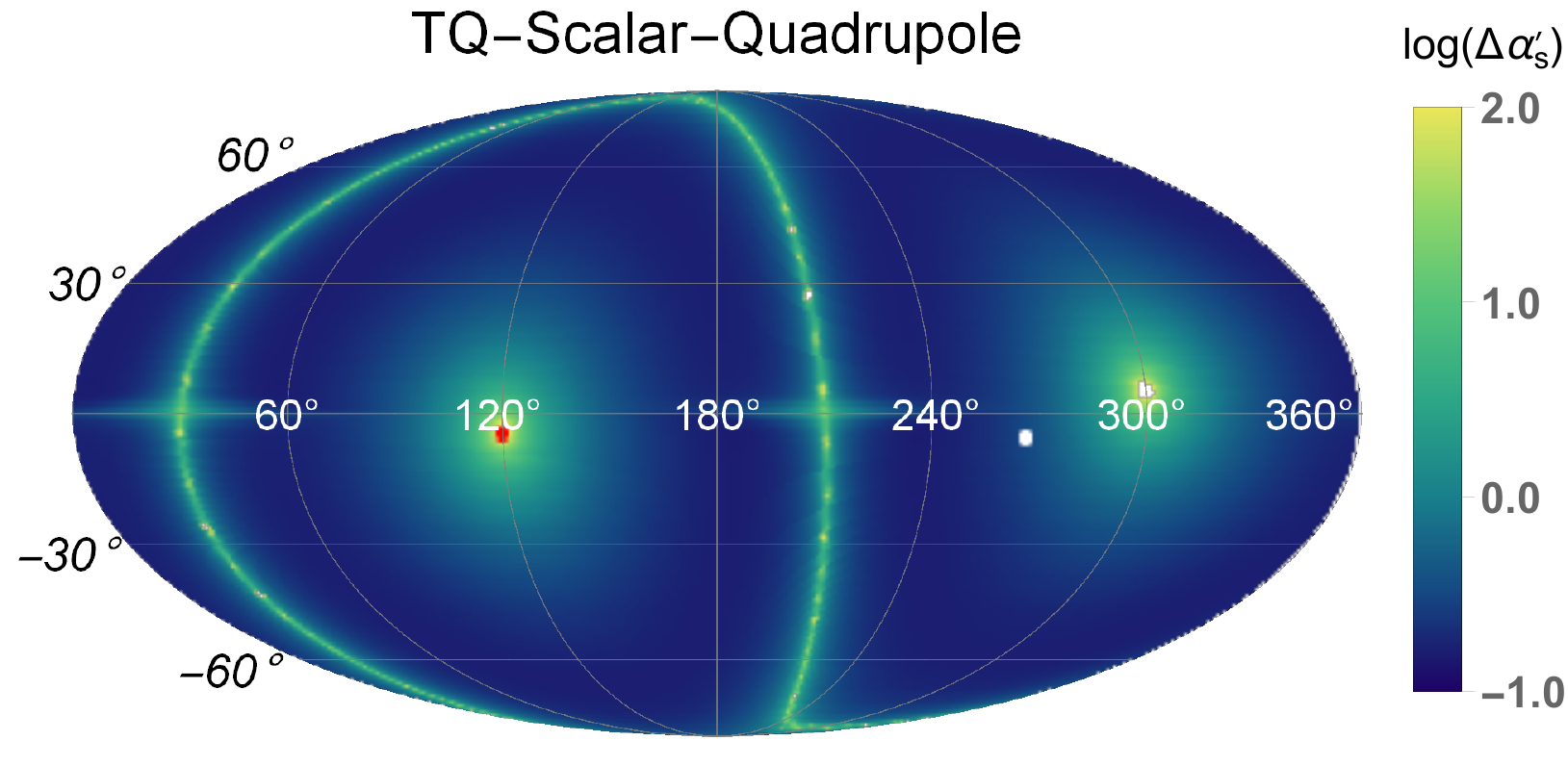}
	\caption{ The distribution of \(\Delta\alpha'_v\) and \(\Delta\alpha'_s\) on the celestial sphere in ecliptic coordinate for TianQin.}
	\label{qT} 
\end{figure}

\begin{figure}[h]
	\includegraphics[width=\linewidth]{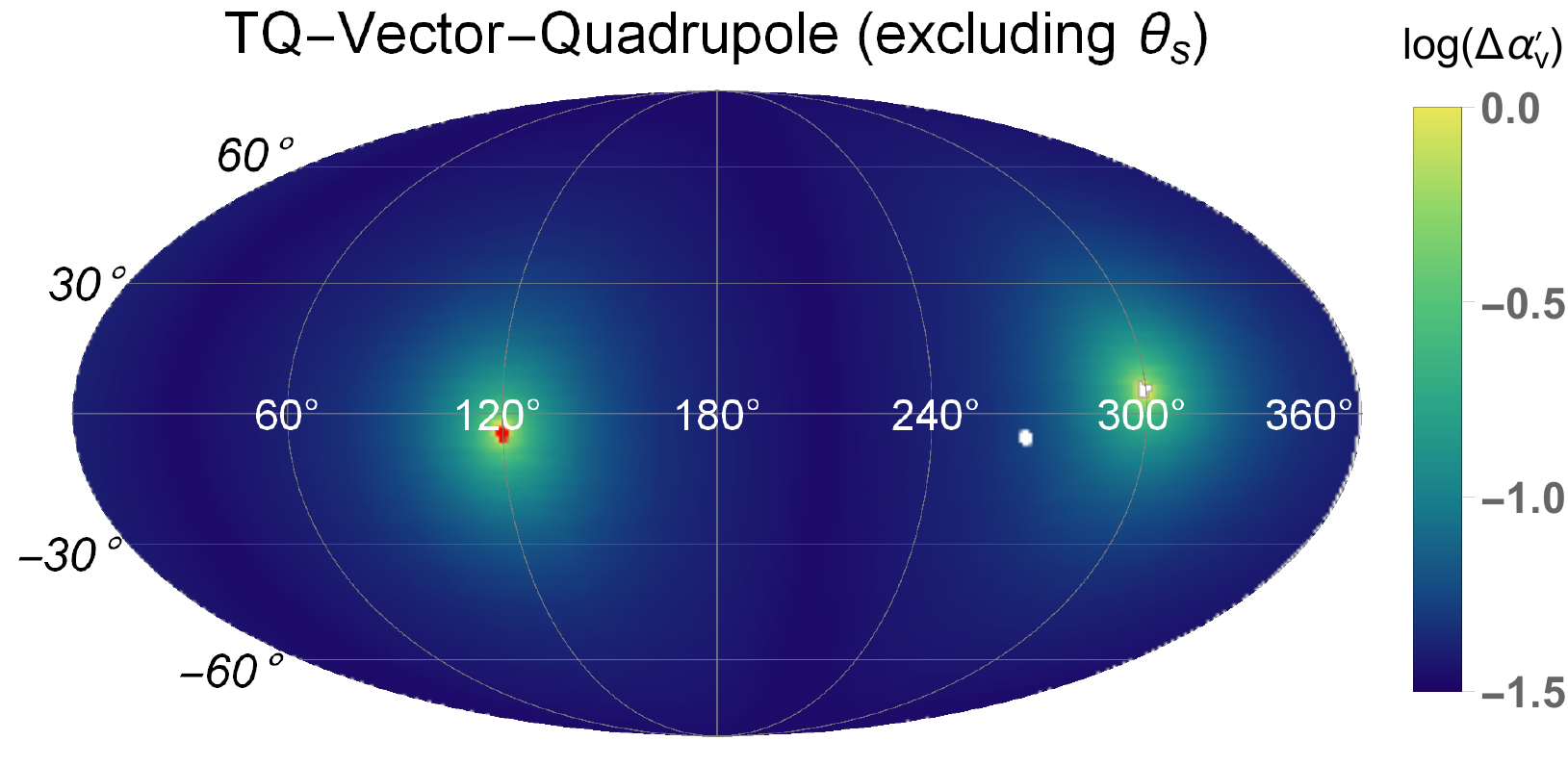}
	\caption{ The distribution of \(\Delta\alpha'_v\) on the celestial sphere for TianQin, with $\theta_s$ excluded in the \ac{PE}. 
		}
	\label{qT-theta} 
\end{figure}

The results for TianQin are shown in Fig.~\ref{qT}. 
We can find that the capability will be a few times worse than the dipole situation,
and it will also have strong position dependence due to the vanish of the response.
Due to the fact that the frequency of the extra modes at quadrupole order will be the same as the tensor mode, the degeneracy will happen between the angles and the relative amplitude of the extra modes.
Thus except the divergent at the orientation of the detector, the result will also be worse on the plane of the detector for scalar modes, and on the direction perpendicular to the orientation and located on the ecliptic plane.
The second additional degeneracy is caused by the degeneracy between $\alpha'_v$ and $\theta_s$.
As we plotted in Fig.~\ref{qT-theta}, if we exclude the parameter $\theta_s$, then the region on the ecliptic plane and perpendicular to the orientation will disappear.  

\begin{figure}[h]
	\includegraphics[width=1\linewidth]{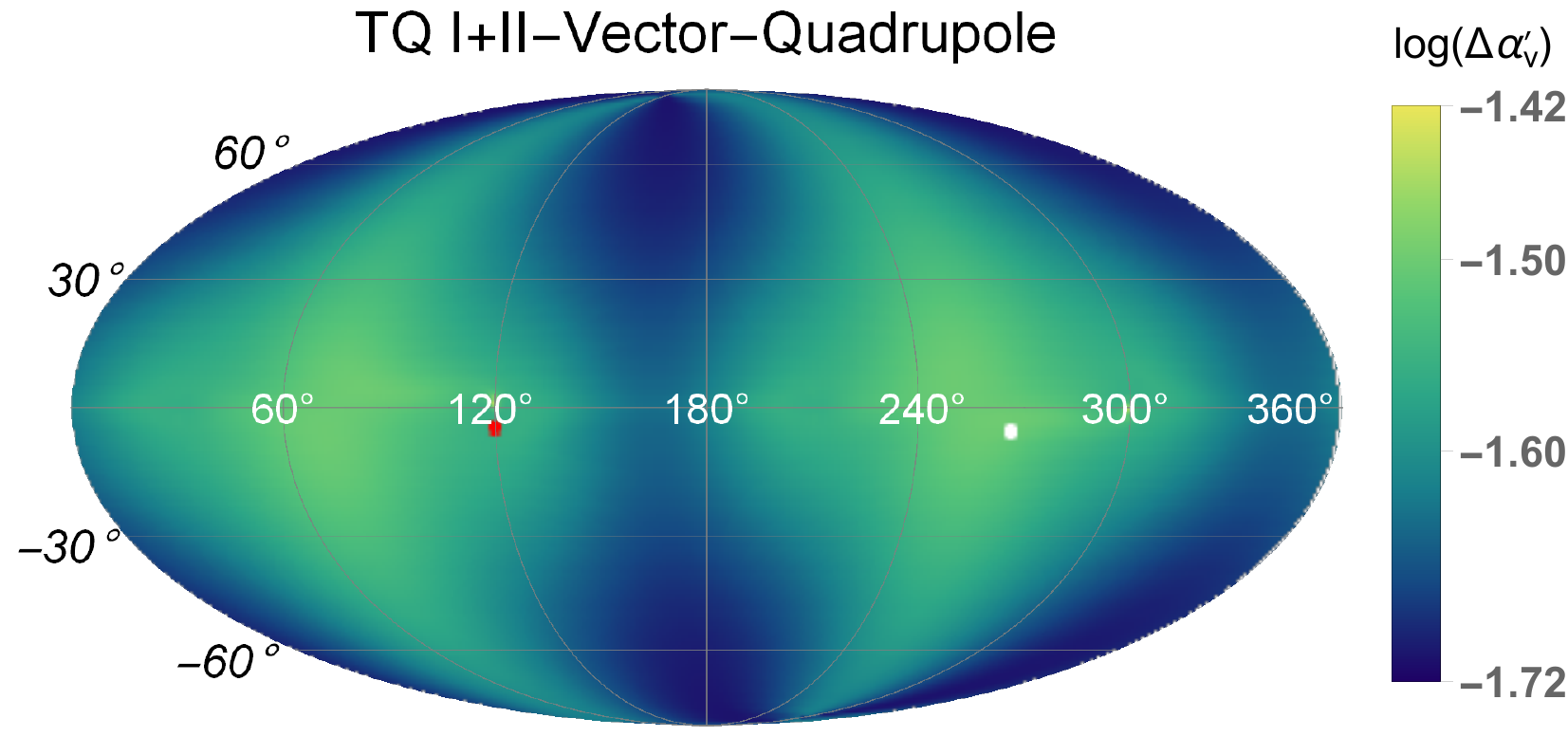}
	\includegraphics[width=1\linewidth]{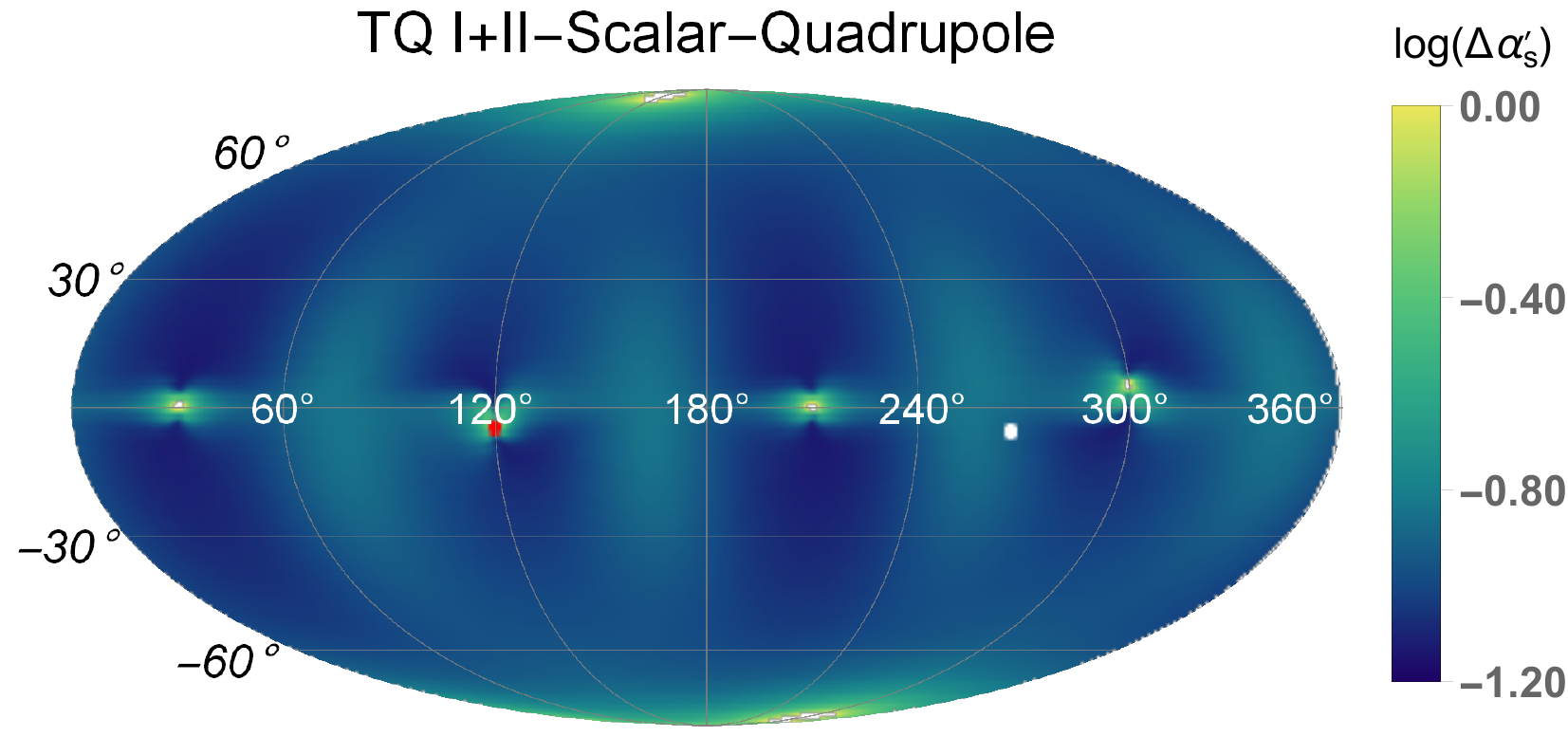}
	\caption{The distribution of \(\Delta\alpha'_v\) and \(\Delta\alpha'_s\) on the celestial sphere in ecliptic coordinate for the TianQin  twin constellation.
	}
	\label{qtT} 
\end{figure}

\begin{figure}[!h]
	\includegraphics[width=\linewidth]{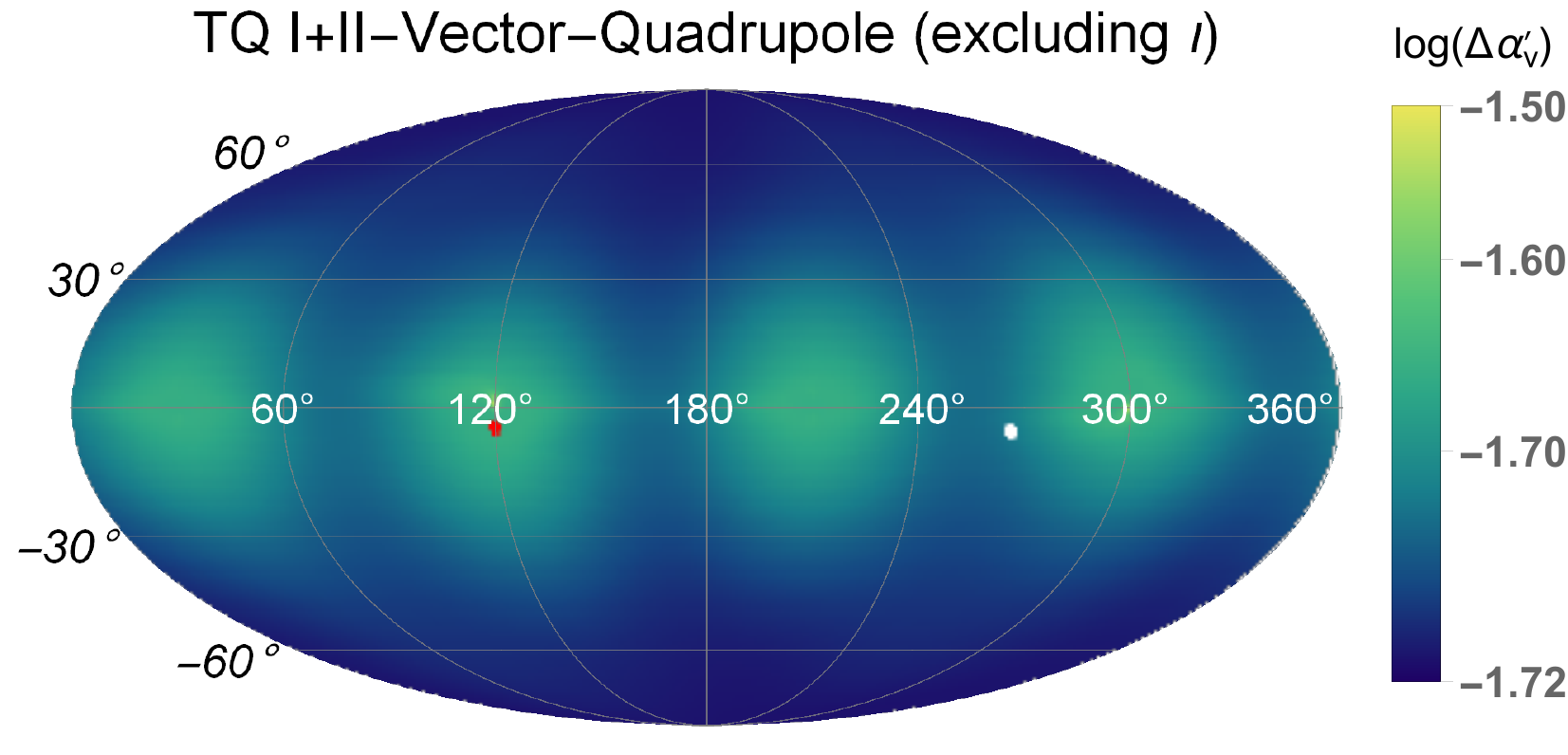}
	\caption{ The distribution of \(\Delta\alpha'_v\) on the celestial sphere for TianQin twin constellation, with $\iota$ excluded in the \ac{PE}. 
	}
	\label{qtT-io} 
\end{figure}

The results for the TianQin twin constellation are displayed in Fig.~\ref{qtT}. 
The position dependence will also be deduced as for the dipole case,
but the shape will be different for the vector mode.
This is caused by the degeneracy between $\alpha'_v$ and $\iota$. 
After removing the inclination angle $\iota$ in the \ac{FIM} analysis, we can find that the result looks like the plot for dipole case (see upper panel of Fig.~\ref{dtT}). 
As a comparison, we show it in Fig.~\ref{qtT-io}. 
The detailed shape is caused by the choice of the value for $\iota$, and it will also change if we change $\iota$.

\begin{figure}[h]
	\includegraphics[width=\linewidth]{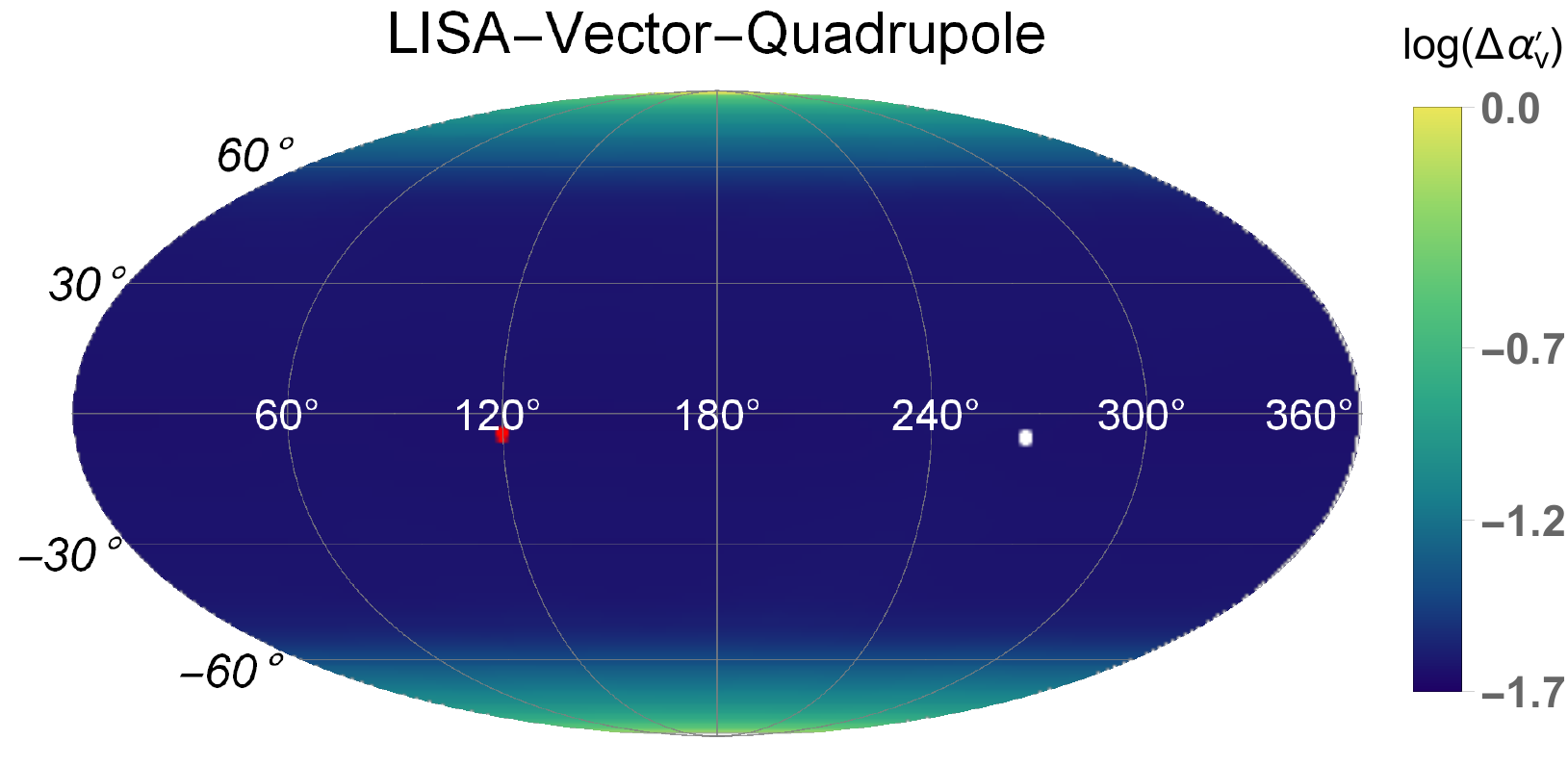}
	\quad
	\includegraphics[width=\linewidth]{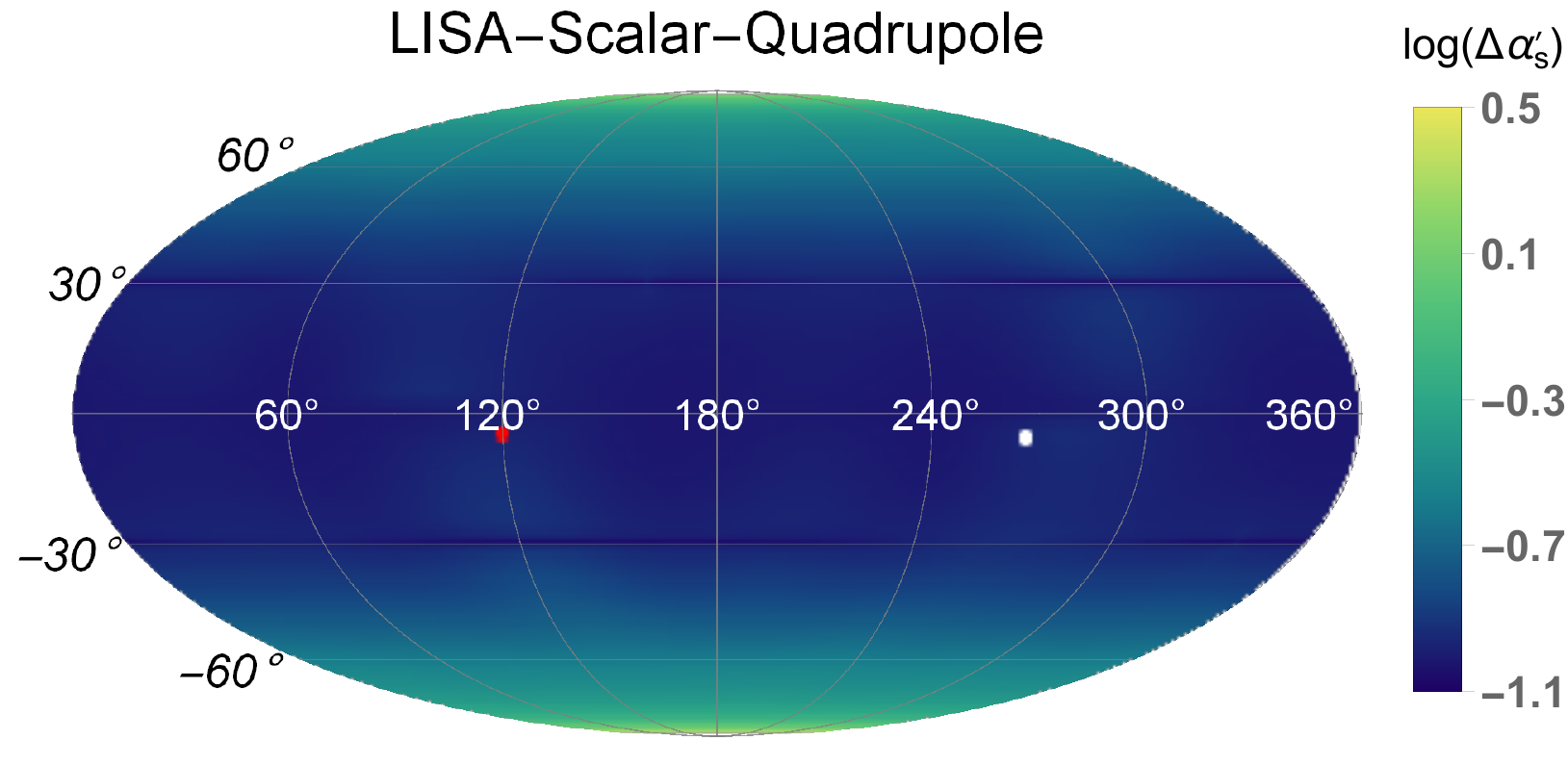}
	\caption{The distribution of \(\Delta\alpha'_{v}\) and \(\Delta\alpha'_{s}\) on the celestial sphere in ecliptic coordinate for LISA.
	}\label{qL} 
\end{figure}

For LISA, we can find that the detecting accuracy is mainly determined by the latitude of the source location, same as the dipole case.  
But this time, the worse part will be the pole region for both vector and scalar modes.
{The result is shown in Fig.~\ref{qL}. }
For the joint detection for LISA with TianQin or TianQin twin constellation, the accuracy will also be improved, and the results are shown in Fig.~\ref{qTL} and Fig.~\ref{qtTL} at Appendix~\ref{figure}.

\begin{figure}[h]
	\includegraphics[width=1\linewidth]{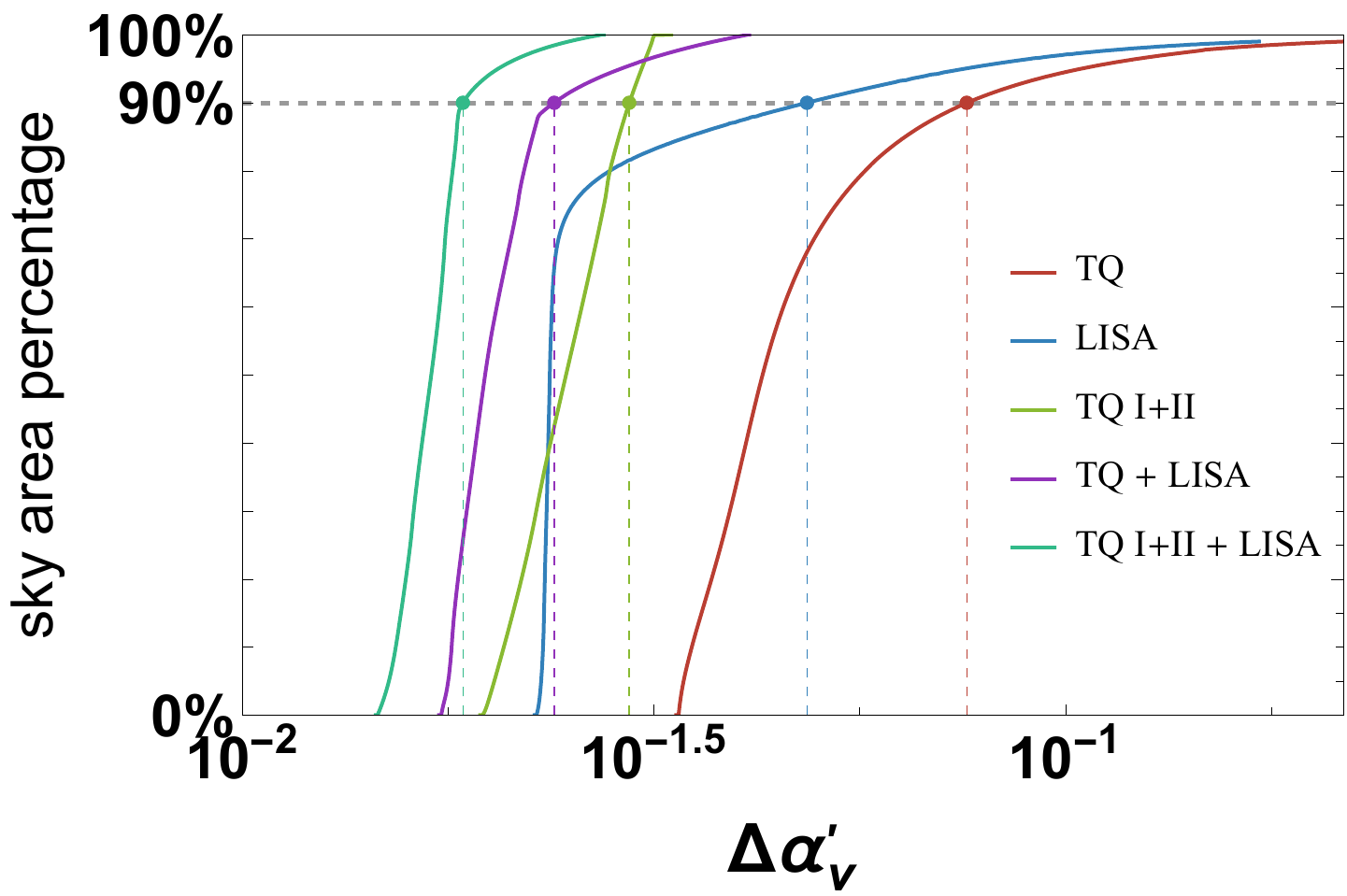}
	\includegraphics[width=1\linewidth]{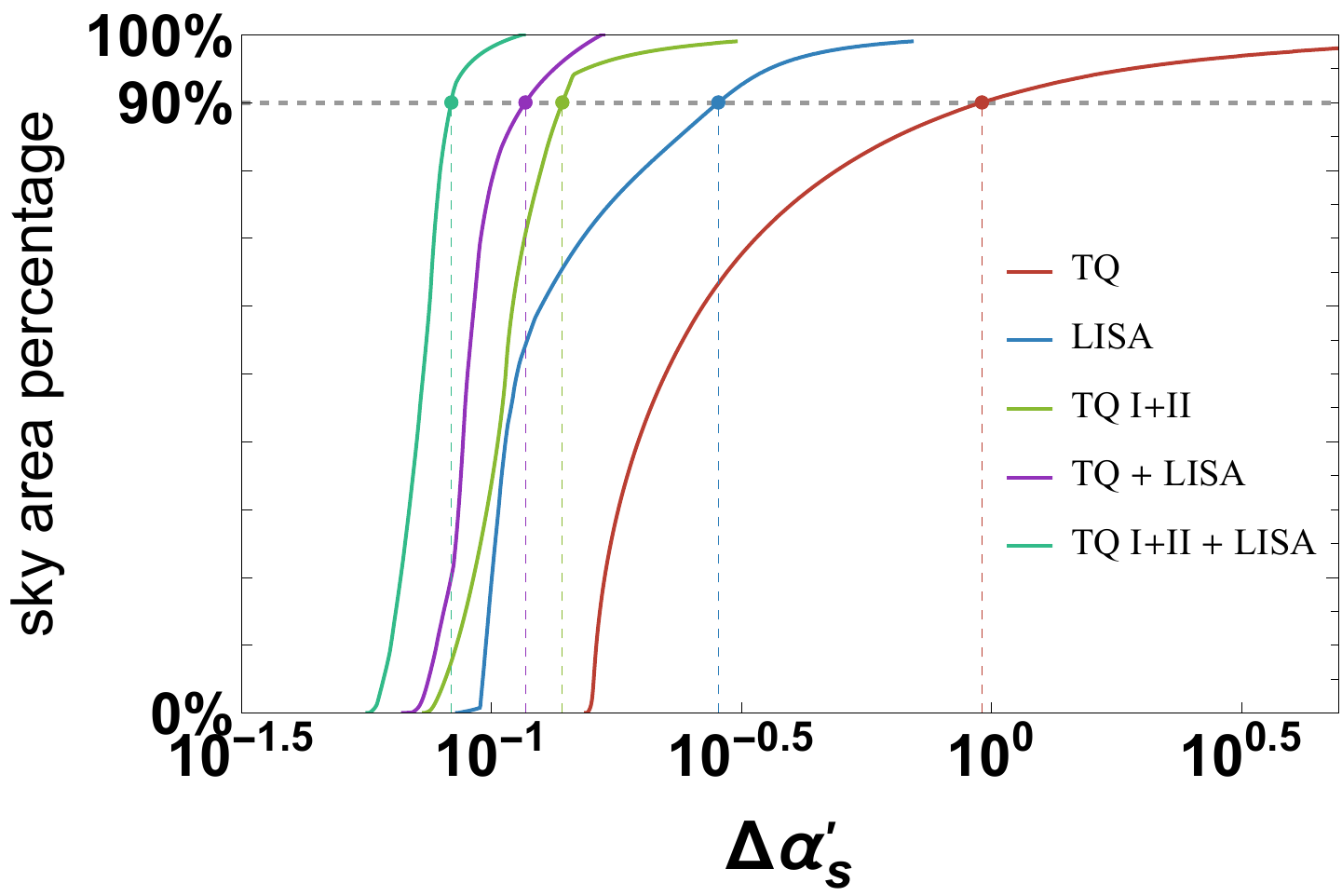}
	\caption{Upper panel: the area ratio of sky region where the $\Delta\alpha'_v$ always takes on a value less than 
	a certain quantity. Bottom panel: same as the former one but for $\Delta\alpha'_s$. 
	The configurations considered include TQ, LISA, TQ I+II, TQ + LISA and TQ I+II + LISA. 
	The dashed lines corresponding to the 90\% threshold.
	}\label{qc} 
\end{figure}

Again, we plot the \ac{CDF} of the accuracy distribution on the sky for each detector configuration in Fig.~\ref{qc}, and list the result for the best value and $90\%$ threshold in Table~\ref{tqc}. 
Except for TianQin, LISA will also have a very wide distribution due to the degeneracy between the relative amplitude and the angles.
But the result for joint detection will still be narrow due to the complementary of different detectors.
Overall, the accuracy for dipole radiation will be a few times better than that for the quadrupole radiation.
Since the amplitude of the leading dipole radiation would also be stronger than the sub-leading quadrepole radiation, this result means that it will be better to constrain the extra modes with dipole radiation, if we ignore the problem of identification of the common origin.

\begin{table}[h]
	\caption{The representative detecting accuracy for the 90\% level and best of $\alpha'_v$ and $\alpha'_s$ for each configuration. 
	}\label{tqc}
	\begin{ruledtabular}
		\begin{tabular}{lccccc}
			&
			\textrm{\footnotesize TQ}&
			\textrm{\footnotesize TQ I+II}&
			\textrm{\footnotesize LISA}&
			\textrm{\footnotesize TQ + LISA}&
			\textrm{\footnotesize TQ I+II + LISA}\\
			\colrule
			$\Delta\alpha'_{v}(90\%)$ & 0.076 & 0.030 & 0.049 & 0.024 & 0.019\\
			$\Delta\alpha'_v(\text{best})$ & 0.034 & 0.019 & 0.023 & 0.018 & 0.014\\
			
			$\Delta\alpha'_{s}(90\%)$ & 0.958 & 0.139 & 0.284 & 0.117 & 0.083\\
			$\Delta\alpha'_s(\text{best})$ & 0.156 & 0.073 & 0.088 & 0.069 & 0.056\\
		\end{tabular}
	\end{ruledtabular}
\end{table}

\begin{figure}[h]
	\includegraphics[width=1\linewidth]{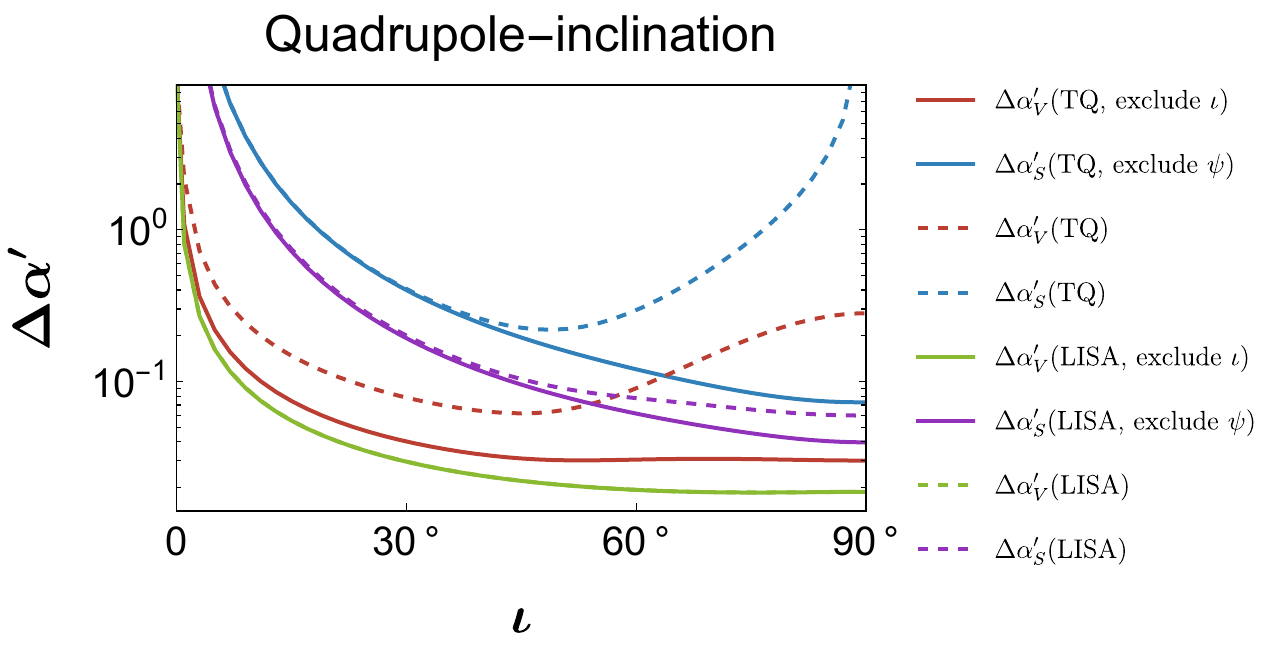}
	\caption{{ The distribution of \(\Delta\alpha'_v\) and \(\Delta\alpha'_s\) with different inclination angle for TianQin and LISA.}}
	\label{qi} 
\end{figure}

{Same as the dipole case, the \ac{PE} accuracy for the sub-leading quadrupole radiation will also be sensitive to inclination angle of the source.
Here we also consider the \ac{DWD} source with different inclination angle by fixing the source position to $\theta_s=30^\circ$ and $\phi_s=60^\circ$. 
The dependence on inclination angle $\iota$ for $\Delta\alpha'_v$ and $\Delta\alpha'_s$ are shown in Fig.~\ref{qi}. 
We find that both $\Delta\alpha'_v$ and $\Delta\alpha'_s$ diverge when $\iota$ becomes zero since all the extra modes both vanish for the face-on \ac{DWD} (cf. Eqs.~(\ref{hxq}-\ref{hlq})). 
Moreover, we can see that only for the TianQin consternation, $\iota \to \pi/2$ can bring the degeneracy between some angle parameters and the extra mode amplitudes (especially $\iota$ and $\alpha'_v$, $\psi$ and $\alpha'_s$). 
When we exclude $\iota$ and $\psi$ respectively in the \ac{PE}, the degeneracy disappears and $\Delta\alpha'_{v,s}$ has the smallest value at $\iota =\pi/2$. 
}

\section{Constraint with verification binaries}\label{sec:4}

As mentioned in \cite{2020Huang}, despite the thousands of \acp{DWD} which could be detected in the future observation, we can also detect several \acp{VB} which have already been detected by \ac{EM} observations.
For \acp{VB}, the position and frequency of the \acp{VB} can be determined by \ac{EM} observation,
and the accuracy of the position is much better than the \ac{GW} observations,
which means that we don't need to consider the \ac{PE} of the two position angles in the \ac{FIM} analysis.
However, since the accuracy of frequency with \ac{EM} observation is worse than the \ac{GW} observation,
the frequency will still be considered in the \ac{FIM} analysis.
Thus the parameters will be reduced to only six while $\theta_s$ and $\phi_s$ will be known parameters.
As we mentioned above, there will be degeneracy between the relative amplitudes and the positions of the source, so if the position is determined through the \ac{EM} observation, the constraint of the extra modes will be better.
{Moreover, for part of \acp{VB}, their inclination angle could also be determined by \ac{EM} observations such as intensity interferometry~\cite{2020Mit} with better precision comparing to the \ac{GW} observation. Thus the parameter $\iota$ can be excluded in the \ac{FIM} analysis.}
The parameters of the \acp{VB} are listed in Table~\ref{t:dwd}. For another part of \acp{VB} there is no direct measurement on the inclination angle through \ac{EM} observation and the angle is assigned based on other information of the binary system~\cite{2020Huang}. We label the estimated values of $\iota$ with the asterisk. Note that we must consider the \ac{PE} of the inclination angle for these special \acp{VB} in \ac{FIM} analysis. 

\begin{table}[t]
    	\caption{\label{t:dwd} The parameters of \acp{VB} which can be detected by TianQin within 5 years. The asterisk after the values means the parameter has no precise \ac{EM} measurement. 
	}
	\begin{ruledtabular}
		\begin{tabular}{llllll}
		\textrm{source} & $\theta$ & $\phi$ & $\iota$ & $f$/mHz & $\mathcal{A}$ \\
		\colrule
		\text{J0806} & 94.7 & 120.4 & 38 & 6.22 & $1.28\times 10^{-22}$ \\
 \text{ZTF J1539} & 23.8 & 205.0 & 84 & 4.82 & $3.68\times 10^{-22}$ \\
 \text{V407 Vul} & 43.2 & 295.0 & 60$^*$ & 3.51 & $2.20\times 10^{-22}$ \\
 \text{ES Cet} & 110.3 & 24.6 & 60$^*$ & 3.22 & $2.14\times 10^{-22}$ \\
 \text{SDSS J0651} & 94.2 & 101.3 & 87 & 2.61 & $3.24\times 10^{-22}$ \\
 \text{SDSS J1351} & 85.5 & 208.4 & 60$^*$ & 2.12 & $1.24\times 10^{-22}$ \\
 \text{AM CVn} & 52.6 & 170.4 & 43 & 1.94 & $5.66\times 10^{-22}$ \\
 \text{SDSS J1908} & 28.5 & 298.2 & 15 & 1.84 & $1.22\times 10^{-22}$ \\
 \text{HP Lib} & 85.0 & 235.1 & 30$^*$ & 1.81 & $3.14\times 10^{-22}$ \\
 \text{SDSS J0935} & 61.9 & 131.0 & 60$^*$ & 1.68 & $5.98\times 10^{-22}$ \\
 \text{SDSS J2322} & 81.5 & 353.4 & 27 & 1.66 & $1.74\times 10^{-22}$ \\
 \text{PTF J0533} & 111.1 & 82.9 & 73 & 1.62 & $1.52\times 10^{-22}$ \\
 \text{CR Boo} & 72.1 & 202.3 & 30$^*$ & 1.36 & $2.58\times 10^{-22}$ \\
 \text{V803 Cen} & 120.3 & 216.2 & 14 & 1.25 & $3.20\times 10^{-22}$ \\
		\end{tabular}
	\end{ruledtabular}
\end{table}

Here we will consider TQ, LISA, TQ I+II, TQ + LISA and TQ I+II + LISA. 
The mission lifetime is 4 yrs for LISA, and 5 yrs for TianQin with 4 years overlap with LISA.The results are shown in Table~\ref{t:mod1} for 
dipole case, and in Table~\ref{t:mod2} for quadrupole case.
The results of J0806 in the first two column are invalid since the estimated errors diverge in the sky location of J0806 for TQ due to the zero response of the extra modes. 
However, due to the suitable frequency and the strong amplitude, the capability for other configurations can still reach to a few percent or less.
{In Tables~\ref{t:mod1} and \ref{t:mod2}, for J0806, ZTF J1539, SDSS J0651, AM CVn, SDSS J1908, SDSS J2322, PTF J0533 and V803 Cen, the \acp{FIM} are calculated with fixed $\iota$. }

\begin{table*}[t]
	\caption{\label{t:mod1} The \ac{PE} accuracy of dipole amplitude $\alpha_v$ and $\alpha_s$ for the \acp{VB}. 
	}
	\begin{ruledtabular}
		\begin{tabular}{lllllllllll}
			&\multicolumn{2}{c}{\textrm{TQ}}&\multicolumn{2}{c}{\textrm{LISA}}&\multicolumn{2}{c}{\textrm{TQ I+II}}&\multicolumn{2}{c}{\textrm{TQ + LISA}}&\multicolumn{2}{c}{\textrm{TQ I+II + LISA}}\\
			&$\Delta\alpha_v$  & $\Delta\alpha_s$  & $\Delta\alpha_v$  & $\Delta\alpha_s$  & $\Delta\alpha_v$  & $\Delta\alpha_s$  & $\Delta\alpha_v$  & $\Delta\alpha_s$  & $\Delta\alpha_v$  & $\Delta\alpha_s$\\
			\colrule
 \text{J0806} & \textbackslash   & \textbackslash   & 0.008  & 0.031  & 0.035  & 0.103  & 0.008  & 0.031  & 0.008  & 0.029   \\
 \text{ZTF J1539} & 0.026  & 0.037  & 0.005  & 0.009  & 0.018  & 0.028  & 0.005  & 0.009  & 0.005  & 0.008  \\
 \text{V407 Vul} & 0.090  & 0.305  & 0.022  & 0.046  & 0.056  & 0.123  & 0.021  & 0.046  & 0.020  & 0.043  \\
 \text{ES Cet} & 0.089  & 0.164  & 0.030  & 0.068  & 0.080  & 0.162  & 0.029  & 0.063  & 0.028  & 0.063  \\
 \text{SDSS J0651} & 0.225  & 1.342  & 0.041  & 0.077  & 0.093  & 0.159  & 0.041  & 0.076  & 0.038  & 0.069  \\
 \text{SDSS J1351} & 0.360  & 0.658  & 0.158  & 0.377  & 0.358  & 0.658  & 0.145  & 0.327  & 0.145  & 0.327  \\
 \text{AM CVn} & 0.090  & 0.285  & 0.038  & 0.113  & 0.063  & 0.217  & 0.035  & 0.105  & 0.032  & 0.100  \\
 \text{SDSS J1908} & 0.433  & 4.538  & 0.167  & 1.526  & 0.286  & 2.477  & 0.155  & 1.445  & 0.144  & 1.299  \\
 \text{HP Lib} & 0.173  & 0.784  & 0.072  & 0.350  & 0.150  & 0.767  & 0.066  & 0.318  & 0.064  & 0.317  \\
 \text{SDSS J0935} & 0.172  & 0.712  & 0.051  & 0.113  & 0.097  & 0.214  & 0.049  & 0.111  & 0.045  & 0.100  \\
 \text{SDSS J2322} & 0.400  & 2.485  & 0.149  & 0.804  & 0.291  & 2.007  & 0.138  & 0.760  & 0.132  & 0.746  \\
 \text{PTF J0533} & 0.686  & 2.065  & 0.230  & 0.439  & 0.445  & 1.070  & 0.218  & 0.429  & 0.204  & 0.406  \\
 \text{CR Boo} & 0.362  & 1.363  & 0.142  & 0.664  & 0.327  & 1.353  & 0.132  & 0.597  & 0.130  & 0.596  \\
 \text{V803 Cen} & 0.326  & 2.787  & 0.124  & 1.252  & 0.270  & 2.709  & 0.116  & 1.142  & 0.113  & 1.136  \\
		\end{tabular}
	\end{ruledtabular}
\end{table*}

\begin{table*}[t]
	\caption{\label{t:mod2} The \ac{PE} accuracy of quadrupole amplitude $\alpha'_v$ and $\alpha'_s$ for the \acp{VB}.
	}
	\begin{ruledtabular}
		\begin{tabular}{lllllllllll}
			&\multicolumn{2}{c}{\textrm{TQ}}&\multicolumn{2}{c}{\textrm{LISA}}&\multicolumn{2}{c}{\textrm{TQ I+II}}&\multicolumn{2}{c}{\textrm{TQ + LISA}}&\multicolumn{2}{c}{\textrm{TQ I+II + LISA}}\\
			&$\Delta\alpha'_{v}$ &$\Delta\alpha'_{s}$ &$\Delta\alpha'_{v}$&$\Delta\alpha'_{s}$&$\Delta\alpha'_{v}$&$\Delta\alpha'_{s}$&$\Delta\alpha'_{v}$&$\Delta\alpha'_{s}$&$\Delta\alpha'_{v}$&$\Delta\alpha'_{s}$\\
			\colrule
 \text{J0806} & \textbackslash & \textbackslash & 0.010 & 0.049 & 0.018 & 0.054 & 0.010 & 0.044 & 0.009 & 0.036 \\
 \text{ZTF J1539} & 0.007 & 0.651 & 0.003 & 0.010 & 0.006 & 0.043 & 0.003 & 0.009 & 0.003 & 0.009 \\
 \text{V407 Vul} & 0.035 & 0.141 & 0.006 & 0.016 & 0.019 & 0.053 & 0.006 & 0.016 & 0.006 & 0.015 \\
 \text{ES Cet} & 0.029 & 0.453 & 0.006 & 0.018 & 0.024 & 0.054 & 0.006 & 0.018 & 0.006 & 0.017 \\
 \text{SDSS J0651} & 0.056 & 3.114 & 0.005 & 0.017 & 0.025 & 0.049 & 0.005 & 0.016 & 0.005 & 0.016 \\
 \text{SDSS J1351} & 0.114 & 4.142 & 0.024 & 0.076 & 0.102 & 0.201 & 0.023 & 0.073 & 0.023 & 0.071 \\
 \text{AM CVn} & 0.033 & 0.172 & 0.008 & 0.030 & 0.023 & 0.123 & 0.008 & 0.029 & 0.008 & 0.029 \\
 \text{SDSS J1908} &0.395 & 6.216&0.107&1.293&0.290&4.088&0.104&1.277&0.100&1.218\\
 \text{HP Lib} & 0.115 & 0.635 & 0.023 & 0.145 & 0.099 & 0.482 & 0.022 & 0.140 & 0.022 & 0.137 \\
 \text{SDSS J0935} & 0.065 & 0.317 & 0.011 & 0.031 & 0.028 & 0.078 & 0.011 & 0.030 & 0.010 & 0.029 \\
 \text{SDSS J2322} & 0.206 & 1.878 & 0.061 & 0.421 & 0.158 & 1.503 & 0.059 & 0.411 & 0.057 & 0.404 \\
 \text{PTF J0533} & 0.230 & 1.132 & 0.047 & 0.127 & 0.220 & 0.792 & 0.046 & 0.125 & 0.044 & 0.123 \\
 \text{CR Boo} & 0.236 & 3.229 & 0.071 & 0.415 & 0.164 & 0.735 & 0.066 & 0.377 & 0.065 & 0.360 \\
 \text{V803 Cen} & 0.342 & 31.95 & 0.146 & 1.802 & 0.286 & 3.360 & 0.135 & 1.633 & 0.130 & 1.584 \\
		\end{tabular}
	\end{ruledtabular}
\end{table*}

We can find that, among all the \acp{VB}, ZTF J1539 can constrain the amplitude $\alpha_v$ and $\alpha_s$ to 0.005 and 0.008 respectively with TQ I+II + LISA, which is the best result. 
That is the combined effect of the higher amplitude of ZTF J1539 with suitable frequency, and the advantage sky location which at the best region as plotted in the previous figures. 
However, the accuracy of quadrupole mode constrained with TianQin will be worse for ZTF J1539, and the reason is that its position located near the detector's plane of TianQin, and we plotted in the bottom panel in Fig. \ref{qT}, those sources will have strong degeneracy with angular parameters.

\section{\label{sec:5}Conclusion and Discussion\protect}

We investigated the capability of TianQin to probe the extra polarizations in GWs.
We considered not only the standard TianQin constellation , but also 
the twin constellations and their joint detection with LISA.
We construct the responded signal with all the six polarizations, and assume that the GW waveform is monochromatic for \acp{DWD}.
Fisher analysis is adopted to estimate the accuracy to measure the parameters.
We defined two parameters, $\alpha_v$ and $\alpha_s$, to describe the 
dipole amplitudes of the extra modes relative to the tensor modes, and $\alpha'_{v}$ and $\alpha'_{s}$ for quadrupole amplitudes. 
Doppler effect due to the periodical motion of the detector around the Sun is also considered in the signal construction. 

We first analyze the behaviour of the \ac{PE} accuracy for the relative amplitudes as the function of the source position, and listed the best accuracy and the accuracy which can be attained by more than 90\% sky zone to describe the overall capability in the polarization detection. 
We find that, for TQ, the \ac{PE} accuracy will divergent at the orientation due to the zero response, and thus have very strong dependency on the sky location of the source, ranging across about 2 orders.
However, the capability of TianQin can still reach to a few percent for most of the regions.
The application of the twin constellation can significantly improve this strong position dependent feature.
A joint detection with LISA will also improve the result for a few times.

Generally, for all the detector configurations and both dipole and quadrupole situations, the capability for vector modes will be better than that for scalar modes.
On the other hand, the capability for dipole radiation will also be better than that for quadrupole radiation.
For the quadrupole case, the amplitude of the extra modes will have strong degeneracy with the position and the direction of the angular momentum.
However, due to the difficulty in the identification of the common origin, it's still very useful to consider the quadrupole radiation.

Finally, we also considered the capabilities with the observation of the \acp{VB}. 
For most of the \acp{VB}, the capabilities are about a few percents.
However, due to the zero response of the extra modes, TianQin could not detect the extra modes with J0806, although its expected to be one of the most loudest \acp{DWD}.
We also find that, for the best source ZTF J1539, the capability will be even less than one percent.

However, our analysis only focus on the monochromatic \ac{DWD} signals.
But the space borne \ac{GW} detectors can also detect many other chirping sources,
such as \acp{BBH} or \acp{EMRI}.
Some modelling works for these sources which focused on the extra polarization modes have already been done in recent years for some specific gravitational theories.
But for a more general analysis, we still need a parametric model.
In the future, we will consider the non-monochromatic sources and including the propagation effect and the modification of the phase.

\begin{acknowledgments}
We thank Changfu Shi for many helpful discussions.
This work is supported by the Guangdong Major Project of Basic and Applied Basic Research (Grant No. 2019B030302001).
Y. H. is supported by the Natural Science Foundation of China (Grants  No. 12173104).

\end{acknowledgments}

\appendix

\section{Results for the joint detection}\label{figure}

To reduce the length of the main text, we plot the results for the joint detection of TianQin and LISA in this appendix.
The analysis of these results is presented in Sec.~\ref{sec:3}, and we will not repeat it here.

\begin{figure}[h]
	\includegraphics[width=1\linewidth]{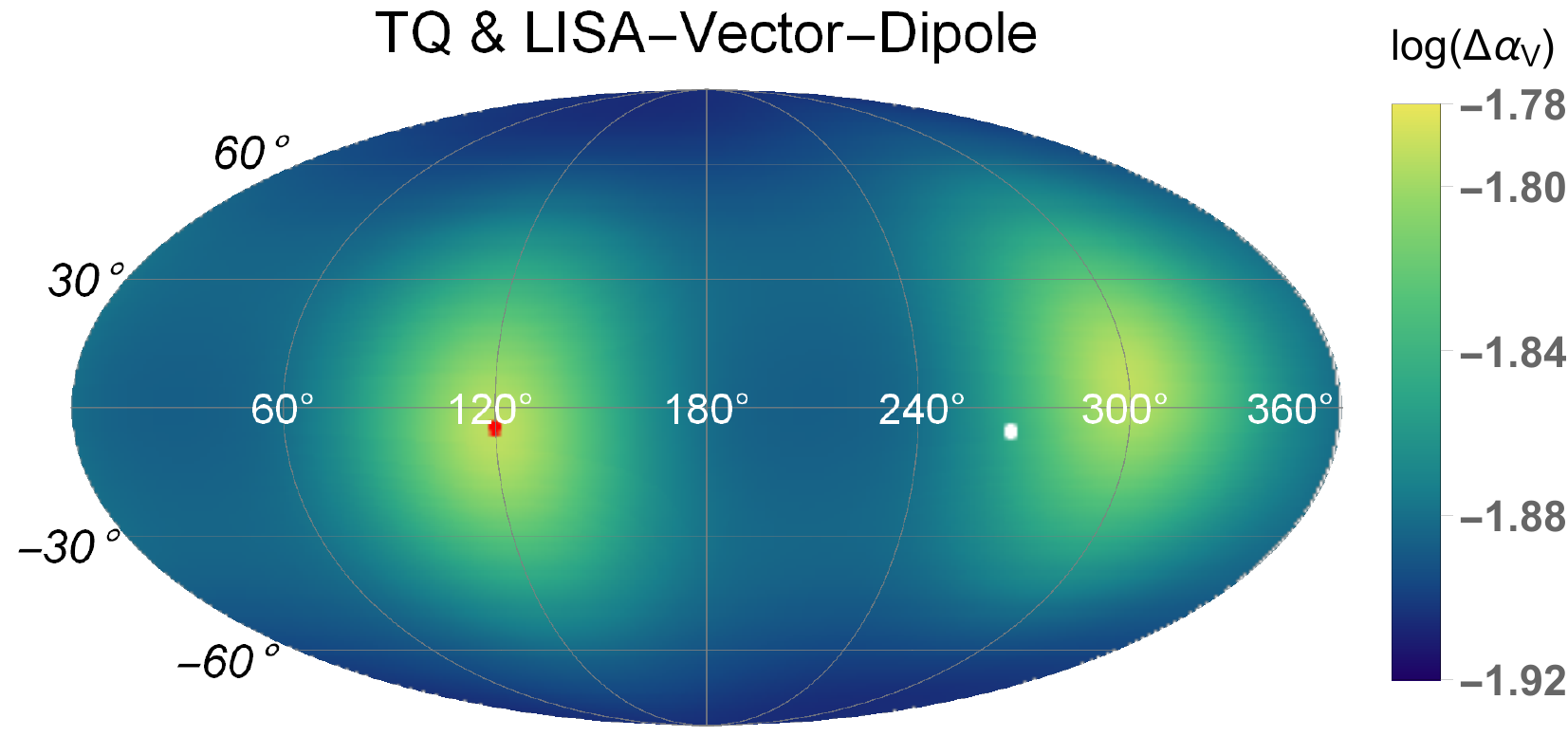}
	\includegraphics[width=1\linewidth]{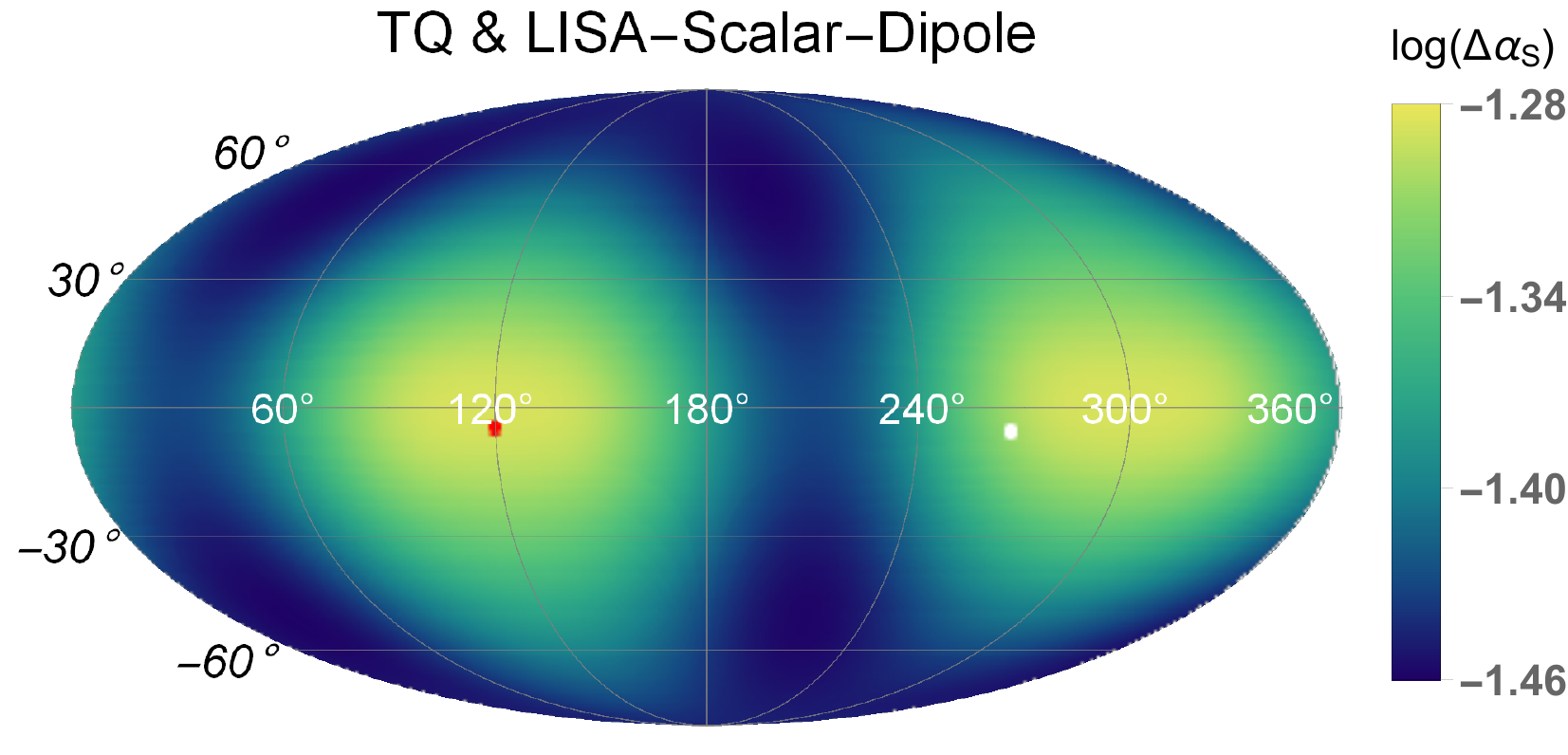}
	\caption{ The distribution of \(\Delta\alpha_v\) and \(\Delta\alpha_s\) on the celestial sphere in ecliptic coordinate for the joint detection of TianQin and LISA.
	}\label{dTL}
\end{figure}

\begin{figure}[h]
	\includegraphics[width=\linewidth]{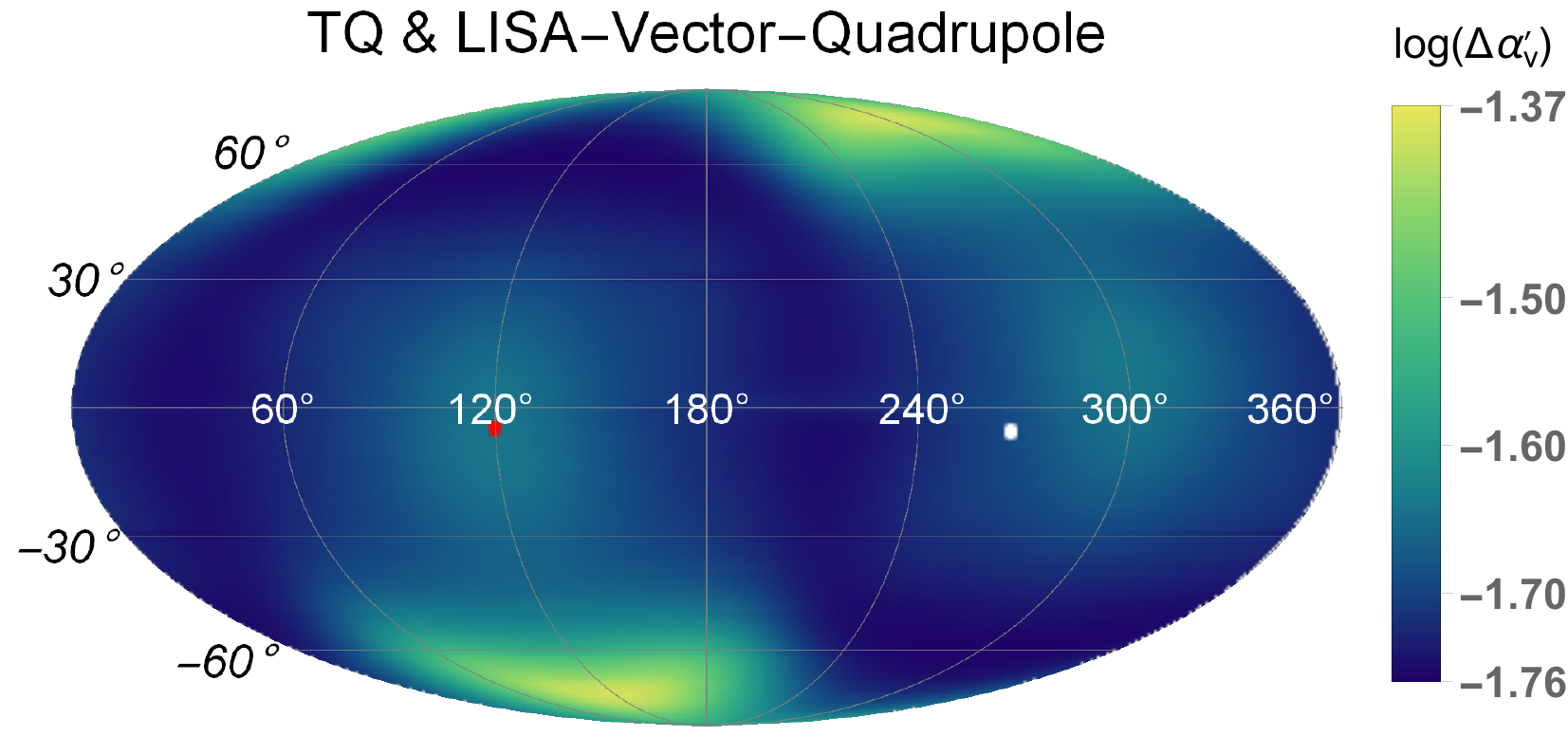}
	\includegraphics[width=\linewidth]{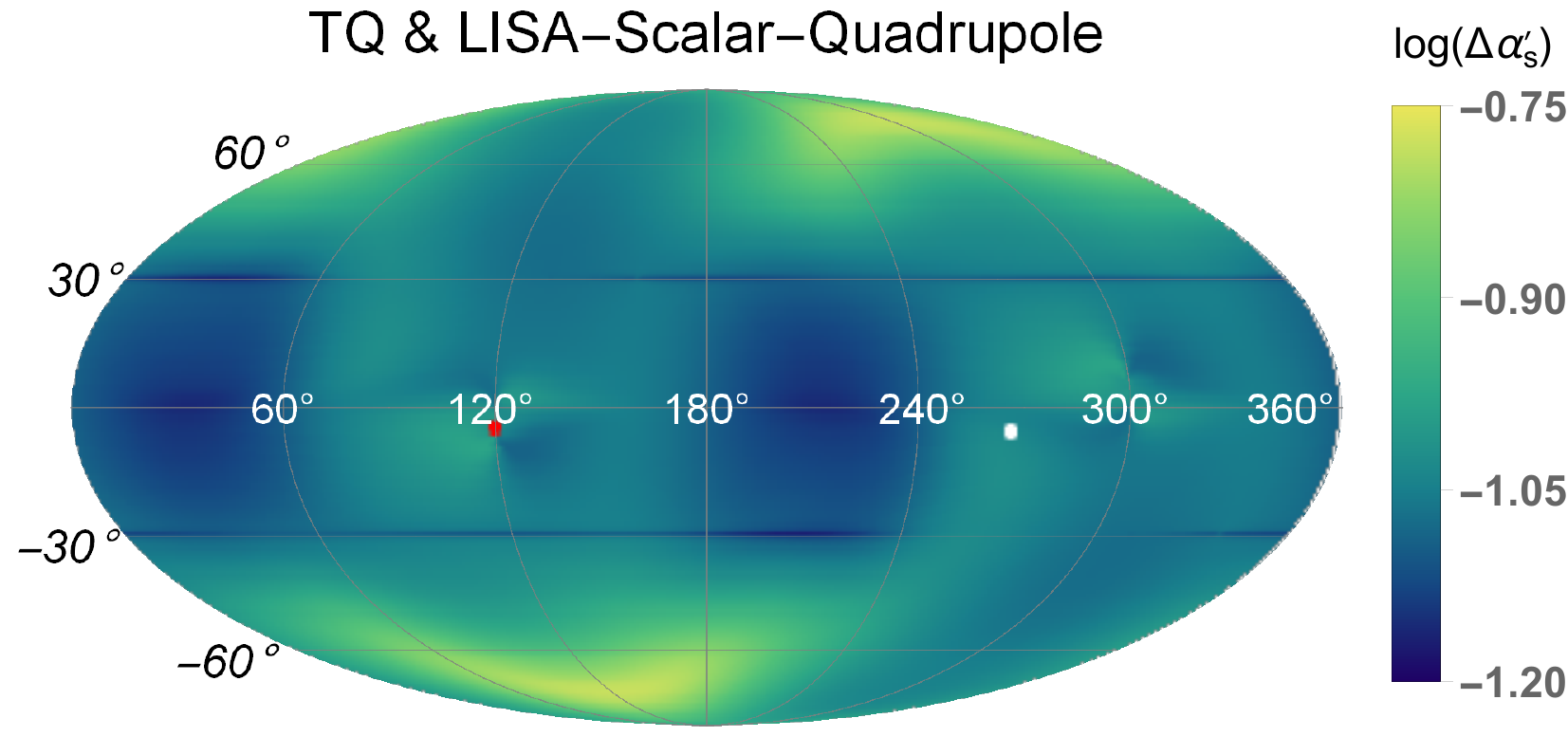}
	\caption{ The distribution of \(\Delta\alpha'_{v}\) and \(\Delta\alpha'_{s}\) on the celestial sphere in ecliptic coordinate for the joint detection of TianQin and LISA.
	}\label{qTL}
\end{figure}

\begin{figure}[h]
	\includegraphics[width=\linewidth]{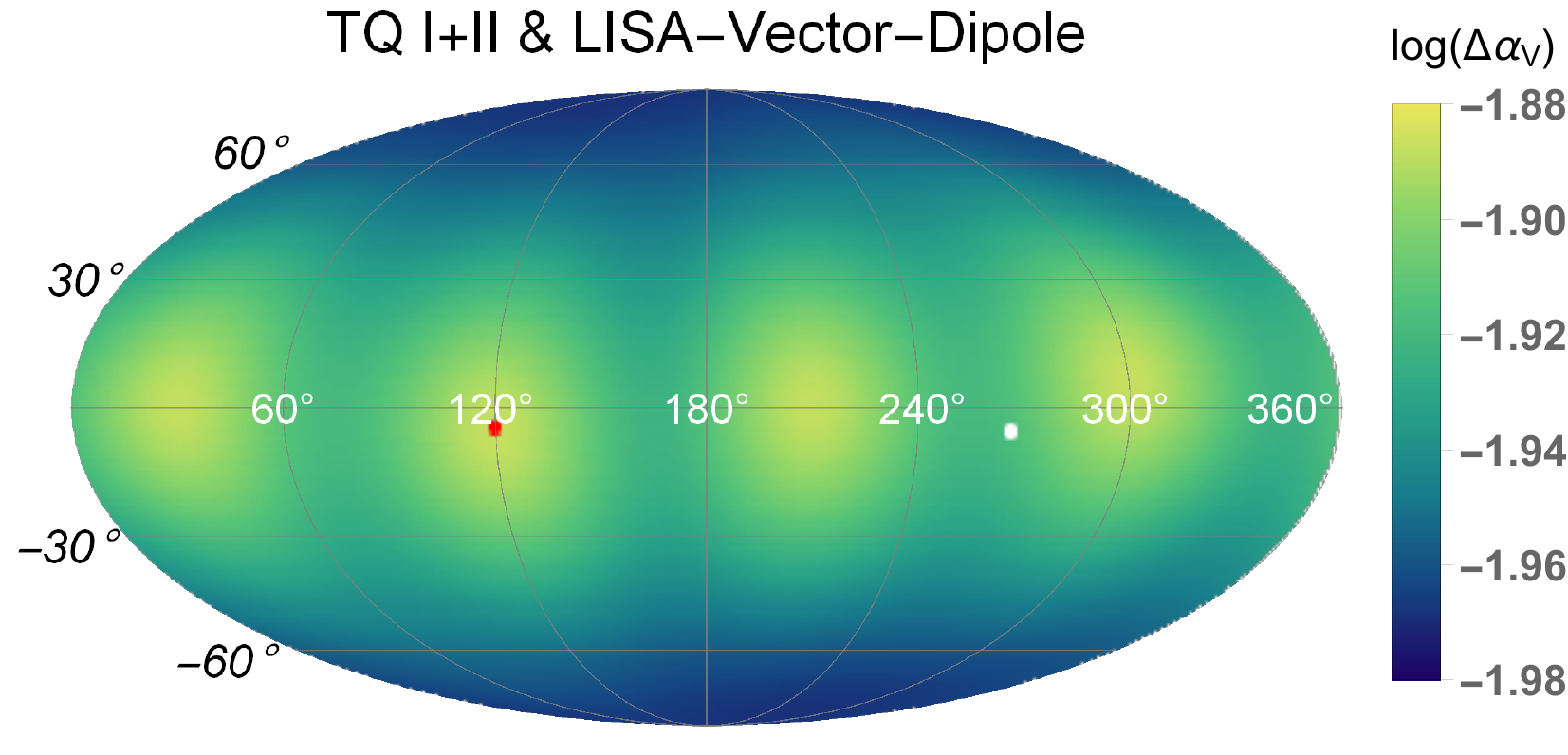}
	\quad
	\includegraphics[width=\linewidth]{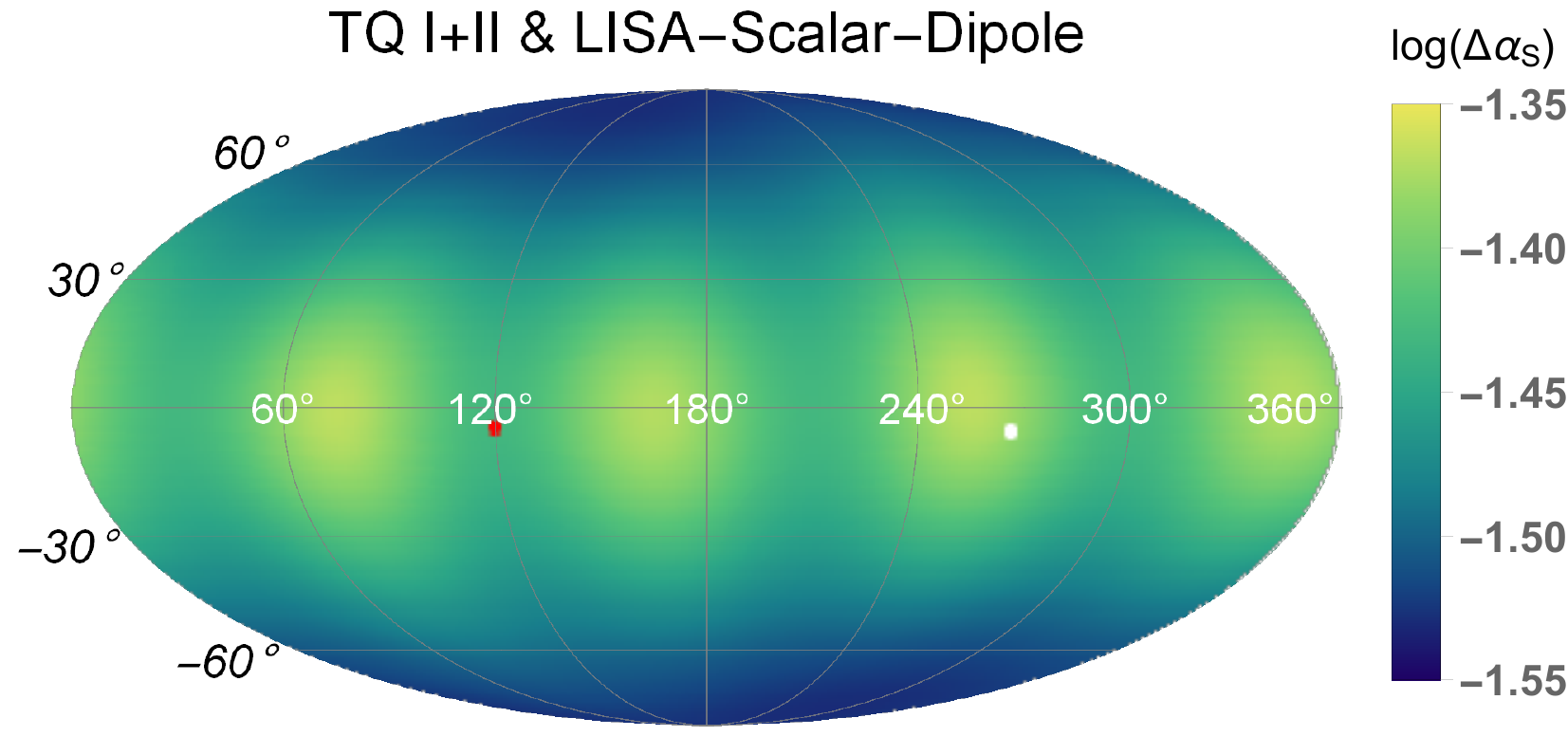}
	\caption{ The distribution of \(\Delta\alpha_v\) and \(\Delta\alpha_s\) on the celestial sphere in ecliptic coordinate for the joint detection of TianQin twin constellation and LISA.
	}\label{dtTL}
\end{figure}

\begin{figure}[h]
	\includegraphics[width=\linewidth]{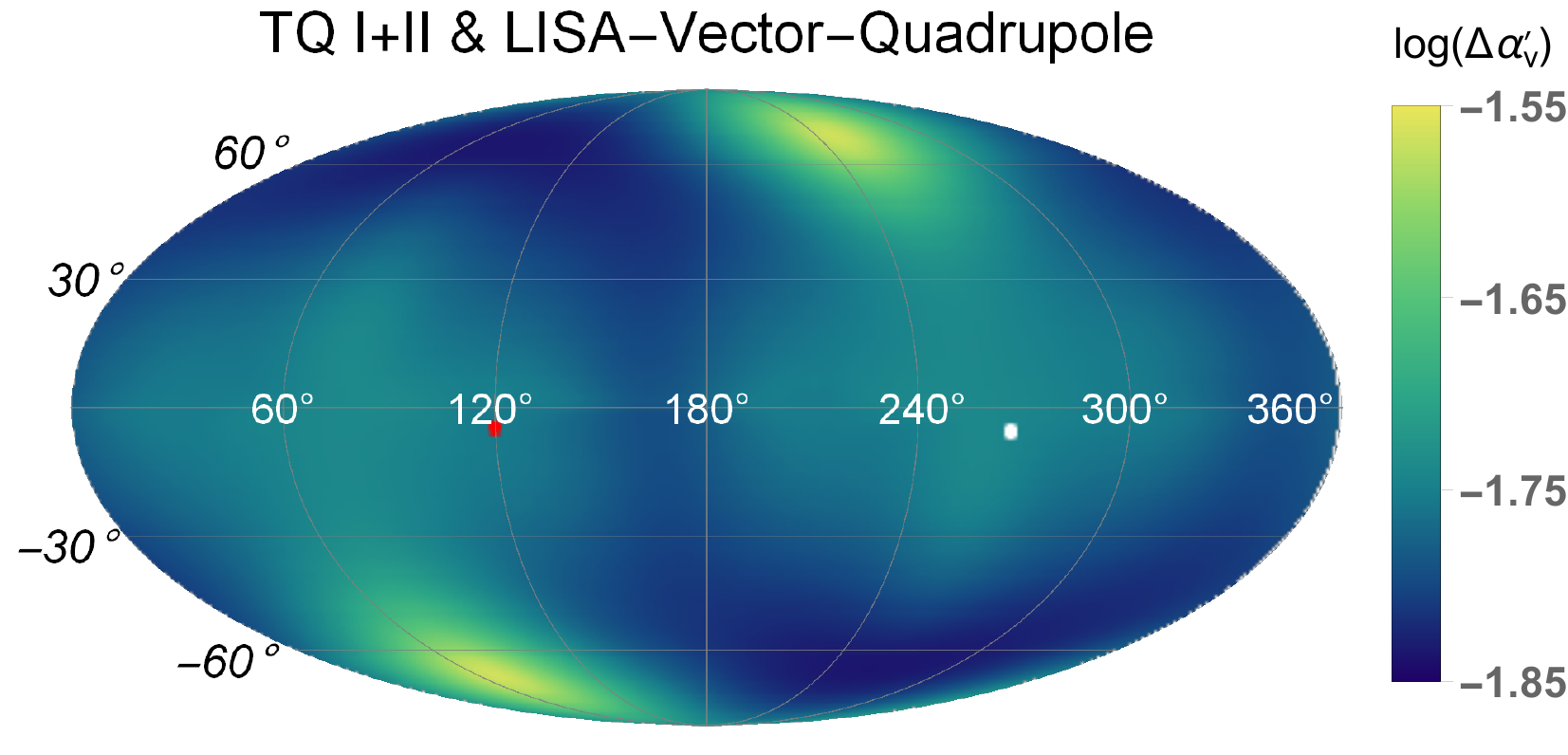}
	\includegraphics[width=\linewidth]{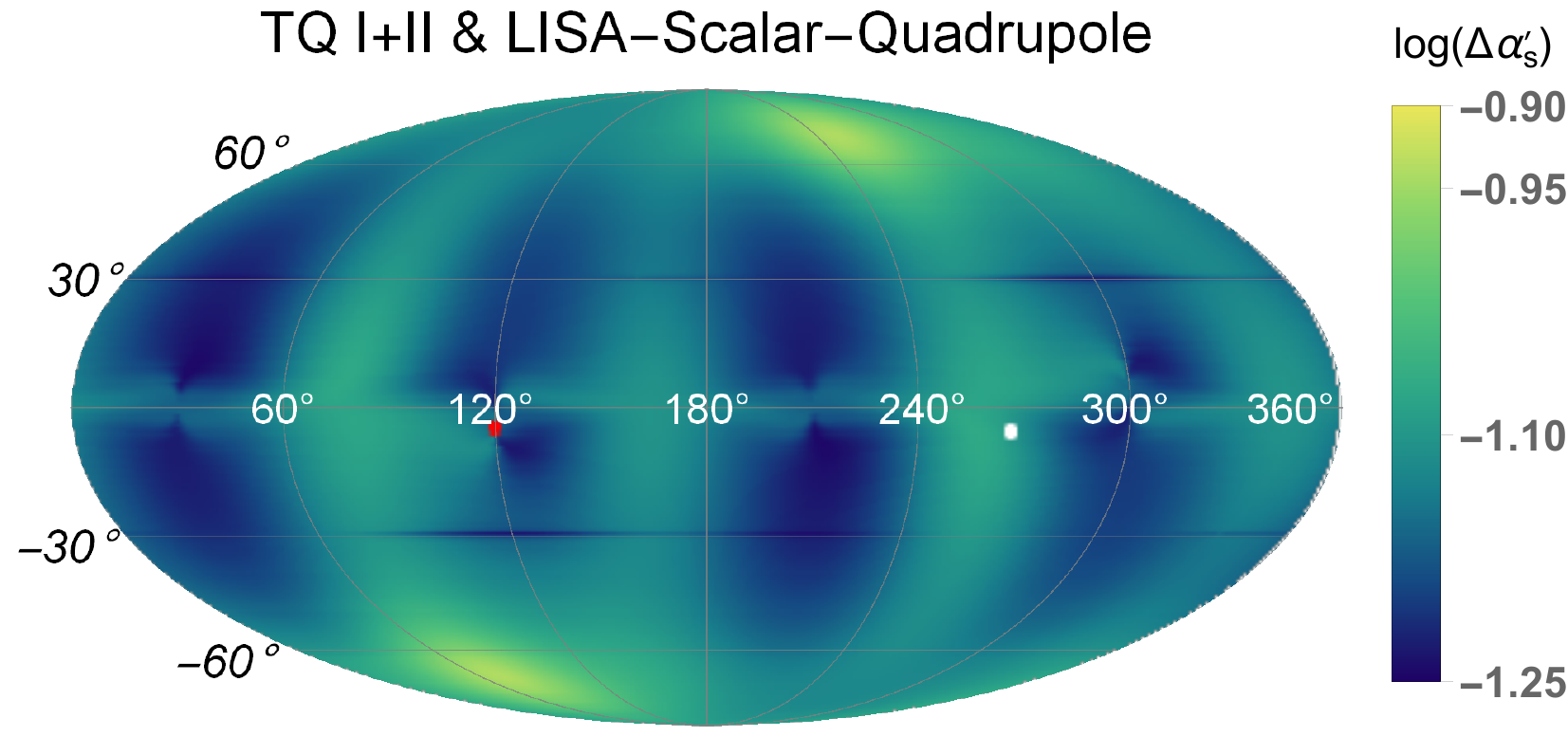}
	\caption{The distribution of \(\Delta\alpha'_{v}\) and \(\Delta\alpha'_{s}\) on the celestial sphere in ecliptic coordinate for the joint detection of TianQin twin constellation and LISA. 
	}\label{qtTL} 
\end{figure}

\bibliography{polarization}

\providecommand{\noopsort}[1]{}\providecommand{\singleletter}[1]{#1}%
\begin{thebibliography}{80}%
\makeatletter
\providecommand \@ifxundefined [1]{%
 \@ifx{#1\undefined}
}%
\providecommand \@ifnum [1]{%
 \ifnum #1\expandafter \@firstoftwo
 \else \expandafter \@secondoftwo
 \fi
}%
\providecommand \@ifx [1]{%
 \ifx #1\expandafter \@firstoftwo
 \else \expandafter \@secondoftwo
 \fi
}%
\providecommand \natexlab [1]{#1}%
\providecommand \enquote  [1]{``#1''}%
\providecommand \bibnamefont  [1]{#1}%
\providecommand \bibfnamefont [1]{#1}%
\providecommand \citenamefont [1]{#1}%
\providecommand \href@noop [0]{\@secondoftwo}%
\providecommand \href [0]{\begingroup \@sanitize@url \@href}%
\providecommand \@href[1]{\@@startlink{#1}\@@href}%
\providecommand \@@href[1]{\endgroup#1\@@endlink}%
\providecommand \@sanitize@url [0]{\catcode `\\12\catcode `\$12\catcode
  `\&12\catcode `\#12\catcode `\^12\catcode `\_12\catcode `\%12\relax}%
\providecommand \@@startlink[1]{}%
\providecommand \@@endlink[0]{}%
\providecommand \url  [0]{\begingroup\@sanitize@url \@url }%
\providecommand \@url [1]{\endgroup\@href {#1}{\urlprefix }}%
\providecommand \urlprefix  [0]{URL }%
\providecommand \Eprint [0]{\href }%
\providecommand \doibase [0]{http://dx.doi.org/}%
\providecommand \selectlanguage [0]{\@gobble}%
\providecommand \bibinfo  [0]{\@secondoftwo}%
\providecommand \bibfield  [0]{\@secondoftwo}%
\providecommand \translation [1]{[#1]}%
\providecommand \BibitemOpen [0]{}%
\providecommand \bibitemStop [0]{}%
\providecommand \bibitemNoStop [0]{.\EOS\space}%
\providecommand \EOS [0]{\spacefactor3000\relax}%
\providecommand \BibitemShut  [1]{\csname bibitem#1\endcsname}%
\let\auto@bib@innerbib\@empty
\bibitem [{\citenamefont {Abbott}\ \emph
  {et~al.}(2016{\natexlab{a}})\citenamefont {Abbott} \emph {et~al.}}]{2016LV}%
  \BibitemOpen
  \bibfield  {author} {\bibinfo {author} {\bibfnamefont {B.}~\bibnamefont
  {Abbott}} \emph {et~al.} (\bibinfo {collaboration} {LIGO Scientific
  Collaboration and Virgo Collaboration}),\ }\href {\doibase
  10.1103/PhysRevLett.116.061102} {\bibfield  {journal} {\bibinfo  {journal}
  {Phys. Rev. Lett.}\ }\textbf {\bibinfo {volume} {116}},\ \bibinfo {pages}
  {061102} (\bibinfo {year} {2016}{\natexlab{a}})},\ \Eprint
  {http://arxiv.org/abs/1602.03837} {arXiv:1602.03837 [gr-qc]} \BibitemShut
  {NoStop}%
\bibitem [{\citenamefont {Abbott}\ \emph
  {et~al.}(2016{\natexlab{b}})\citenamefont {Abbott} \emph
  {et~al.}}]{2016LIGO}%
  \BibitemOpen
  \bibfield  {author} {\bibinfo {author} {\bibfnamefont {B.~P.}\ \bibnamefont
  {Abbott}} \emph {et~al.} (\bibinfo {collaboration} {LIGO Scientific,
  Virgo}),\ }\href {\doibase 10.1103/PhysRevX.6.041015} {\bibfield  {journal}
  {\bibinfo  {journal} {Phys. Rev. X}\ }\textbf {\bibinfo {volume} {6}},\
  \bibinfo {pages} {041015} (\bibinfo {year} {2016}{\natexlab{b}})},\ \Eprint
  {http://arxiv.org/abs/1606.04856} {arXiv:1606.04856 [gr-qc]} \BibitemShut
  {NoStop}%
\bibitem [{\citenamefont {Abbott}\ \emph
  {et~al.}(2019{\natexlab{a}})\citenamefont {Abbott} \emph
  {et~al.}}]{2019LIGO}%
  \BibitemOpen
  \bibfield  {author} {\bibinfo {author} {\bibfnamefont {B.~P.}\ \bibnamefont
  {Abbott}} \emph {et~al.} (\bibinfo {collaboration} {LIGO Scientific,
  Virgo}),\ }\href {\doibase 10.1103/PhysRevX.9.031040} {\bibfield  {journal}
  {\bibinfo  {journal} {Phys. Rev. X}\ }\textbf {\bibinfo {volume} {9}},\
  \bibinfo {pages} {031040} (\bibinfo {year} {2019}{\natexlab{a}})},\ \Eprint
  {http://arxiv.org/abs/1811.12907} {arXiv:1811.12907 [astro-ph.HE]}
  \BibitemShut {NoStop}%
\bibitem [{\citenamefont {Abbott}\ \emph
  {et~al.}(2021{\natexlab{a}})\citenamefont {Abbott} \emph
  {et~al.}}]{2020LIGO}%
  \BibitemOpen
  \bibfield  {author} {\bibinfo {author} {\bibfnamefont {R.}~\bibnamefont
  {Abbott}} \emph {et~al.} (\bibinfo {collaboration} {LIGO Scientific,
  Virgo}),\ }\href {\doibase 10.1103/PhysRevX.11.021053} {\bibfield  {journal}
  {\bibinfo  {journal} {Phys. Rev. X}\ }\textbf {\bibinfo {volume} {11}},\
  \bibinfo {pages} {021053} (\bibinfo {year} {2021}{\natexlab{a}})},\ \Eprint
  {http://arxiv.org/abs/2010.14527} {arXiv:2010.14527 [gr-qc]} \BibitemShut
  {NoStop}%
\bibitem [{\citenamefont {Abbott}\ \emph
  {et~al.}(2021{\natexlab{b}})\citenamefont {Abbott} \emph
  {et~al.}}]{2021LIGOE}%
  \BibitemOpen
  \bibfield  {author} {\bibinfo {author} {\bibfnamefont {R.}~\bibnamefont
  {Abbott}} \emph {et~al.} (\bibinfo {collaboration} {LIGO Scientific,
  VIRGO}),\ }\href@noop {} {\  (\bibinfo {year} {2021}{\natexlab{b}})},\
  \Eprint {http://arxiv.org/abs/2108.01045} {arXiv:2108.01045 [gr-qc]}
  \BibitemShut {NoStop}%
\bibitem [{\citenamefont {Abbott}\ \emph
  {et~al.}(2021{\natexlab{c}})\citenamefont {Abbott} \emph
  {et~al.}}]{2021LIGO}%
  \BibitemOpen
  \bibfield  {author} {\bibinfo {author} {\bibfnamefont {R.}~\bibnamefont
  {Abbott}} \emph {et~al.} (\bibinfo {collaboration} {LIGO Scientific, VIRGO,
  KAGRA}),\ }\href@noop {} {\  (\bibinfo {year} {2021}{\natexlab{c}})},\
  \Eprint {http://arxiv.org/abs/2111.03606} {arXiv:2111.03606 [gr-qc]}
  \BibitemShut {NoStop}%
\bibitem [{\citenamefont {Abbott}\ \emph
  {et~al.}(2016{\natexlab{c}})\citenamefont {Abbott} \emph
  {et~al.}}]{GW150914}%
  \BibitemOpen
  \bibfield  {author} {\bibinfo {author} {\bibfnamefont {B.~P.}\ \bibnamefont
  {Abbott}} \emph {et~al.} (\bibinfo {collaboration} {LIGO Scientific,
  Virgo}),\ }\href {\doibase 10.1103/PhysRevLett.116.221101} {\bibfield
  {journal} {\bibinfo  {journal} {Phys. Rev. Lett.}\ }\textbf {\bibinfo
  {volume} {116}},\ \bibinfo {pages} {221101} (\bibinfo {year}
  {2016}{\natexlab{c}})},\ \bibinfo {note} {[Erratum: Phys.Rev.Lett. 121,
  129902 (2018)]},\ \Eprint {http://arxiv.org/abs/1602.03841} {arXiv:1602.03841
  [gr-qc]} \BibitemShut {NoStop}%
\bibitem [{\citenamefont {Abbott}\ \emph
  {et~al.}(2019{\natexlab{b}})\citenamefont {Abbott} \emph
  {et~al.}}]{GW170817}%
  \BibitemOpen
  \bibfield  {author} {\bibinfo {author} {\bibfnamefont {B.}~\bibnamefont
  {Abbott}} \emph {et~al.} (\bibinfo {collaboration} {LIGO Scientific,
  Virgo}),\ }\href {\doibase 10.1103/PhysRevLett.123.011102} {\bibfield
  {journal} {\bibinfo  {journal} {Phys. Rev. Lett.}\ }\textbf {\bibinfo
  {volume} {123}},\ \bibinfo {pages} {011102} (\bibinfo {year}
  {2019}{\natexlab{b}})},\ \Eprint {http://arxiv.org/abs/1811.00364}
  {arXiv:1811.00364 [gr-qc]} \BibitemShut {NoStop}%
\bibitem [{\citenamefont {Abbott}\ \emph
  {et~al.}(2019{\natexlab{c}})\citenamefont {Abbott} \emph {et~al.}}]{GWTC}%
  \BibitemOpen
  \bibfield  {author} {\bibinfo {author} {\bibfnamefont {B.}~\bibnamefont
  {Abbott}} \emph {et~al.} (\bibinfo {collaboration} {LIGO Scientific,
  Virgo}),\ }\href {\doibase 10.1103/PhysRevD.100.104036} {\bibfield  {journal}
  {\bibinfo  {journal} {Phys. Rev. D}\ }\textbf {\bibinfo {volume} {100}},\
  \bibinfo {pages} {104036} (\bibinfo {year} {2019}{\natexlab{c}})},\ \Eprint
  {http://arxiv.org/abs/1903.04467} {arXiv:1903.04467 [gr-qc]} \BibitemShut
  {NoStop}%
\bibitem [{\citenamefont {Abbott}\ \emph
  {et~al.}(2021{\natexlab{d}})\citenamefont {Abbott} \emph
  {et~al.}}]{2020GWTC2}%
  \BibitemOpen
  \bibfield  {author} {\bibinfo {author} {\bibfnamefont {R.}~\bibnamefont
  {Abbott}} \emph {et~al.} (\bibinfo {collaboration} {LIGO Scientific,
  Virgo}),\ }\href {\doibase 10.1103/PhysRevD.103.122002} {\bibfield  {journal}
  {\bibinfo  {journal} {Phys. Rev. D}\ }\textbf {\bibinfo {volume} {103}},\
  \bibinfo {pages} {122002} (\bibinfo {year} {2021}{\natexlab{d}})},\ \Eprint
  {http://arxiv.org/abs/2010.14529} {arXiv:2010.14529 [gr-qc]} \BibitemShut
  {NoStop}%
\bibitem [{\citenamefont {Abbott}\ \emph
  {et~al.}(2021{\natexlab{e}})\citenamefont {Abbott} \emph {et~al.}}]{GWTC3}%
  \BibitemOpen
  \bibfield  {author} {\bibinfo {author} {\bibfnamefont {R.}~\bibnamefont
  {Abbott}} \emph {et~al.} (\bibinfo {collaboration} {LIGO Scientific, VIRGO,
  KAGRA}),\ }\href@noop {} {\  (\bibinfo {year} {2021}{\natexlab{e}})},\
  \Eprint {http://arxiv.org/abs/2112.06861} {arXiv:2112.06861 [gr-qc]}
  \BibitemShut {NoStop}%
\bibitem [{\citenamefont {Punturo}\ \emph {et~al.}(2010)\citenamefont {Punturo}
  \emph {et~al.}}]{2010ET}%
  \BibitemOpen
  \bibfield  {author} {\bibinfo {author} {\bibfnamefont {M.}~\bibnamefont
  {Punturo}} \emph {et~al.},\ }\href {\doibase 10.1088/0264-9381/27/19/194002}
  {\bibfield  {journal} {\bibinfo  {journal} {Class. Quant. Grav.}\ }\textbf
  {\bibinfo {volume} {27}},\ \bibinfo {pages} {194002} (\bibinfo {year}
  {2010})}\BibitemShut {NoStop}%
\bibitem [{\citenamefont {Abbott}\ \emph
  {et~al.}(2017{\natexlab{a}})\citenamefont {Abbott} \emph {et~al.}}]{2016CE}%
  \BibitemOpen
  \bibfield  {author} {\bibinfo {author} {\bibfnamefont {B.~P.}\ \bibnamefont
  {Abbott}} \emph {et~al.} (\bibinfo {collaboration} {LIGO Scientific}),\
  }\href {\doibase 10.1088/1361-6382/aa51f4} {\bibfield  {journal} {\bibinfo
  {journal} {Class. Quant. Grav.}\ }\textbf {\bibinfo {volume} {34}},\ \bibinfo
  {pages} {044001} (\bibinfo {year} {2017}{\natexlab{a}})},\ \Eprint
  {http://arxiv.org/abs/1607.08697} {arXiv:1607.08697 [astro-ph.IM]}
  \BibitemShut {NoStop}%
\bibitem [{\citenamefont {Luo}\ \emph {et~al.}(2016)\citenamefont {Luo} \emph
  {et~al.}}]{2016TQ}%
  \BibitemOpen
  \bibfield  {author} {\bibinfo {author} {\bibfnamefont {J.}~\bibnamefont
  {Luo}} \emph {et~al.} (\bibinfo {collaboration} {TianQin}),\ }\href {\doibase
  10.1088/0264-9381/33/3/035010} {\bibfield  {journal} {\bibinfo  {journal}
  {Class. Quant. Grav.}\ }\textbf {\bibinfo {volume} {33}},\ \bibinfo {pages}
  {035010} (\bibinfo {year} {2016})},\ \Eprint
  {http://arxiv.org/abs/1512.02076} {arXiv:1512.02076 [astro-ph.IM]}
  \BibitemShut {NoStop}%
\bibitem [{\citenamefont {Amaro-Seoane}\ \emph {et~al.}(2017)\citenamefont
  {Amaro-Seoane} \emph {et~al.}}]{2017LISA}%
  \BibitemOpen
  \bibfield  {author} {\bibinfo {author} {\bibfnamefont {P.}~\bibnamefont
  {Amaro-Seoane}} \emph {et~al.} (\bibinfo {collaboration} {LISA}),\
  }\href@noop {} {\enquote {\bibinfo {title} {{Laser Interferometer Space
  Antenna}},}\ } (\bibinfo {year} {2017}),\ \Eprint
  {http://arxiv.org/abs/1702.00786} {arXiv:1702.00786 [astro-ph.IM]}
  \BibitemShut {NoStop}%
\bibitem [{\citenamefont {Eardley}\ \emph
  {et~al.}(1973{\natexlab{a}})\citenamefont {Eardley}, \citenamefont {Lee},
  \citenamefont {Lightman}, \citenamefont {Wagoner},\ and\ \citenamefont
  {Will}}]{1973LightmanLee}%
  \BibitemOpen
  \bibfield  {author} {\bibinfo {author} {\bibfnamefont {D.}~\bibnamefont
  {Eardley}}, \bibinfo {author} {\bibfnamefont {D.}~\bibnamefont {Lee}},
  \bibinfo {author} {\bibfnamefont {A.}~\bibnamefont {Lightman}}, \bibinfo
  {author} {\bibfnamefont {R.}~\bibnamefont {Wagoner}}, \ and\ \bibinfo
  {author} {\bibfnamefont {C.}~\bibnamefont {Will}},\ }\href {\doibase
  10.1103/PhysRevLett.30.884} {\bibfield  {journal} {\bibinfo  {journal} {Phys.
  Rev. Lett.}\ }\textbf {\bibinfo {volume} {30}},\ \bibinfo {pages} {884}
  (\bibinfo {year} {1973}{\natexlab{a}})}\BibitemShut {NoStop}%
\bibitem [{\citenamefont {Eardley}\ \emph
  {et~al.}(1973{\natexlab{b}})\citenamefont {Eardley}, \citenamefont {Lee},\
  and\ \citenamefont {Lightman}}]{1973L2}%
  \BibitemOpen
  \bibfield  {author} {\bibinfo {author} {\bibfnamefont {D.}~\bibnamefont
  {Eardley}}, \bibinfo {author} {\bibfnamefont {D.}~\bibnamefont {Lee}}, \ and\
  \bibinfo {author} {\bibfnamefont {A.}~\bibnamefont {Lightman}},\ }\href
  {\doibase 10.1103/PhysRevD.8.3308} {\bibfield  {journal} {\bibinfo  {journal}
  {Phys. Rev. D}\ }\textbf {\bibinfo {volume} {8}},\ \bibinfo {pages} {3308}
  (\bibinfo {year} {1973}{\natexlab{b}})}\BibitemShut {NoStop}%
\bibitem [{\citenamefont {Gong}\ and\ \citenamefont
  {Hou}(2018{\natexlab{a}})}]{Gong}%
  \BibitemOpen
  \bibfield  {author} {\bibinfo {author} {\bibfnamefont {Y.}~\bibnamefont
  {Gong}}\ and\ \bibinfo {author} {\bibfnamefont {S.}~\bibnamefont {Hou}},\
  }\href {\doibase 10.3390/universe4080085} {\bibfield  {journal} {\bibinfo
  {journal} {Universe}\ }\textbf {\bibinfo {volume} {4}},\ \bibinfo {pages}
  {85} (\bibinfo {year} {2018}{\natexlab{a}})},\ \Eprint
  {http://arxiv.org/abs/1806.04027} {arXiv:1806.04027 [gr-qc]} \BibitemShut
  {NoStop}%
\bibitem [{\citenamefont {Will}(2014)}]{WillLiv}%
  \BibitemOpen
  \bibfield  {author} {\bibinfo {author} {\bibfnamefont {C.~M.}\ \bibnamefont
  {Will}},\ }\href {\doibase 10.12942/lrr-2014-4} {\bibfield  {journal}
  {\bibinfo  {journal} {Living Rev. Rel.}\ }\textbf {\bibinfo {volume} {17}},\
  \bibinfo {pages} {4} (\bibinfo {year} {2014})},\ \Eprint
  {http://arxiv.org/abs/1403.7377} {arXiv:1403.7377 [gr-qc]} \BibitemShut
  {NoStop}%
\bibitem [{\citenamefont {Maggiore}\ and\ \citenamefont
  {Nicolis}(2000)}]{2000Magg}%
  \BibitemOpen
  \bibfield  {author} {\bibinfo {author} {\bibfnamefont {M.}~\bibnamefont
  {Maggiore}}\ and\ \bibinfo {author} {\bibfnamefont {A.}~\bibnamefont
  {Nicolis}},\ }\href {\doibase 10.1103/PhysRevD.62.024004} {\bibfield
  {journal} {\bibinfo  {journal} {Phys. Rev. D}\ }\textbf {\bibinfo {volume}
  {62}},\ \bibinfo {pages} {024004} (\bibinfo {year} {2000})},\ \Eprint
  {http://arxiv.org/abs/gr-qc/9907055} {arXiv:gr-qc/9907055} \BibitemShut
  {NoStop}%
\bibitem [{\citenamefont {Capozziello}\ and\ \citenamefont
  {Corda}(2006)}]{2006Cap}%
  \BibitemOpen
  \bibfield  {author} {\bibinfo {author} {\bibfnamefont {S.}~\bibnamefont
  {Capozziello}}\ and\ \bibinfo {author} {\bibfnamefont {C.}~\bibnamefont
  {Corda}},\ }\href {\doibase 10.1142/S0218271806008814} {\bibfield  {journal}
  {\bibinfo  {journal} {Int. J. Mod. Phys. D}\ }\textbf {\bibinfo {volume}
  {15}},\ \bibinfo {pages} {1119} (\bibinfo {year} {2006})}\BibitemShut
  {NoStop}%
\bibitem [{\citenamefont {Rizwana~Kausar}\ \emph {et~al.}(2016)\citenamefont
  {Rizwana~Kausar}, \citenamefont {Philippoz},\ and\ \citenamefont
  {Jetzer}}]{2016Kausar}%
  \BibitemOpen
  \bibfield  {author} {\bibinfo {author} {\bibfnamefont {H.}~\bibnamefont
  {Rizwana~Kausar}}, \bibinfo {author} {\bibfnamefont {L.}~\bibnamefont
  {Philippoz}}, \ and\ \bibinfo {author} {\bibfnamefont {P.}~\bibnamefont
  {Jetzer}},\ }\href {\doibase 10.1103/PhysRevD.93.124071} {\bibfield
  {journal} {\bibinfo  {journal} {Phys. Rev. D}\ }\textbf {\bibinfo {volume}
  {93}},\ \bibinfo {pages} {124071} (\bibinfo {year} {2016})},\ \Eprint
  {http://arxiv.org/abs/1606.07000} {arXiv:1606.07000 [gr-qc]} \BibitemShut
  {NoStop}%
\bibitem [{\citenamefont {Gong}\ and\ \citenamefont
  {Hou}(2018{\natexlab{b}})}]{2018Gong}%
  \BibitemOpen
  \bibfield  {author} {\bibinfo {author} {\bibfnamefont {Y.}~\bibnamefont
  {Gong}}\ and\ \bibinfo {author} {\bibfnamefont {S.}~\bibnamefont {Hou}},\
  }\href {\doibase 10.1051/epjconf/201816801003} {\bibfield  {journal}
  {\bibinfo  {journal} {EPJ Web Conf.}\ }\textbf {\bibinfo {volume} {168}},\
  \bibinfo {pages} {01003} (\bibinfo {year} {2018}{\natexlab{b}})},\ \Eprint
  {http://arxiv.org/abs/1709.03313} {arXiv:1709.03313 [gr-qc]} \BibitemShut
  {NoStop}%
\bibitem [{\citenamefont {Katsuragawa}\ \emph {et~al.}(2019)\citenamefont
  {Katsuragawa}, \citenamefont {Nakamura}, \citenamefont {Ikeda},\ and\
  \citenamefont {Capozziello}}]{2019Katsuragawa}%
  \BibitemOpen
  \bibfield  {author} {\bibinfo {author} {\bibfnamefont {T.}~\bibnamefont
  {Katsuragawa}}, \bibinfo {author} {\bibfnamefont {T.}~\bibnamefont
  {Nakamura}}, \bibinfo {author} {\bibfnamefont {T.}~\bibnamefont {Ikeda}}, \
  and\ \bibinfo {author} {\bibfnamefont {S.}~\bibnamefont {Capozziello}},\
  }\href {\doibase 10.1103/PhysRevD.99.124050} {\bibfield  {journal} {\bibinfo
  {journal} {Phys. Rev. D}\ }\textbf {\bibinfo {volume} {99}},\ \bibinfo
  {pages} {124050} (\bibinfo {year} {2019})},\ \Eprint
  {http://arxiv.org/abs/1902.02494} {arXiv:1902.02494 [gr-qc]} \BibitemShut
  {NoStop}%
\bibitem [{\citenamefont {Moretti}\ \emph {et~al.}(2019)\citenamefont
  {Moretti}, \citenamefont {Bombacigno},\ and\ \citenamefont
  {Montani}}]{2019Moretti}%
  \BibitemOpen
  \bibfield  {author} {\bibinfo {author} {\bibfnamefont {F.}~\bibnamefont
  {Moretti}}, \bibinfo {author} {\bibfnamefont {F.}~\bibnamefont {Bombacigno}},
  \ and\ \bibinfo {author} {\bibfnamefont {G.}~\bibnamefont {Montani}},\ }\href
  {\doibase 10.1103/PhysRevD.100.084014} {\bibfield  {journal} {\bibinfo
  {journal} {Phys. Rev. D}\ }\textbf {\bibinfo {volume} {100}},\ \bibinfo
  {pages} {084014} (\bibinfo {year} {2019})},\ \Eprint
  {http://arxiv.org/abs/1906.01899} {arXiv:1906.01899 [gr-qc]} \BibitemShut
  {NoStop}%
\bibitem [{\citenamefont {Gong}\ \emph {et~al.}(2018)\citenamefont {Gong},
  \citenamefont {Hou}, \citenamefont {Liang},\ and\ \citenamefont
  {Papantonopoulos}}]{2018Gong2}%
  \BibitemOpen
  \bibfield  {author} {\bibinfo {author} {\bibfnamefont {Y.}~\bibnamefont
  {Gong}}, \bibinfo {author} {\bibfnamefont {S.}~\bibnamefont {Hou}}, \bibinfo
  {author} {\bibfnamefont {D.}~\bibnamefont {Liang}}, \ and\ \bibinfo {author}
  {\bibfnamefont {E.}~\bibnamefont {Papantonopoulos}},\ }\href {\doibase
  10.1103/PhysRevD.97.084040} {\bibfield  {journal} {\bibinfo  {journal} {Phys.
  Rev. D}\ }\textbf {\bibinfo {volume} {97}},\ \bibinfo {pages} {084040}
  (\bibinfo {year} {2018})},\ \Eprint {http://arxiv.org/abs/1801.03382}
  {arXiv:1801.03382 [gr-qc]} \BibitemShut {NoStop}%
\bibitem [{\citenamefont {Sagi}(2010)}]{2010Sagi}%
  \BibitemOpen
  \bibfield  {author} {\bibinfo {author} {\bibfnamefont {E.}~\bibnamefont
  {Sagi}},\ }\href {\doibase 10.1103/PhysRevD.81.064031} {\bibfield  {journal}
  {\bibinfo  {journal} {Phys. Rev. D}\ }\textbf {\bibinfo {volume} {81}},\
  \bibinfo {pages} {064031} (\bibinfo {year} {2010})},\ \Eprint
  {http://arxiv.org/abs/1001.1555} {arXiv:1001.1555 [gr-qc]} \BibitemShut
  {NoStop}%
\bibitem [{\citenamefont {de~Paula}\ \emph {et~al.}(2004)\citenamefont
  {de~Paula}, \citenamefont {Miranda},\ and\ \citenamefont
  {Marinho}}]{2004Paula}%
  \BibitemOpen
  \bibfield  {author} {\bibinfo {author} {\bibfnamefont {W.~L.}\ \bibnamefont
  {de~Paula}}, \bibinfo {author} {\bibfnamefont {O.~D.}\ \bibnamefont
  {Miranda}}, \ and\ \bibinfo {author} {\bibfnamefont {R.~M.}\ \bibnamefont
  {Marinho}},\ }\href {\doibase 10.1088/0264-9381/21/19/008} {\bibfield
  {journal} {\bibinfo  {journal} {Class. Quant. Grav.}\ }\textbf {\bibinfo
  {volume} {21}},\ \bibinfo {pages} {4595} (\bibinfo {year} {2004})},\ \Eprint
  {http://arxiv.org/abs/gr-qc/0409041} {arXiv:gr-qc/0409041} \BibitemShut
  {NoStop}%
\bibitem [{\citenamefont {Abbott}\ \emph
  {et~al.}(2017{\natexlab{b}})\citenamefont {Abbott} \emph
  {et~al.}}]{2017BHGW}%
  \BibitemOpen
  \bibfield  {author} {\bibinfo {author} {\bibfnamefont {B.}~\bibnamefont
  {Abbott}} \emph {et~al.} (\bibinfo {collaboration} {LIGO Scientific,
  Virgo}),\ }\href {\doibase 10.1103/PhysRevLett.119.141101} {\bibfield
  {journal} {\bibinfo  {journal} {Phys. Rev. Lett.}\ }\textbf {\bibinfo
  {volume} {119}},\ \bibinfo {pages} {141101} (\bibinfo {year}
  {2017}{\natexlab{b}})},\ \Eprint {http://arxiv.org/abs/1709.09660}
  {arXiv:1709.09660 [gr-qc]} \BibitemShut {NoStop}%
\bibitem [{\citenamefont {Guersel}\ and\ \citenamefont
  {Tinto}(1989)}]{1989Guersel}%
  \BibitemOpen
  \bibfield  {author} {\bibinfo {author} {\bibfnamefont {Y.}~\bibnamefont
  {Guersel}}\ and\ \bibinfo {author} {\bibfnamefont {M.}~\bibnamefont
  {Tinto}},\ }\href {\doibase 10.1103/PhysRevD.40.3884} {\bibfield  {journal}
  {\bibinfo  {journal} {Phys. Rev. D}\ }\textbf {\bibinfo {volume} {40}},\
  \bibinfo {pages} {3884} (\bibinfo {year} {1989})}\BibitemShut {NoStop}%
\bibitem [{\citenamefont {Wen}\ and\ \citenamefont {Schutz}(2005)}]{2005Wen}%
  \BibitemOpen
  \bibfield  {author} {\bibinfo {author} {\bibfnamefont {L.}~\bibnamefont
  {Wen}}\ and\ \bibinfo {author} {\bibfnamefont {B.~F.}\ \bibnamefont
  {Schutz}},\ }\href {\doibase 10.1088/0264-9381/22/18/S46} {\bibfield
  {journal} {\bibinfo  {journal} {Class. Quant. Grav.}\ }\textbf {\bibinfo
  {volume} {22}},\ \bibinfo {pages} {S1321} (\bibinfo {year} {2005})},\ \Eprint
  {http://arxiv.org/abs/gr-qc/0508042} {arXiv:gr-qc/0508042} \BibitemShut
  {NoStop}%
\bibitem [{\citenamefont {Wen}(2008)}]{2007Wen}%
  \BibitemOpen
  \bibfield  {author} {\bibinfo {author} {\bibfnamefont {L.}~\bibnamefont
  {Wen}},\ }\href {\doibase 10.1142/S0218271808012723} {\bibfield  {journal}
  {\bibinfo  {journal} {Int. J. Mod. Phys. D}\ }\textbf {\bibinfo {volume}
  {17}},\ \bibinfo {pages} {1095} (\bibinfo {year} {2008})},\ \Eprint
  {http://arxiv.org/abs/gr-qc/0702096} {arXiv:gr-qc/0702096} \BibitemShut
  {NoStop}%
\bibitem [{\citenamefont {Chatterji}\ \emph {et~al.}(2006)\citenamefont
  {Chatterji}, \citenamefont {Lazzarini}, \citenamefont {Stein}, \citenamefont
  {Sutton}, \citenamefont {Searle},\ and\ \citenamefont
  {Tinto}}]{2006Chatterji}%
  \BibitemOpen
  \bibfield  {author} {\bibinfo {author} {\bibfnamefont {S.}~\bibnamefont
  {Chatterji}}, \bibinfo {author} {\bibfnamefont {A.}~\bibnamefont
  {Lazzarini}}, \bibinfo {author} {\bibfnamefont {L.}~\bibnamefont {Stein}},
  \bibinfo {author} {\bibfnamefont {P.~J.}\ \bibnamefont {Sutton}}, \bibinfo
  {author} {\bibfnamefont {A.}~\bibnamefont {Searle}}, \ and\ \bibinfo {author}
  {\bibfnamefont {M.}~\bibnamefont {Tinto}},\ }\href {\doibase
  10.1103/PhysRevD.74.082005} {\bibfield  {journal} {\bibinfo  {journal} {Phys.
  Rev. D}\ }\textbf {\bibinfo {volume} {74}},\ \bibinfo {pages} {082005}
  (\bibinfo {year} {2006})},\ \Eprint {http://arxiv.org/abs/gr-qc/0605002}
  {arXiv:gr-qc/0605002} \BibitemShut {NoStop}%
\bibitem [{\citenamefont {Hagihara}\ \emph {et~al.}(2018)\citenamefont
  {Hagihara}, \citenamefont {Era}, \citenamefont {Iikawa},\ and\ \citenamefont
  {Asada}}]{2018Hagihara}%
  \BibitemOpen
  \bibfield  {author} {\bibinfo {author} {\bibfnamefont {Y.}~\bibnamefont
  {Hagihara}}, \bibinfo {author} {\bibfnamefont {N.}~\bibnamefont {Era}},
  \bibinfo {author} {\bibfnamefont {D.}~\bibnamefont {Iikawa}}, \ and\ \bibinfo
  {author} {\bibfnamefont {H.}~\bibnamefont {Asada}},\ }\href {\doibase
  10.1103/PhysRevD.98.064035} {\bibfield  {journal} {\bibinfo  {journal} {Phys.
  Rev. D}\ }\textbf {\bibinfo {volume} {98}},\ \bibinfo {pages} {064035}
  (\bibinfo {year} {2018})},\ \Eprint {http://arxiv.org/abs/1807.07234}
  {arXiv:1807.07234 [gr-qc]} \BibitemShut {NoStop}%
\bibitem [{\citenamefont {Hagihara}\ \emph {et~al.}(2019)\citenamefont
  {Hagihara}, \citenamefont {Era}, \citenamefont {Iikawa}, \citenamefont
  {Nishizawa},\ and\ \citenamefont {Asada}}]{2019Hagihara}%
  \BibitemOpen
  \bibfield  {author} {\bibinfo {author} {\bibfnamefont {Y.}~\bibnamefont
  {Hagihara}}, \bibinfo {author} {\bibfnamefont {N.}~\bibnamefont {Era}},
  \bibinfo {author} {\bibfnamefont {D.}~\bibnamefont {Iikawa}}, \bibinfo
  {author} {\bibfnamefont {A.}~\bibnamefont {Nishizawa}}, \ and\ \bibinfo
  {author} {\bibfnamefont {H.}~\bibnamefont {Asada}},\ }\href {\doibase
  10.1103/PhysRevD.100.064010} {\bibfield  {journal} {\bibinfo  {journal}
  {Phys. Rev. D}\ }\textbf {\bibinfo {volume} {100}},\ \bibinfo {pages}
  {064010} (\bibinfo {year} {2019})},\ \Eprint
  {http://arxiv.org/abs/1904.02300} {arXiv:1904.02300 [gr-qc]} \BibitemShut
  {NoStop}%
\bibitem [{\citenamefont {Hagihara}\ \emph {et~al.}(2020)\citenamefont
  {Hagihara}, \citenamefont {Era}, \citenamefont {Iikawa}, \citenamefont
  {Takeda},\ and\ \citenamefont {Asada}}]{2019HagiharaY}%
  \BibitemOpen
  \bibfield  {author} {\bibinfo {author} {\bibfnamefont {Y.}~\bibnamefont
  {Hagihara}}, \bibinfo {author} {\bibfnamefont {N.}~\bibnamefont {Era}},
  \bibinfo {author} {\bibfnamefont {D.}~\bibnamefont {Iikawa}}, \bibinfo
  {author} {\bibfnamefont {N.}~\bibnamefont {Takeda}}, \ and\ \bibinfo {author}
  {\bibfnamefont {H.}~\bibnamefont {Asada}},\ }\href {\doibase
  10.1103/PhysRevD.101.041501} {\bibfield  {journal} {\bibinfo  {journal}
  {Phys. Rev. D}\ }\textbf {\bibinfo {volume} {101}},\ \bibinfo {pages}
  {041501} (\bibinfo {year} {2020})},\ \Eprint
  {http://arxiv.org/abs/1912.06340} {arXiv:1912.06340 [gr-qc]} \BibitemShut
  {NoStop}%
\bibitem [{\citenamefont {Pang}\ \emph {et~al.}(2020)\citenamefont {Pang},
  \citenamefont {Lo}, \citenamefont {Wong}, \citenamefont {Li},\ and\
  \citenamefont {Van Den~Broeck}}]{2020ns}%
  \BibitemOpen
  \bibfield  {author} {\bibinfo {author} {\bibfnamefont {P.~T.}\ \bibnamefont
  {Pang}}, \bibinfo {author} {\bibfnamefont {R.~K.}\ \bibnamefont {Lo}},
  \bibinfo {author} {\bibfnamefont {I.~C.}\ \bibnamefont {Wong}}, \bibinfo
  {author} {\bibfnamefont {T.~G.}\ \bibnamefont {Li}}, \ and\ \bibinfo {author}
  {\bibfnamefont {C.}~\bibnamefont {Van Den~Broeck}},\ }\href {\doibase
  10.1103/PhysRevD.101.104055} {\bibfield  {journal} {\bibinfo  {journal}
  {Phys. Rev. D}\ }\textbf {\bibinfo {volume} {101}},\ \bibinfo {pages}
  {104055} (\bibinfo {year} {2020})},\ \Eprint
  {http://arxiv.org/abs/2003.07375} {arXiv:2003.07375 [gr-qc]} \BibitemShut
  {NoStop}%
\bibitem [{\citenamefont {Wong}\ \emph {et~al.}(2021)\citenamefont {Wong},
  \citenamefont {Pang}, \citenamefont {Lo}, \citenamefont {Li},\ and\
  \citenamefont {Broeck}}]{2021ns}%
  \BibitemOpen
  \bibfield  {author} {\bibinfo {author} {\bibfnamefont {I.~C.~F.}\
  \bibnamefont {Wong}}, \bibinfo {author} {\bibfnamefont {P.~T.~H.}\
  \bibnamefont {Pang}}, \bibinfo {author} {\bibfnamefont {R.~K.~L.}\
  \bibnamefont {Lo}}, \bibinfo {author} {\bibfnamefont {T.~G.~F.}\ \bibnamefont
  {Li}}, \ and\ \bibinfo {author} {\bibfnamefont {C.~V.~D.}\ \bibnamefont
  {Broeck}},\ }\href@noop {} {\enquote {\bibinfo {title} {Null-stream-based
  bayesian unmodeled framework to probe generic gravitational-wave
  polarizations},}\ } (\bibinfo {year} {2021}),\ \Eprint
  {http://arxiv.org/abs/2105.09485} {arXiv:2105.09485 [gr-qc]} \BibitemShut
  {NoStop}%
\bibitem [{\citenamefont {Takeda}\ \emph {et~al.}(2022)\citenamefont {Takeda},
  \citenamefont {Morisaki},\ and\ \citenamefont {Nishizawa}}]{2021Takeda}%
  \BibitemOpen
  \bibfield  {author} {\bibinfo {author} {\bibfnamefont {H.}~\bibnamefont
  {Takeda}}, \bibinfo {author} {\bibfnamefont {S.}~\bibnamefont {Morisaki}}, \
  and\ \bibinfo {author} {\bibfnamefont {A.}~\bibnamefont {Nishizawa}},\ }\href
  {\doibase 10.1103/PhysRevD.105.084019} {\bibfield  {journal} {\bibinfo
  {journal} {Phys. Rev. D}\ }\textbf {\bibinfo {volume} {105}},\ \bibinfo
  {pages} {084019} (\bibinfo {year} {2022})},\ \Eprint
  {http://arxiv.org/abs/2105.00253} {arXiv:2105.00253 [gr-qc]} \BibitemShut
  {NoStop}%
\bibitem [{\citenamefont {Chatziioannou}\ \emph {et~al.}(2012)\citenamefont
  {Chatziioannou}, \citenamefont {Yunes},\ and\ \citenamefont
  {Cornish}}]{2012Chatziioannou}%
  \BibitemOpen
  \bibfield  {author} {\bibinfo {author} {\bibfnamefont {K.}~\bibnamefont
  {Chatziioannou}}, \bibinfo {author} {\bibfnamefont {N.}~\bibnamefont
  {Yunes}}, \ and\ \bibinfo {author} {\bibfnamefont {N.}~\bibnamefont
  {Cornish}},\ }\href {\doibase 10.1103/PhysRevD.86.022004} {\bibfield
  {journal} {\bibinfo  {journal} {Phys. Rev. D}\ }\textbf {\bibinfo {volume}
  {86}},\ \bibinfo {pages} {022004} (\bibinfo {year} {2012})},\ \bibinfo {note}
  {[Erratum: Phys.Rev.D 95, 129901 (2017)]},\ \Eprint
  {http://arxiv.org/abs/1204.2585} {arXiv:1204.2585 [gr-qc]} \BibitemShut
  {NoStop}%
\bibitem [{\citenamefont {Takeda}\ \emph {et~al.}(2018)\citenamefont {Takeda},
  \citenamefont {Nishizawa}, \citenamefont {Michimura}, \citenamefont {Nagano},
  \citenamefont {Komori}, \citenamefont {Ando},\ and\ \citenamefont
  {Hayama}}]{2018Takeda}%
  \BibitemOpen
  \bibfield  {author} {\bibinfo {author} {\bibfnamefont {H.}~\bibnamefont
  {Takeda}}, \bibinfo {author} {\bibfnamefont {A.}~\bibnamefont {Nishizawa}},
  \bibinfo {author} {\bibfnamefont {Y.}~\bibnamefont {Michimura}}, \bibinfo
  {author} {\bibfnamefont {K.}~\bibnamefont {Nagano}}, \bibinfo {author}
  {\bibfnamefont {K.}~\bibnamefont {Komori}}, \bibinfo {author} {\bibfnamefont
  {M.}~\bibnamefont {Ando}}, \ and\ \bibinfo {author} {\bibfnamefont
  {K.}~\bibnamefont {Hayama}},\ }\href {\doibase 10.1103/PhysRevD.98.022008}
  {\bibfield  {journal} {\bibinfo  {journal} {Phys. Rev. D}\ }\textbf {\bibinfo
  {volume} {98}},\ \bibinfo {pages} {022008} (\bibinfo {year} {2018})},\
  \Eprint {http://arxiv.org/abs/1806.02182} {arXiv:1806.02182 [gr-qc]}
  \BibitemShut {NoStop}%
\bibitem [{\citenamefont {Isi}\ \emph {et~al.}(2015)\citenamefont {Isi},
  \citenamefont {Weinstein}, \citenamefont {Mead},\ and\ \citenamefont
  {Pitkin}}]{2015Isi}%
  \BibitemOpen
  \bibfield  {author} {\bibinfo {author} {\bibfnamefont {M.}~\bibnamefont
  {Isi}}, \bibinfo {author} {\bibfnamefont {A.~J.}\ \bibnamefont {Weinstein}},
  \bibinfo {author} {\bibfnamefont {C.}~\bibnamefont {Mead}}, \ and\ \bibinfo
  {author} {\bibfnamefont {M.}~\bibnamefont {Pitkin}},\ }\href {\doibase
  10.1103/PhysRevD.91.082002} {\bibfield  {journal} {\bibinfo  {journal} {Phys.
  Rev. D}\ }\textbf {\bibinfo {volume} {91}},\ \bibinfo {pages} {082002}
  (\bibinfo {year} {2015})},\ \Eprint {http://arxiv.org/abs/1502.00333}
  {arXiv:1502.00333 [gr-qc]} \BibitemShut {NoStop}%
\bibitem [{\citenamefont {Isi}\ \emph {et~al.}(2017)\citenamefont {Isi},
  \citenamefont {Pitkin},\ and\ \citenamefont {Weinstein}}]{2017Isi}%
  \BibitemOpen
  \bibfield  {author} {\bibinfo {author} {\bibfnamefont {M.}~\bibnamefont
  {Isi}}, \bibinfo {author} {\bibfnamefont {M.}~\bibnamefont {Pitkin}}, \ and\
  \bibinfo {author} {\bibfnamefont {A.~J.}\ \bibnamefont {Weinstein}},\ }\href
  {\doibase 10.1103/PhysRevD.96.042001} {\bibfield  {journal} {\bibinfo
  {journal} {Phys. Rev. D}\ }\textbf {\bibinfo {volume} {96}},\ \bibinfo
  {pages} {042001} (\bibinfo {year} {2017})},\ \Eprint
  {http://arxiv.org/abs/1703.07530} {arXiv:1703.07530 [gr-qc]} \BibitemShut
  {NoStop}%
\bibitem [{\citenamefont {Abbott}\ \emph {et~al.}(2018)\citenamefont {Abbott}
  \emph {et~al.}}]{2017Abbott}%
  \BibitemOpen
  \bibfield  {author} {\bibinfo {author} {\bibfnamefont {B.~P.}\ \bibnamefont
  {Abbott}} \emph {et~al.} (\bibinfo {collaboration} {LIGO Scientific,
  Virgo}),\ }\href {\doibase 10.1103/PhysRevLett.120.031104} {\bibfield
  {journal} {\bibinfo  {journal} {Phys. Rev. Lett.}\ }\textbf {\bibinfo
  {volume} {120}},\ \bibinfo {pages} {031104} (\bibinfo {year} {2018})},\
  \Eprint {http://arxiv.org/abs/1709.09203} {arXiv:1709.09203 [gr-qc]}
  \BibitemShut {NoStop}%
\bibitem [{\citenamefont {Isi}\ \emph {et~al.}(2020)\citenamefont {Isi},
  \citenamefont {Mastrogiovanni}, \citenamefont {Pitkin},\ and\ \citenamefont
  {Piccinni}}]{2020puls}%
  \BibitemOpen
  \bibfield  {author} {\bibinfo {author} {\bibfnamefont {M.}~\bibnamefont
  {Isi}}, \bibinfo {author} {\bibfnamefont {S.}~\bibnamefont {Mastrogiovanni}},
  \bibinfo {author} {\bibfnamefont {M.}~\bibnamefont {Pitkin}}, \ and\ \bibinfo
  {author} {\bibfnamefont {O.~J.}\ \bibnamefont {Piccinni}},\ }\href {\doibase
  10.1103/PhysRevD.102.123027} {\bibfield  {journal} {\bibinfo  {journal}
  {Phys. Rev. D}\ }\textbf {\bibinfo {volume} {102}},\ \bibinfo {pages}
  {123027} (\bibinfo {year} {2020})},\ \Eprint
  {http://arxiv.org/abs/2010.12612} {arXiv:2010.12612 [gr-qc]} \BibitemShut
  {NoStop}%
\bibitem [{\citenamefont {da~Silva~Alves}\ and\ \citenamefont
  {Tinto}(2011)}]{2011daSilvaAlves}%
  \BibitemOpen
  \bibfield  {author} {\bibinfo {author} {\bibfnamefont {M.~E.}\ \bibnamefont
  {da~Silva~Alves}}\ and\ \bibinfo {author} {\bibfnamefont {M.}~\bibnamefont
  {Tinto}},\ }\href {\doibase 10.1103/PhysRevD.83.123529} {\bibfield  {journal}
  {\bibinfo  {journal} {Phys. Rev. D}\ }\textbf {\bibinfo {volume} {83}},\
  \bibinfo {pages} {123529} (\bibinfo {year} {2011})},\ \Eprint
  {http://arxiv.org/abs/1102.4824} {arXiv:1102.4824 [gr-qc]} \BibitemShut
  {NoStop}%
\bibitem [{\citenamefont {Lee}\ \emph {et~al.}(2008)\citenamefont {Lee},
  \citenamefont {Jenet},\ and\ \citenamefont {Price}}]{2008Lee}%
  \BibitemOpen
  \bibfield  {author} {\bibinfo {author} {\bibfnamefont {K.~J.}\ \bibnamefont
  {Lee}}, \bibinfo {author} {\bibfnamefont {F.~A.}\ \bibnamefont {Jenet}}, \
  and\ \bibinfo {author} {\bibfnamefont {R.~H.}\ \bibnamefont {Price}},\ }\href
  {\doibase 10.1086/591080} {\bibfield  {journal} {\bibinfo  {journal} {The
  Astrophysical Journal}\ }\textbf {\bibinfo {volume} {685}},\ \bibinfo {pages}
  {1304} (\bibinfo {year} {2008})}\BibitemShut {NoStop}%
\bibitem [{\citenamefont {Niu}\ and\ \citenamefont {Zhao}(2019)}]{2018Zhao}%
  \BibitemOpen
  \bibfield  {author} {\bibinfo {author} {\bibfnamefont {R.}~\bibnamefont
  {Niu}}\ and\ \bibinfo {author} {\bibfnamefont {W.}~\bibnamefont {Zhao}},\
  }\href {\doibase 10.1007/s11433-018-9340-6} {\bibfield  {journal} {\bibinfo
  {journal} {Sci. China Phys. Mech. Astron.}\ }\textbf {\bibinfo {volume}
  {62}},\ \bibinfo {pages} {970411} (\bibinfo {year} {2019})},\ \Eprint
  {http://arxiv.org/abs/1812.00208} {arXiv:1812.00208 [gr-qc]} \BibitemShut
  {NoStop}%
\bibitem [{\citenamefont {O'Beirne}\ \emph {et~al.}(2019)\citenamefont
  {O'Beirne}, \citenamefont {Cornish}, \citenamefont {Vigeland},\ and\
  \citenamefont {Taylor}}]{2019OBeirne}%
  \BibitemOpen
  \bibfield  {author} {\bibinfo {author} {\bibfnamefont {L.}~\bibnamefont
  {O'Beirne}}, \bibinfo {author} {\bibfnamefont {N.~J.}\ \bibnamefont
  {Cornish}}, \bibinfo {author} {\bibfnamefont {S.~J.}\ \bibnamefont
  {Vigeland}}, \ and\ \bibinfo {author} {\bibfnamefont {S.~R.}\ \bibnamefont
  {Taylor}},\ }\href {\doibase 10.1103/PhysRevD.99.124039} {\bibfield
  {journal} {\bibinfo  {journal} {Phys. Rev. D}\ }\textbf {\bibinfo {volume}
  {99}},\ \bibinfo {pages} {124039} (\bibinfo {year} {2019})},\ \Eprint
  {http://arxiv.org/abs/1904.02744} {arXiv:1904.02744 [gr-qc]} \BibitemShut
  {NoStop}%
\bibitem [{\citenamefont {Bo\^\i{}tier}\ \emph {et~al.}(2020)\citenamefont
  {Bo\^\i{}tier}, \citenamefont {Tiwari}, \citenamefont {Philippoz},\ and\
  \citenamefont {Jetzer}}]{2020Boitier}%
  \BibitemOpen
  \bibfield  {author} {\bibinfo {author} {\bibfnamefont {A.}~\bibnamefont
  {Bo\^\i{}tier}}, \bibinfo {author} {\bibfnamefont {S.}~\bibnamefont
  {Tiwari}}, \bibinfo {author} {\bibfnamefont {L.}~\bibnamefont {Philippoz}}, \
  and\ \bibinfo {author} {\bibfnamefont {P.}~\bibnamefont {Jetzer}},\ }\href
  {\doibase 10.1103/PhysRevD.102.064051} {\bibfield  {journal} {\bibinfo
  {journal} {Phys. Rev. D}\ }\textbf {\bibinfo {volume} {102}},\ \bibinfo
  {pages} {064051} (\bibinfo {year} {2020})},\ \Eprint
  {http://arxiv.org/abs/2008.13520} {arXiv:2008.13520 [gr-qc]} \BibitemShut
  {NoStop}%
\bibitem [{\citenamefont {Arzoumanian}\ \emph {et~al.}(2020)\citenamefont
  {Arzoumanian} \emph {et~al.}}]{NANOG}%
  \BibitemOpen
  \bibfield  {author} {\bibinfo {author} {\bibfnamefont {Z.}~\bibnamefont
  {Arzoumanian}} \emph {et~al.} (\bibinfo {collaboration} {NANOGrav}),\ }\href
  {\doibase 10.3847/2041-8213/abd401} {\bibfield  {journal} {\bibinfo
  {journal} {Astrophys. J. Lett.}\ }\textbf {\bibinfo {volume} {905}},\
  \bibinfo {pages} {L34} (\bibinfo {year} {2020})},\ \Eprint
  {http://arxiv.org/abs/2009.04496} {arXiv:2009.04496 [astro-ph.HE]}
  \BibitemShut {NoStop}%
\bibitem [{\citenamefont {Chen}\ \emph {et~al.}(2021)\citenamefont {Chen},
  \citenamefont {Yuan},\ and\ \citenamefont {Huang}}]{2021chen}%
  \BibitemOpen
  \bibfield  {author} {\bibinfo {author} {\bibfnamefont {Z.-C.}\ \bibnamefont
  {Chen}}, \bibinfo {author} {\bibfnamefont {C.}~\bibnamefont {Yuan}}, \ and\
  \bibinfo {author} {\bibfnamefont {Q.-G.}\ \bibnamefont {Huang}},\ }\href
  {\doibase 10.1007/s11433-021-1797-y} {\bibfield  {journal} {\bibinfo
  {journal} {Sci. China Phys. Mech. Astron.}\ }\textbf {\bibinfo {volume}
  {64}},\ \bibinfo {pages} {120412} (\bibinfo {year} {2021})},\ \Eprint
  {http://arxiv.org/abs/2101.06869} {arXiv:2101.06869 [astro-ph.CO]}
  \BibitemShut {NoStop}%
\bibitem [{\citenamefont {Arzoumanian}\ \emph {et~al.}(2021)\citenamefont
  {Arzoumanian} \emph {et~al.}}]{2021NANOGrav}%
  \BibitemOpen
  \bibfield  {author} {\bibinfo {author} {\bibfnamefont {Z.}~\bibnamefont
  {Arzoumanian}} \emph {et~al.} (\bibinfo {collaboration} {NANOGrav}),\ }\href
  {\doibase 10.3847/2041-8213/ac401c} {\bibfield  {journal} {\bibinfo
  {journal} {Astrophys. J. Lett.}\ }\textbf {\bibinfo {volume} {923}},\
  \bibinfo {pages} {L22} (\bibinfo {year} {2021})},\ \Eprint
  {http://arxiv.org/abs/2109.14706} {arXiv:2109.14706 [gr-qc]} \BibitemShut
  {NoStop}%
\bibitem [{\citenamefont {Klein}\ \emph {et~al.}(2016)\citenamefont {Klein}
  \emph {et~al.}}]{2016enrico}%
  \BibitemOpen
  \bibfield  {author} {\bibinfo {author} {\bibfnamefont {A.}~\bibnamefont
  {Klein}} \emph {et~al.},\ }\href {\doibase 10.1103/PhysRevD.93.024003}
  {\bibfield  {journal} {\bibinfo  {journal} {Phys. Rev. D}\ }\textbf {\bibinfo
  {volume} {93}},\ \bibinfo {pages} {024003} (\bibinfo {year} {2016})},\
  \Eprint {http://arxiv.org/abs/1511.05581} {arXiv:1511.05581 [gr-qc]}
  \BibitemShut {NoStop}%
\bibitem [{\citenamefont {Feng}\ \emph {et~al.}(2019)\citenamefont {Feng},
  \citenamefont {Wang}, \citenamefont {Hu}, \citenamefont {Hu},\ and\
  \citenamefont {Wang}}]{2019Feng}%
  \BibitemOpen
  \bibfield  {author} {\bibinfo {author} {\bibfnamefont {W.-F.}\ \bibnamefont
  {Feng}}, \bibinfo {author} {\bibfnamefont {H.-T.}\ \bibnamefont {Wang}},
  \bibinfo {author} {\bibfnamefont {X.-C.}\ \bibnamefont {Hu}}, \bibinfo
  {author} {\bibfnamefont {Y.-M.}\ \bibnamefont {Hu}}, \ and\ \bibinfo {author}
  {\bibfnamefont {Y.}~\bibnamefont {Wang}},\ }\href {\doibase
  10.1103/PhysRevD.99.123002} {\bibfield  {journal} {\bibinfo  {journal} {Phys.
  Rev. D}\ }\textbf {\bibinfo {volume} {99}},\ \bibinfo {pages} {123002}
  (\bibinfo {year} {2019})},\ \Eprint {http://arxiv.org/abs/1901.02159}
  {arXiv:1901.02159 [astro-ph.IM]} \BibitemShut {NoStop}%
\bibitem [{\citenamefont {Wang}\ \emph {et~al.}(2019)\citenamefont {Wang} \emph
  {et~al.}}]{2019Wang}%
  \BibitemOpen
  \bibfield  {author} {\bibinfo {author} {\bibfnamefont {H.-T.}\ \bibnamefont
  {Wang}} \emph {et~al.},\ }\href {\doibase 10.1103/PhysRevD.100.043003}
  {\bibfield  {journal} {\bibinfo  {journal} {Phys. Rev. D}\ }\textbf {\bibinfo
  {volume} {100}},\ \bibinfo {pages} {043003} (\bibinfo {year} {2019})},\
  \Eprint {http://arxiv.org/abs/1902.04423} {arXiv:1902.04423 [astro-ph.HE]}
  \BibitemShut {NoStop}%
\bibitem [{\citenamefont {Babak}\ \emph {et~al.}(2017)\citenamefont {Babak},
  \citenamefont {Gair}, \citenamefont {Sesana}, \citenamefont {Barausse},
  \citenamefont {Sopuerta}, \citenamefont {Berry}, \citenamefont {Berti},
  \citenamefont {Amaro-Seoane}, \citenamefont {Petiteau},\ and\ \citenamefont
  {Klein}}]{2017Babak}%
  \BibitemOpen
  \bibfield  {author} {\bibinfo {author} {\bibfnamefont {S.}~\bibnamefont
  {Babak}}, \bibinfo {author} {\bibfnamefont {J.}~\bibnamefont {Gair}},
  \bibinfo {author} {\bibfnamefont {A.}~\bibnamefont {Sesana}}, \bibinfo
  {author} {\bibfnamefont {E.}~\bibnamefont {Barausse}}, \bibinfo {author}
  {\bibfnamefont {C.~F.}\ \bibnamefont {Sopuerta}}, \bibinfo {author}
  {\bibfnamefont {C.~P.}\ \bibnamefont {Berry}}, \bibinfo {author}
  {\bibfnamefont {E.}~\bibnamefont {Berti}}, \bibinfo {author} {\bibfnamefont
  {P.}~\bibnamefont {Amaro-Seoane}}, \bibinfo {author} {\bibfnamefont
  {A.}~\bibnamefont {Petiteau}}, \ and\ \bibinfo {author} {\bibfnamefont
  {A.}~\bibnamefont {Klein}},\ }\href {\doibase 10.1103/PhysRevD.95.103012}
  {\bibfield  {journal} {\bibinfo  {journal} {Phys. Rev. D}\ }\textbf {\bibinfo
  {volume} {95}},\ \bibinfo {pages} {103012} (\bibinfo {year} {2017})},\
  \Eprint {http://arxiv.org/abs/1703.09722} {arXiv:1703.09722 [gr-qc]}
  \BibitemShut {NoStop}%
\bibitem [{\citenamefont {Fan}\ \emph {et~al.}(2020)\citenamefont {Fan},
  \citenamefont {Hu}, \citenamefont {Barausse}, \citenamefont {Sesana},
  \citenamefont {Zhang}, \citenamefont {Zhang}, \citenamefont {Zi},\ and\
  \citenamefont {Mei}}]{2020Fan}%
  \BibitemOpen
  \bibfield  {author} {\bibinfo {author} {\bibfnamefont {H.-M.}\ \bibnamefont
  {Fan}}, \bibinfo {author} {\bibfnamefont {Y.-M.}\ \bibnamefont {Hu}},
  \bibinfo {author} {\bibfnamefont {E.}~\bibnamefont {Barausse}}, \bibinfo
  {author} {\bibfnamefont {A.}~\bibnamefont {Sesana}}, \bibinfo {author}
  {\bibfnamefont {J.-d.}\ \bibnamefont {Zhang}}, \bibinfo {author}
  {\bibfnamefont {X.}~\bibnamefont {Zhang}}, \bibinfo {author} {\bibfnamefont
  {T.-G.}\ \bibnamefont {Zi}}, \ and\ \bibinfo {author} {\bibfnamefont
  {J.}~\bibnamefont {Mei}},\ }\href {\doibase 10.1103/PhysRevD.102.063016}
  {\bibfield  {journal} {\bibinfo  {journal} {Phys. Rev. D}\ }\textbf {\bibinfo
  {volume} {102}},\ \bibinfo {pages} {063016} (\bibinfo {year} {2020})},\
  \Eprint {http://arxiv.org/abs/2005.08212} {arXiv:2005.08212 [astro-ph.HE]}
  \BibitemShut {NoStop}%
\bibitem [{\citenamefont {Robson}\ \emph {et~al.}(2018)\citenamefont {Robson},
  \citenamefont {Cornish}, \citenamefont {Tamanini},\ and\ \citenamefont
  {Toonen}}]{2018Cornish2}%
  \BibitemOpen
  \bibfield  {author} {\bibinfo {author} {\bibfnamefont {T.}~\bibnamefont
  {Robson}}, \bibinfo {author} {\bibfnamefont {N.~J.}\ \bibnamefont {Cornish}},
  \bibinfo {author} {\bibfnamefont {N.}~\bibnamefont {Tamanini}}, \ and\
  \bibinfo {author} {\bibfnamefont {S.}~\bibnamefont {Toonen}},\ }\href
  {\doibase 10.1103/PhysRevD.98.064012} {\bibfield  {journal} {\bibinfo
  {journal} {Phys. Rev. D}\ }\textbf {\bibinfo {volume} {98}},\ \bibinfo
  {pages} {064012} (\bibinfo {year} {2018})},\ \Eprint
  {http://arxiv.org/abs/1806.00500} {arXiv:1806.00500 [gr-qc]} \BibitemShut
  {NoStop}%
\bibitem [{\citenamefont {Lau}\ \emph {et~al.}(2020)\citenamefont {Lau},
  \citenamefont {Mandel}, \citenamefont {Vigna-G\'omez}, \citenamefont
  {Neijssel}, \citenamefont {Stevenson},\ and\ \citenamefont
  {Sesana}}]{2019Sesana}%
  \BibitemOpen
  \bibfield  {author} {\bibinfo {author} {\bibfnamefont {M.~Y.}\ \bibnamefont
  {Lau}}, \bibinfo {author} {\bibfnamefont {I.}~\bibnamefont {Mandel}},
  \bibinfo {author} {\bibfnamefont {A.}~\bibnamefont {Vigna-G\'omez}}, \bibinfo
  {author} {\bibfnamefont {C.~J.}\ \bibnamefont {Neijssel}}, \bibinfo {author}
  {\bibfnamefont {S.}~\bibnamefont {Stevenson}}, \ and\ \bibinfo {author}
  {\bibfnamefont {A.}~\bibnamefont {Sesana}},\ }\href {\doibase
  10.1093/mnras/staa002} {\bibfield  {journal} {\bibinfo  {journal} {Mon. Not.
  Roy. Astron. Soc.}\ }\textbf {\bibinfo {volume} {492}},\ \bibinfo {pages}
  {3061} (\bibinfo {year} {2020})},\ \Eprint {http://arxiv.org/abs/1910.12422}
  {arXiv:1910.12422 [astro-ph.HE]} \BibitemShut {NoStop}%
\bibitem [{\citenamefont {Liu}\ \emph {et~al.}(2020)\citenamefont {Liu},
  \citenamefont {Hu}, \citenamefont {Zhang},\ and\ \citenamefont
  {Mei}}]{2020Liu}%
  \BibitemOpen
  \bibfield  {author} {\bibinfo {author} {\bibfnamefont {S.}~\bibnamefont
  {Liu}}, \bibinfo {author} {\bibfnamefont {Y.-M.}\ \bibnamefont {Hu}},
  \bibinfo {author} {\bibfnamefont {J.-d.}\ \bibnamefont {Zhang}}, \ and\
  \bibinfo {author} {\bibfnamefont {J.}~\bibnamefont {Mei}},\ }\href {\doibase
  10.1103/PhysRevD.101.103027} {\bibfield  {journal} {\bibinfo  {journal}
  {Phys. Rev. D}\ }\textbf {\bibinfo {volume} {101}},\ \bibinfo {pages}
  {103027} (\bibinfo {year} {2020})},\ \Eprint
  {http://arxiv.org/abs/2004.14242} {arXiv:2004.14242 [astro-ph.HE]}
  \BibitemShut {NoStop}%
\bibitem [{\citenamefont {Huang}\ \emph {et~al.}(2020)\citenamefont {Huang},
  \citenamefont {Hu}, \citenamefont {Korol}, \citenamefont {Li}, \citenamefont
  {Liang}, \citenamefont {Lu}, \citenamefont {Wang}, \citenamefont {Yu},\ and\
  \citenamefont {Mei}}]{2020Huang}%
  \BibitemOpen
  \bibfield  {author} {\bibinfo {author} {\bibfnamefont {S.-J.}\ \bibnamefont
  {Huang}}, \bibinfo {author} {\bibfnamefont {Y.-M.}\ \bibnamefont {Hu}},
  \bibinfo {author} {\bibfnamefont {V.}~\bibnamefont {Korol}}, \bibinfo
  {author} {\bibfnamefont {P.-C.}\ \bibnamefont {Li}}, \bibinfo {author}
  {\bibfnamefont {Z.-C.}\ \bibnamefont {Liang}}, \bibinfo {author}
  {\bibfnamefont {Y.}~\bibnamefont {Lu}}, \bibinfo {author} {\bibfnamefont
  {H.-T.}\ \bibnamefont {Wang}}, \bibinfo {author} {\bibfnamefont
  {S.}~\bibnamefont {Yu}}, \ and\ \bibinfo {author} {\bibfnamefont
  {J.}~\bibnamefont {Mei}},\ }\href {\doibase 10.1103/PhysRevD.102.063021}
  {\bibfield  {journal} {\bibinfo  {journal} {Phys. Rev. D}\ }\textbf {\bibinfo
  {volume} {102}},\ \bibinfo {pages} {063021} (\bibinfo {year} {2020})},\
  \Eprint {http://arxiv.org/abs/2005.07889} {arXiv:2005.07889 [astro-ph.HE]}
  \BibitemShut {NoStop}%
\bibitem [{\citenamefont {Romano}\ and\ \citenamefont
  {Cornish}(2017)}]{2016Cornish}%
  \BibitemOpen
  \bibfield  {author} {\bibinfo {author} {\bibfnamefont {J.~D.}\ \bibnamefont
  {Romano}}\ and\ \bibinfo {author} {\bibfnamefont {N.~J.}\ \bibnamefont
  {Cornish}},\ }\href {\doibase 10.1007/s41114-017-0004-1} {\bibfield
  {journal} {\bibinfo  {journal} {Living Rev. Rel.}\ }\textbf {\bibinfo
  {volume} {20}},\ \bibinfo {pages} {2} (\bibinfo {year} {2017})},\ \Eprint
  {http://arxiv.org/abs/1608.06889} {arXiv:1608.06889 [gr-qc]} \BibitemShut
  {NoStop}%
\bibitem [{\citenamefont {Liang}\ \emph {et~al.}(2022)\citenamefont {Liang},
  \citenamefont {Hu}, \citenamefont {Jiang}, \citenamefont {Cheng},
  \citenamefont {Zhang},\ and\ \citenamefont {Mei}}]{2021Liang}%
  \BibitemOpen
  \bibfield  {author} {\bibinfo {author} {\bibfnamefont {Z.-C.}\ \bibnamefont
  {Liang}}, \bibinfo {author} {\bibfnamefont {Y.-M.}\ \bibnamefont {Hu}},
  \bibinfo {author} {\bibfnamefont {Y.}~\bibnamefont {Jiang}}, \bibinfo
  {author} {\bibfnamefont {J.}~\bibnamefont {Cheng}}, \bibinfo {author}
  {\bibfnamefont {J.-d.}\ \bibnamefont {Zhang}}, \ and\ \bibinfo {author}
  {\bibfnamefont {J.}~\bibnamefont {Mei}},\ }\href {\doibase
  10.1103/PhysRevD.105.022001} {\bibfield  {journal} {\bibinfo  {journal}
  {Phys. Rev. D}\ }\textbf {\bibinfo {volume} {105}},\ \bibinfo {pages}
  {022001} (\bibinfo {year} {2022})},\ \Eprint
  {http://arxiv.org/abs/2107.08643} {arXiv:2107.08643 [astro-ph.CO]}
  \BibitemShut {NoStop}%
\bibitem [{\citenamefont {Burdge}\ \emph {et~al.}(2019)\citenamefont {Burdge}
  \emph {et~al.}}]{2019Na}%
  \BibitemOpen
  \bibfield  {author} {\bibinfo {author} {\bibfnamefont {K.~B.}\ \bibnamefont
  {Burdge}} \emph {et~al.},\ }\href {\doibase 10.1038/s41586-019-1403-0}
  {\bibfield  {journal} {\bibinfo  {journal} {Nature}\ }\textbf {\bibinfo
  {volume} {571}},\ \bibinfo {pages} {528} (\bibinfo {year} {2019})},\ \Eprint
  {http://arxiv.org/abs/1907.11291} {arXiv:1907.11291 [astro-ph.SR]}
  \BibitemShut {NoStop}%
\bibitem [{\citenamefont {Will}(1994)}]{1994Will}%
  \BibitemOpen
  \bibfield  {author} {\bibinfo {author} {\bibfnamefont {C.~M.}\ \bibnamefont
  {Will}},\ }\href {\doibase 10.1103/PhysRevD.50.6058} {\bibfield  {journal}
  {\bibinfo  {journal} {Phys. Rev. D}\ }\textbf {\bibinfo {volume} {50}},\
  \bibinfo {pages} {6058} (\bibinfo {year} {1994})},\ \Eprint
  {http://arxiv.org/abs/gr-qc/9406022} {arXiv:gr-qc/9406022} \BibitemShut
  {NoStop}%
\bibitem [{\citenamefont {Sennett}\ \emph {et~al.}(2016)\citenamefont
  {Sennett}, \citenamefont {Marsat},\ and\ \citenamefont
  {Buonanno}}]{2016Sennett}%
  \BibitemOpen
  \bibfield  {author} {\bibinfo {author} {\bibfnamefont {N.}~\bibnamefont
  {Sennett}}, \bibinfo {author} {\bibfnamefont {S.}~\bibnamefont {Marsat}}, \
  and\ \bibinfo {author} {\bibfnamefont {A.}~\bibnamefont {Buonanno}},\ }\href
  {\doibase 10.1103/PhysRevD.94.084003} {\bibfield  {journal} {\bibinfo
  {journal} {Phys. Rev. D}\ }\textbf {\bibinfo {volume} {94}},\ \bibinfo
  {pages} {084003} (\bibinfo {year} {2016})},\ \Eprint
  {http://arxiv.org/abs/1607.01420} {arXiv:1607.01420 [gr-qc]} \BibitemShut
  {NoStop}%
\bibitem [{\citenamefont {Poisson}\ and\ \citenamefont
  {Will}(2014)}]{2014gravi}%
  \BibitemOpen
  \bibfield  {author} {\bibinfo {author} {\bibfnamefont {E.}~\bibnamefont
  {Poisson}}\ and\ \bibinfo {author} {\bibfnamefont {C.~M.}\ \bibnamefont
  {Will}},\ }\href@noop {} {\emph {\bibinfo {title} {Gravity: Newtonian,
  post-newtonian, relativistic}}}\ (\bibinfo  {publisher} {Cambridge University
  Press},\ \bibinfo {year} {2014})\BibitemShut {NoStop}%
\bibitem [{\citenamefont {Roelofs}\ \emph {et~al.}(2010)\citenamefont
  {Roelofs}, \citenamefont {Rau}, \citenamefont {Marsh}, \citenamefont
  {Steeghs}, \citenamefont {Groot},\ and\ \citenamefont
  {Nelemans}}]{2010Roelofs}%
  \BibitemOpen
  \bibfield  {author} {\bibinfo {author} {\bibfnamefont {G.~H.~A.}\
  \bibnamefont {Roelofs}}, \bibinfo {author} {\bibfnamefont {A.}~\bibnamefont
  {Rau}}, \bibinfo {author} {\bibfnamefont {T.~R.}\ \bibnamefont {Marsh}},
  \bibinfo {author} {\bibfnamefont {D.}~\bibnamefont {Steeghs}}, \bibinfo
  {author} {\bibfnamefont {P.~J.}\ \bibnamefont {Groot}}, \ and\ \bibinfo
  {author} {\bibfnamefont {G.}~\bibnamefont {Nelemans}},\ }\href {\doibase
  10.1088/2041-8205/711/2/L138} {\bibfield  {journal} {\bibinfo  {journal}
  {Astrophys. J. Lett.}\ }\textbf {\bibinfo {volume} {711}},\ \bibinfo {pages}
  {L138} (\bibinfo {year} {2010})},\ \Eprint {http://arxiv.org/abs/1003.0658}
  {arXiv:1003.0658 [astro-ph.SR]} \BibitemShut {NoStop}%
\bibitem [{\citenamefont {Peters}\ and\ \citenamefont
  {Mathews}(1963)}]{1963PM}%
  \BibitemOpen
  \bibfield  {author} {\bibinfo {author} {\bibfnamefont {P.~C.}\ \bibnamefont
  {Peters}}\ and\ \bibinfo {author} {\bibfnamefont {J.}~\bibnamefont
  {Mathews}},\ }\href {\doibase 10.1103/PhysRev.131.435} {\bibfield  {journal}
  {\bibinfo  {journal} {Phys. Rev.}\ }\textbf {\bibinfo {volume} {131}},\
  \bibinfo {pages} {435} (\bibinfo {year} {1963})}\BibitemShut {NoStop}%
\bibitem [{\citenamefont {Nelemans}\ \emph {et~al.}(2001)\citenamefont
  {Nelemans}, \citenamefont {Yungelson},\ and\ \citenamefont
  {Portegies~Zwart}}]{2001Nelemans}%
  \BibitemOpen
  \bibfield  {author} {\bibinfo {author} {\bibfnamefont {G.}~\bibnamefont
  {Nelemans}}, \bibinfo {author} {\bibfnamefont {L.~R.}\ \bibnamefont
  {Yungelson}}, \ and\ \bibinfo {author} {\bibfnamefont {S.~F.}\ \bibnamefont
  {Portegies~Zwart}},\ }\href {\doibase 10.1051/0004-6361:20010683} {\bibfield
  {journal} {\bibinfo  {journal} {Astron. Astrophys.}\ }\textbf {\bibinfo
  {volume} {375}},\ \bibinfo {pages} {890} (\bibinfo {year} {2001})},\ \Eprint
  {http://arxiv.org/abs/astro-ph/0105221} {arXiv:astro-ph/0105221} \BibitemShut
  {NoStop}%
\bibitem [{\citenamefont {Zhang}\ \emph {et~al.}(2019)\citenamefont {Zhang},
  \citenamefont {Gao}, \citenamefont {Gong}, \citenamefont {Liang},
  \citenamefont {Weinstein},\ and\ \citenamefont {Zhang}}]{2019Zhang}%
  \BibitemOpen
  \bibfield  {author} {\bibinfo {author} {\bibfnamefont {C.}~\bibnamefont
  {Zhang}}, \bibinfo {author} {\bibfnamefont {Q.}~\bibnamefont {Gao}}, \bibinfo
  {author} {\bibfnamefont {Y.}~\bibnamefont {Gong}}, \bibinfo {author}
  {\bibfnamefont {D.}~\bibnamefont {Liang}}, \bibinfo {author} {\bibfnamefont
  {A.~J.}\ \bibnamefont {Weinstein}}, \ and\ \bibinfo {author} {\bibfnamefont
  {C.}~\bibnamefont {Zhang}},\ }\href {\doibase 10.1103/PhysRevD.100.064033}
  {\bibfield  {journal} {\bibinfo  {journal} {Phys. Rev. D}\ }\textbf {\bibinfo
  {volume} {100}},\ \bibinfo {pages} {064033} (\bibinfo {year} {2019})},\
  \Eprint {http://arxiv.org/abs/1906.10901} {arXiv:1906.10901 [gr-qc]}
  \BibitemShut {NoStop}%
\bibitem [{\citenamefont {Hu}\ \emph {et~al.}(2018)\citenamefont {Hu},
  \citenamefont {Li}, \citenamefont {Wang}, \citenamefont {Feng}, \citenamefont
  {Zhou}, \citenamefont {Hu}, \citenamefont {Hu}, \citenamefont {Mei},\ and\
  \citenamefont {Shao}}]{2018Hu}%
  \BibitemOpen
  \bibfield  {author} {\bibinfo {author} {\bibfnamefont {X.-C.}\ \bibnamefont
  {Hu}}, \bibinfo {author} {\bibfnamefont {X.-H.}\ \bibnamefont {Li}}, \bibinfo
  {author} {\bibfnamefont {Y.}~\bibnamefont {Wang}}, \bibinfo {author}
  {\bibfnamefont {W.-F.}\ \bibnamefont {Feng}}, \bibinfo {author}
  {\bibfnamefont {M.-Y.}\ \bibnamefont {Zhou}}, \bibinfo {author}
  {\bibfnamefont {Y.-M.}\ \bibnamefont {Hu}}, \bibinfo {author} {\bibfnamefont
  {S.-C.}\ \bibnamefont {Hu}}, \bibinfo {author} {\bibfnamefont {J.-W.}\
  \bibnamefont {Mei}}, \ and\ \bibinfo {author} {\bibfnamefont {C.-G.}\
  \bibnamefont {Shao}},\ }\href {\doibase 10.1088/1361-6382/aab52f} {\bibfield
  {journal} {\bibinfo  {journal} {Class. Quant. Grav.}\ }\textbf {\bibinfo
  {volume} {35}},\ \bibinfo {pages} {095008} (\bibinfo {year} {2018})},\
  \Eprint {http://arxiv.org/abs/1803.03368} {arXiv:1803.03368 [gr-qc]}
  \BibitemShut {NoStop}%
\bibitem [{\citenamefont {Zhang}\ \emph {et~al.}(2020)\citenamefont {Zhang},
  \citenamefont {Gao}, \citenamefont {Gong}, \citenamefont {Wang},
  \citenamefont {Weinstein},\ and\ \citenamefont {Zhang}}]{2020Zhang}%
  \BibitemOpen
  \bibfield  {author} {\bibinfo {author} {\bibfnamefont {C.}~\bibnamefont
  {Zhang}}, \bibinfo {author} {\bibfnamefont {Q.}~\bibnamefont {Gao}}, \bibinfo
  {author} {\bibfnamefont {Y.}~\bibnamefont {Gong}}, \bibinfo {author}
  {\bibfnamefont {B.}~\bibnamefont {Wang}}, \bibinfo {author} {\bibfnamefont
  {A.~J.}\ \bibnamefont {Weinstein}}, \ and\ \bibinfo {author} {\bibfnamefont
  {C.}~\bibnamefont {Zhang}},\ }\href {\doibase 10.1103/PhysRevD.101.124027}
  {\bibfield  {journal} {\bibinfo  {journal} {Phys. Rev. D}\ }\textbf {\bibinfo
  {volume} {101}},\ \bibinfo {pages} {124027} (\bibinfo {year} {2020})},\
  \Eprint {http://arxiv.org/abs/2003.01441} {arXiv:2003.01441 [gr-qc]}
  \BibitemShut {NoStop}%
\bibitem [{\citenamefont {Rubbo}\ \emph {et~al.}(2004)\citenamefont {Rubbo},
  \citenamefont {Cornish},\ and\ \citenamefont {Poujade}}]{2004Rubbo}%
  \BibitemOpen
  \bibfield  {author} {\bibinfo {author} {\bibfnamefont {L.~J.}\ \bibnamefont
  {Rubbo}}, \bibinfo {author} {\bibfnamefont {N.~J.}\ \bibnamefont {Cornish}},
  \ and\ \bibinfo {author} {\bibfnamefont {O.}~\bibnamefont {Poujade}},\ }\href
  {\doibase 10.1103/PhysRevD.69.082003} {\bibfield  {journal} {\bibinfo
  {journal} {Phys. Rev. D}\ }\textbf {\bibinfo {volume} {69}},\ \bibinfo
  {pages} {082003} (\bibinfo {year} {2004})},\ \Eprint
  {http://arxiv.org/abs/gr-qc/0311069} {arXiv:gr-qc/0311069} \BibitemShut
  {NoStop}%
\bibitem [{\citenamefont {Vallisneri}(2008)}]{2007FIM}%
  \BibitemOpen
  \bibfield  {author} {\bibinfo {author} {\bibfnamefont {M.}~\bibnamefont
  {Vallisneri}},\ }\href {\doibase 10.1103/PhysRevD.77.042001} {\bibfield
  {journal} {\bibinfo  {journal} {Phys. Rev. D}\ }\textbf {\bibinfo {volume}
  {77}},\ \bibinfo {pages} {042001} (\bibinfo {year} {2008})},\ \Eprint
  {http://arxiv.org/abs/gr-qc/0703086} {arXiv:gr-qc/0703086} \BibitemShut
  {NoStop}%
\bibitem [{\citenamefont {Finn}(1992)}]{1992Finn}%
  \BibitemOpen
  \bibfield  {author} {\bibinfo {author} {\bibfnamefont {L.~S.}\ \bibnamefont
  {Finn}},\ }\href {\doibase 10.1103/PhysRevD.46.5236} {\bibfield  {journal}
  {\bibinfo  {journal} {Phys. Rev. D}\ }\textbf {\bibinfo {volume} {46}},\
  \bibinfo {pages} {5236} (\bibinfo {year} {1992})},\ \Eprint
  {http://arxiv.org/abs/gr-qc/9209010} {arXiv:gr-qc/9209010} \BibitemShut
  {NoStop}%
\bibitem [{\citenamefont {Cutler}\ and\ \citenamefont
  {Flanagan}(1994)}]{1994Cutler}%
  \BibitemOpen
  \bibfield  {author} {\bibinfo {author} {\bibfnamefont {C.}~\bibnamefont
  {Cutler}}\ and\ \bibinfo {author} {\bibfnamefont {E.~E.}\ \bibnamefont
  {Flanagan}},\ }\href {\doibase 10.1103/PhysRevD.49.2658} {\bibfield
  {journal} {\bibinfo  {journal} {Phys. Rev. D}\ }\textbf {\bibinfo {volume}
  {49}},\ \bibinfo {pages} {2658} (\bibinfo {year} {1994})},\ \Eprint
  {http://arxiv.org/abs/gr-qc/9402014} {arXiv:gr-qc/9402014} \BibitemShut
  {NoStop}%
\bibitem [{\citenamefont {Robson}\ \emph {et~al.}(2019)\citenamefont {Robson},
  \citenamefont {Cornish},\ and\ \citenamefont {Liu}}]{2018Cornish}%
  \BibitemOpen
  \bibfield  {author} {\bibinfo {author} {\bibfnamefont {T.}~\bibnamefont
  {Robson}}, \bibinfo {author} {\bibfnamefont {N.~J.}\ \bibnamefont {Cornish}},
  \ and\ \bibinfo {author} {\bibfnamefont {C.}~\bibnamefont {Liu}},\ }\href
  {\doibase 10.1088/1361-6382/ab1101} {\bibfield  {journal} {\bibinfo
  {journal} {Class. Quant. Grav.}\ }\textbf {\bibinfo {volume} {36}},\ \bibinfo
  {pages} {105011} (\bibinfo {year} {2019})},\ \Eprint
  {http://arxiv.org/abs/1803.01944} {arXiv:1803.01944 [astro-ph.HE]}
  \BibitemShut {NoStop}%
\bibitem [{\citenamefont {Baumgartner}\ \emph {et~al.}(2020)\citenamefont
  {Baumgartner}, \citenamefont {Bernardini}, \citenamefont {Canivete~Cuissa},
  \citenamefont {de~Laroussilhe}, \citenamefont {Mitchell}, \citenamefont
  {Neuenschwander}, \citenamefont {Saha}, \citenamefont {Schaeffer},
  \citenamefont {Soyuer},\ and\ \citenamefont {Zwick}}]{2020Mit}%
  \BibitemOpen
  \bibfield  {author} {\bibinfo {author} {\bibfnamefont {S.}~\bibnamefont
  {Baumgartner}}, \bibinfo {author} {\bibfnamefont {M.}~\bibnamefont
  {Bernardini}}, \bibinfo {author} {\bibfnamefont {J.~R.}\ \bibnamefont
  {Canivete~Cuissa}}, \bibinfo {author} {\bibfnamefont {H.}~\bibnamefont
  {de~Laroussilhe}}, \bibinfo {author} {\bibfnamefont {A.~M.~W.}\ \bibnamefont
  {Mitchell}}, \bibinfo {author} {\bibfnamefont {B.~A.}\ \bibnamefont
  {Neuenschwander}}, \bibinfo {author} {\bibfnamefont {P.}~\bibnamefont
  {Saha}}, \bibinfo {author} {\bibfnamefont {T.}~\bibnamefont {Schaeffer}},
  \bibinfo {author} {\bibfnamefont {D.}~\bibnamefont {Soyuer}}, \ and\ \bibinfo
  {author} {\bibfnamefont {L.}~\bibnamefont {Zwick}},\ }\href {\doibase
  10.1093/mnras/staa2638} {\bibfield  {journal} {\bibinfo  {journal} {Mon. Not.
  Roy. Astron. Soc.}\ }\textbf {\bibinfo {volume} {498}},\ \bibinfo {pages}
  {4577} (\bibinfo {year} {2020})},\ \Eprint {http://arxiv.org/abs/2008.11538}
  {arXiv:2008.11538 [astro-ph.IM]} \BibitemShut {NoStop}%
\end{thebibliography}%

\end{document}